\documentclass[a4paper,11pt]{article}
\pdfoutput=1 

\usepackage{jheppub} 
\usepackage{tikz}
\usepackage{tikz-cd}

\newcommand{\bea}{\begin{eqnarray}}
\newcommand{\eea}{\end{eqnarray}}

\usepackage[utf8]{inputenc}
\usepackage{hyperref}
\usepackage{amsfonts}
\usepackage{amsmath}
\usepackage{amssymb}
\usepackage{mathrsfs}  
\usepackage{color}
\usepackage{physics}
\usepackage{multicol}
\usepackage{tikz}

\usepackage{amsmath,bbm,array,amsfonts,graphicx,wrapfig,lscape,float,mathtools,multirow,longtable,amsthm}
\usepackage{stackrel}
\usepackage[all]{xy}
\usepackage{caption}
\usepackage{subcaption}
\usepackage{multirow}
\usepackage[bottom]{footmisc}
\usepackage{mathrsfs}
\usepackage{tikz}
\usepackage{bm}
\usepackage{color}
\usepackage{enumerate}
\usetikzlibrary{decorations.pathreplacing}
\usepackage{diagbox}
\numberwithin{equation}{section}
\numberwithin{figure}{section}
\numberwithin{table}{section}
\usepackage{pgfplots}
\pgfplotsset{compat=1.14}
\usepackage{makecell}
\usepackage{lscape}
\usepackage[utf8]{inputenc}
\usepackage{mathtools}
\usepackage{todonotes}
\usepackage{epigraph}
\definecolor{bananayellow}{rgb}{1.0, 0.88, 0.21}

\usepackage[autostyle=true]{csquotes}

\usepackage[toc,page]{appendix}
\usepackage[inline]{enumitem}

\newcommand{\commentout}[1]{}

\title{Crystal Melting, BPS Quivers and Plethystics}

\author[a,b]{Jiakang Bao,}
\author[b,a,c,d]{Yang-Hui He,}
\author[e,b]{Ali Zahabi }

	\affiliation[a]{
		Department of Mathematics, City, University of London, EC1V 0HB, UK}
	\affiliation[b]{
	    London Institute for Mathematical Sciences, Royal Institution, London W1S 4BS, UK}
	\affiliation[c]{
		Merton College, University of Oxford, OX1 4JD, UK}
	\affiliation[d]{
		School of Physics, NanKai University, Tianjin, 300071, P.R. China}

    \affiliation[e]{Institut de Math\'ematiques de Bourgogne, Universit\'e Bourgogne Franche-Comt\'e, France}

	\emailAdd{jiakang.bao@city.ac.uk}
	\emailAdd{hey@maths.ox.ac.uk}
    \emailAdd{zahabi.ali@gmail.com}

\preprint{
		\begin{flushright}
			LIMS-2022-010
		\end{flushright}
	}

\abstract{We study the refined and unrefined crystal/BPS partition functions of D6-D2-D0 brane bound states for all toric Calabi-Yau threefolds without compact 4-cycles and some non-toric examples. They can be written as products of (generalized) MacMahon functions. We check our expressions and use them as vacuum characters to study the gluings. We then consider the wall crossings and discuss possible crystal descriptions for different chambers. We also express the partition functions in terms of plethystic exponentials. For $\mathbb{C}^3$ and tripled affine quivers, we find their connections to nilpotent Kac polynomials. Similarly, the partition functions of D4-D2-D0 brane bound states can be obtained by replacing the (generalized) MacMahon functions with the inverse of (generalized) Euler functions.
}

\begin{document} 
\maketitle

\section{Introduction and Summary}\label{intro}
Studying the BPS spectrum \cite{Bogomolny:1975de,Prasad:1975kr} of particles has been an important topic in quantum field theory and string theory. Although there is little known for the case of compact Calabi-Yau (CY) manifolds, the techniques have been greatly developed in the context of non-compact, or local, CYs, especially when they afford a toric description.
As the lattice polygons nicely encode combinatorial information from the toric CY threefolds, crystal melting \cite{Okounkov:2003sp,Iqbal:2003ds,Ooguri:2009ijd,Yamazaki:2010fz,Dimofte:2010wxa} and quivers \cite{Nakajima:1994nid,Douglas:1996sw} have become extremely useful tools in BPS counting.

Mathematically, BPS counting has a close relation with Donaldson-Thomas (DT) invariants\footnote{In the usual canonical crystal melting setting, we are working in the non-commutative DT (NCDT) chamber.}, and are hence also connected to Gromov-Witten and many other geometric invariants. Going one step further, we would also like to understand more about the Hilbert space of the BPS states, which can be recast as the cohomology of chain complexes. This then leads to the categorification of BPS indices and wall crossings \cite{Kontsevich:2008fj,kiem2012categorification,Gaiotto:2015aoa,Gaiotto:2015zna}. Although we will not discuss such categorification in this paper, they should be intimately related to the algebraic structure of BPS states.

BPS algebras have been of great interest since \cite{Hanany:2005ve}. In particular, cohomological Hall algebras (COHAs) were introduced in \cite{Kontsevich:2010px} as a mathematical description for the BPS algebras. The study of quiver quantum mechanics and relevant quantum algebras has now become an active area. For instance, with the utility of crystal melting, we recently have a better understanding on the quiver Yangians $\mathtt{Y}$ \cite{Li:2020rij}. Given a quiver Yangian, the character of its vacuum module is precisely the BPS partition function for the corresponding CY. One can then translate it into the crystal generating function for the associated 3-dimensional partition.

In this paper, we study both the refined and unrefined expressions for those partition functions by speculating on their patterns for all the toric CYs without compact 4-cycles as well as (tripled) quivers from affine type (including non-toric ones). For toric cases, their toric diagrams are lattice polygons without internal points. It is then clear that they include generalized conifolds which are trapezia (including triangles) of height one plus an exceptional triangle $\mathbb{C}^3/(\mathbb{Z}_2\times\mathbb{Z}_2)$ (we will draw these explicitly in \S\ref{examples}). The crystal/BPS partition functions for triangles and the conifold have been obtained in the literature such as \cite{Szendroi:2007nu,Young:2008hn,Cirafici:2010bd,Cirafici:2012qc}. The other examples can also be obtained from topological strings following \cite{Aganagic:2003db,Iqbal:2004ne}. They were also studied in \cite{Mozgovoy:2020has,mozgovoy2021donaldson} recently. One may check that our expressions agree with these results. All of them can be expressed using (generalized) MacMahon functions\footnote{For refined partition functions, we will use refined (generalized) MacMahon functions as in \S\ref{refined}.}:
\begin{equation}
    M(p,q):=\prod_{k=1}^\infty\frac{1}{(1-pq^k)^k},\quad M(q):=M(1,q),\quad\widetilde{M}(p,q):=M(p,q)M(p^{-1},q).
\end{equation}
In the above, $M(q)$ is the standard MacMahon function \cite{macmahon2001combinatory}. We can then also use these expressions to study the gluing process beyond two trivalent vertices in the web diagram and identify the bosonic and fermionic generators.

When studying wall crossings, it is convenient to introduce the shorthand notation
\begin{equation}
    M_\wedge(p,q;k_0):=\prod_{k=k_0}^\infty\frac{1}{(1-pq^k)^k},\quad  M^\wedge(p,q;k_0):=\prod_{k=1}^{k_0}\frac{1}{(1-pq^k)^k}
\end{equation}
as the truncated MacMahon functions from below and above. For different chambers separated by the walls of marginal stability, we shall discuss their possible crystal descriptions. For chambers $\widetilde{C}$ described by $M^\wedge$, the model could be constructed by combining a union of (sub-)crystals. For chambers $C$ described by $M_\wedge$, the model could be constructed by peeling semi-infinite faces off the crystal.

We will also write these generating functions in terms of plethystic exponential\footnote{The reader is also referred to \cite{Benvenuti:2006qr,Feng:2007ur} for a plethystic programme for counting BPS operators in quiver gauge theories, though we emphasize that what we study here is in a different context. For further disscussions on PE, see \cite{fulton2013representation,florentino2021plethystic}.} (PE) of a multi-variable analytic function $f(t_1,\dots,t_r)$:
\begin{equation}
    \text{PE}[f(t_1,\dots,t_r)]=\exp\left(\sum_{k=1}^\infty\frac{f(t_1^k,\dots,t_r^k)-f(0,\dots,0)}{k}\right) \ .
\end{equation}
As the PE computes the character of the symmetric algebra, this indicates that the quiver Yangians are symmetric algebras. They can then be endowed with Hopf algebra structures as one may expect.

For some cases, namely the $\mathbb{C}^3$ and tripled affine quivers, we shall also discuss the PE expressions in the context of (nilpotent) Kac polynomials \cite{kac1980infinite} and consider the connections to different quantum algebras. More specifically, for $\mathbb{C}^3$, the partition function agrees with the Poincar\'e polynomial encoded by Kac polynomials for some nilpotent (sub)stack. For (tripled) affine quiver cases, the double of such Poincar\'e polynomial contains the partition function as a factor, and it seems that there exists some subalgebra structure. All these will be checked for both unrefined and refined expressions. It could be possible that the other cases may as well have certain interpretations in their PE expressions.

The above discussions may be summarized schematically as
\[
\mbox{
    \tikzset{every picture/.style={line width=0.75pt}} 
\begin{tikzpicture}[x=0.75pt,y=0.75pt,yscale=-1,xscale=1]
\draw   (115.3,79.45) -- (261.9,79.45) -- (261.9,159.45) -- (115.3,159.45) -- cycle ;
\draw    (261.5,110.5) -- (316.3,110.35) ;
\draw   (317,80.95) -- (517.7,80.95) -- (517.7,139.7) -- (317,139.7) -- cycle ;
\draw    (189.4,159.2) -- (189,174.95) ;
\draw   (115.2,175.45) -- (407.3,175.45) -- (407.3,249.95) -- (115.2,249.95) -- cycle ;
\draw (189.45,124.6) node    {$Z_{\text{BPS/crystal}} =\chi _{\text{vac(\texttt{Y})}}$};
\draw (419.64,124.2) node    {$\text{Poincar\'e} =\text{PE}\left[\text{(nilp.) Kac}\right]$};
\draw (189.64,97.7) node   [align=left] {Quiver $\mathcal{Q}_A$};
\draw (410.14,96.2) node   [align=left] {Quiver $\mathcal{Q}_B$};
\draw (186.72,148.7) node   [align=left] {$=$ ``gluing of $\mathcal{W}_{1+\infty}$''};
\draw (261.93,220.7) node   [align=left] {$\widetilde{C}$: gluing/merging crystals;\\$C$: peeling semi-infinite faces off the crystal};
\draw (210.8,180.7) node [anchor=north west][inner sep=0.75pt]   [align=left] {Wall crossings:};
\draw (263,90) node [anchor=north west][inner sep=0.75pt]  [font=\tiny] [align=left] {{\tiny $\quad\mathcal{Q}_A=$} \\ {\tiny Tripled $\mathcal{Q}_B$}};
\draw (270,115) node [anchor=north west][inner sep=0.75pt]  [font=\tiny] [align=left] {{\tiny Any $\mathcal{Q}_A$:} \\ {\tiny Unkown}};
\end{tikzpicture}
}.
\]

The paper is organized as follows. In \S\ref{yangian}, we give a brief review on crystal melting and quiver Yangians. In \S\ref{examples}, we discuss various implications of the partition functions for all toric CY$_3$ without compact 4-cycles and some non-toric cases (DE singularities).
We study the wall crossing phenomena and their crystals in \S\ref{wallcrossing}, along with the refinement of partition functions. Similar results are also mentioned in \S\ref{D4D2D0} for D4-D2-D0 bound states. In \S\ref{outlook}, we mention a few future directions.

\section{Quiver Yangians and Related Concepts}\label{yangian}
We start with a toric diagram $\mathfrak{D}$, which for us is a convex polygon with all its vertices on the lattice $\mathbb{Z}^2$.
From this we can construct a non-compact, or local, Calabi-Yau 3-fold, CY$_3$.
We can think of  CY$_3$ as an affine complex cone over a base, compact, toric surface whose toric fan is given by a star triangulation of $\mathfrak{D}$.
This cone is in general singular and is called a Gorenstein singularity. 
Our CY$_3$ is toric, so the lattice polygon $\mathfrak{D}$ encodes certain combinatorial-geometric information. For instance, the lattice points in the polygon correspond to the divisors (of complex codimension 1). In particular, internal lattice points represent compact 4-cycles while boundary points give non-compact ones.

\subsection{Crystal Melting}\label{crystal}
For type IIA string theory compactified on a general toric CY$_3$, the BPS states are the bound states formed by D$p$-branes wrapping holomorphic $p$-cycles therein. Here, we shall focus on the following setting: (i) a single D6 wrapping the whole CY$_3$; (ii) D0-branes supported on points which are trivially compact in the CY; (iii) D2-/D4-branes wrapping either compact or non-compact 2-/4-cycles. The compact D-branes are then light BPS particles that are dynamical. In contrast, non-compact D-branes are heavy line operators which become non-dynamical in our compactified theory. As we are considering toric diagrams without internal points in this paper, we will count the D2 and D0 states bound to a single D6.

As the dimensional reduction from 4d $\mathcal{N}=1$ gauge theory, the effective supersymmetric quantum mechanics on the D-branes is a quiver theory. A quiver $\mathcal{Q}$ is a graph $(\mathcal{Q}_0,\mathcal{Q}_1)$ with $\mathcal{Q}_0$ denoting the set of nodes and $\mathcal{Q}_1$ its edges. In particular, the edges $X_{ab}$ are all oriented here, emanating from node $a$ and ending at node $b$. Each quiver also has an associated superpotential $W$. For toric CYs, the superpotential is fully determined. The general algorithm involves the technique of brane tilings (aka dimer models). See \cite{Hanany:2005ve,Franco:2005rj,Franco:2005sm,Feng:2005gw,Yamazaki:2008bt} for details.

The brane tiling is the dual graph of the quiver on the 2-torus. As a result, the quiver is also periodic. The crystal model can then be thought of as a 3-dimensional uplift of the periodic quiver, where each atom in the crystal corresponds to a gauge node $a$ in the quiver while the arrows are the chemical bonds. Remarkably, BPS states can be constructed by removing atoms from the crystal model. More precisely, each molten crystal configuration corresponds to a BPS state.

In the crystal, the atoms from different gauge nodes are of different ``colours''. They correspond to D2s stretched between NS5-branes in different regions on the tiling. To construct the crystal, we shall choose an initial atom $\mathfrak{o}$ in the periodic quiver. Then all the other atoms are placed at the nodes in the periodic quiver level by level following the arrows/chemical bonds. Any path from $\mathfrak{o}$ to an atom $\mathfrak{a}$ is of form $p_{\mathfrak{oa}}\omega^n$ modulo F-term relations $\partial W/\partial X_{ab}=0$, where $p_{\mathfrak{oa}}$ is one of the shortest paths from $\mathfrak{o}$ to $\mathfrak{a}$ and $\omega$ is a loop along any face in the periodic diagram \cite{mozgovoy2010noncommutative}. Then the atom $\mathfrak{a}$ is placed at level $n$ in the crystal. Clear illustrations can be found in \cite[Figure 5 and 6]{Li:2020rij}. Mathematically, the F-term relations form an ideal of $\mathbb{C}\mathcal{Q}$, and hence define the path algebra $\mathbb{C}\mathcal{Q}/\langle\partial W\rangle$.

The BPS states can then be obtained by the crystal melting rule, which states that an atom $\mathfrak{j}$ is in the molten crystal $\mathfrak{C}$ (i.e., removed from the initial complete crystal) if there exists an arrow $X$ such that $X\cdot\mathfrak{j}\in\mathfrak{C}$. This means that the complement of $\mathfrak{C}$ is an ideal in the path algebra.

We can then write the crystal generating function to enumerate the possible configurations:
\begin{equation}
    Z_\text{crystal}(q_j)=\sum_{\mathfrak{C}}\prod_{j\in\mathcal{Q}_0}q_j^{|\mathfrak{C}(j)|},
\end{equation}
where $|\mathfrak{C}(j)|$ denotes the number of atoms with colour $j$ in $\mathfrak{C}$. For BPS states counting, we have the BPS partition function
\begin{equation}
    Z_\text{BPS}(q,\bm{Q})=\sum_{n_0,\bm{n_2}}\Omega(n_0,\bm{n_2})q^{n_0}\prod_{i=1}^{|\mathcal{Q}_0|-1}Q_i^{n_{2,i}},
\end{equation}
where $\Omega$ is the Witten index for the bound states of $n_0$ D0s and $\bm{n_2}$ D2s inside a single non-compact D6 with $n_{2,i}$ the number of D2's wrapping the $i^\text{th}$ 2-cycle. Note that $n_0\in\mathbb{Z}_{\geq0}$ is a non-negative integer and $\bm{n_2}=(n_{2,i})\in\mathbb{Z}_{\geq0}^{|\mathcal{Q}_0|-1}$ is a vector, where $|\mathcal{Q}_0|-1$ is the number of compact 2-cycles in the CY$_3$. In topological strings, these fugacities $q$ and $\bm{Q}=(Q_i)$ are related to string coupling $g_s$ and K\"ahler moduli respectively \cite{Ooguri:2009ri}. Moreover, $Z_\text{BPS}$ is equivalent to $Z_\text{crystal}$ modulo signs. See Appendix \ref{crystal2bps} for further details.

\paragraph{Quiver Yangians} As we have an infinite number of BPS degeneracies with some structures therein, it is natural to expect a BPS algebra acting on the BPS states \cite{Harvey:1996gc}. In the $\mathbb{C}^3$ case which has been extensively studied in literature, the affine Yangian of $\mathfrak{gl}_1$, $\mathtt{Y}\left(\widehat{\mathfrak{gl}_1}\right)$, acts on the plane partition and it enumerates the BPS states \cite{schiffmann2013cherednik,maulik2019quantum,Tsymbaliuk2017affine,Prochazka:2015deb,Rapcak:2021hdh}. In particular, the BPS partition function is the character for the vacuum module of $\mathtt{Y}\left(\widehat{\mathfrak{gl}_1}\right)$. This affine Yangian is also the universal enveloping algebra of the $\mathcal{W}_{1+\infty}$-algebra.  Recently, such BPS algebras were also constructed for general toric CY 3-folds in \cite{Li:2020rij,Galakhov:2020vyb}. These infinite-dimensional algebras, known as the quiver Yangians $\mathtt{Y}$, can be ``bootstrapped'' from the structure of molten crystals. For instance, the BPS algebra for generalized conifold $xy=z^mw^n$ is expected to be the affine Yangian of $\mathfrak{gl}_{m|n}$. Moreover, the corresponding BPS partition function should also be identified with the vacuum character of the algebra.

Each quiver Yangian is generated by three sets of operators: $e_n^{(a)}$, $\psi_n^{(a)}$ and $f_n^{(a)}$ for $n\in\mathbb{Z}_{\geq0}$, where $a$ still denotes the quiver nodes. As the quiver Yangian acts on the BPS states, the generators $e_n^{(a)}$ are the creation operators while $f_n^{(a)}$ are annihilation ones. The charges are given by the Cartan part $\psi_n^{(a)}$. Therefore, when acting for instance $e_n^{(a)}$ to a state $|\mathfrak{C}\rangle$, it essentially adds more atoms to the molten crystal $\mathfrak{C}$ following the melting rule. Since the ways of arranging these operators acting on the states would give rise to much more possible combinations than the number of actual BPS states, the generators are constrained by certain (anti-)commutation relations and Serre relations. See \cite{Li:2020rij} for the complete lists.

In fact, the generators $e_n^{(a)}$ form the positive part $\mathtt{Y}^+$ of $\mathtt{Y}$. Likewise, $f_n^{(a)}$ give the negative copy $\mathtt{Y}^-$, and $\psi_n^{(a)}$ generate the subalgebra $\mathtt{Y}^0$. It is conjectured that the Drinfeld double $D(\mathtt{Y}^+)=\mathtt{Y}^+\otimes\mathtt{Y}^{+*}=\mathtt{Y}^+\otimes\hom_\mathbb{C}(\mathtt{Y}^+,\mathbb{C})$ is isomorphic to the quiver Yangian $\mathtt{Y}$ \cite{Rapcak:2018nsl}.

Moreover, multiplication should induce an isomorphism as vector spaces, $m:\mathtt{Y}^+\otimes\mathtt{Y}^0\otimes\mathtt{Y}^-\rightarrow\mathtt{Y}$. We may also consider the Borel subalgebra $\mathtt{Y}^{\geq}$ ($\mathtt{Y}^{\leq}$) generated by $\mathtt{Y}^+$ ($\mathtt{Y}^-$) and $\mathtt{Y}^0$. This should be isomorphic to $\mathcal{H}^{(\mathcal{Q},W)}$, which is the COHA generated by the dimension vectors of the representations of the quiver. The COHA $\mathcal{H}^{(\mathcal{Q},W)}$ has a subalgebra known as the spherical COHA $\mathcal{SH}^{(\mathcal{Q},W)}$ generated by the dimension vectors $\bm{e_j}=(\delta_{ij})$. Then we should have $\mathtt{Y}^+\cong\mathcal{SH}^{(\mathcal{Q},W)}$. For the $\mathbb{C}^3$ case, these were already proven in \cite{Rapcak:2018nsl}. More propositions for $\mathtt{Y}\left(\widehat{\mathfrak{gl}_1}\right)$ can be found for example in \cite{Tsymbaliuk2017affine}. It would be natural to expect that these could be extended to any general quiver Yangians.

\subsection{Kac Polynomials}\label{kacppoly}
As our partition functions can also be expressed in terms of PE, we would like to see whether they could be related to Kac polynomials. Given a locally finite quiver $Q=(Q_0,Q_1)$, the Kac polynomial $A_{\bm{d}}(\mathbb{F}_p)$ is the number of absolutely indecomposable representations of the quiver over a finite field $\mathbb{F}_p$ of dimension $\bm{d}\in\mathbb{Z}_{\geq0}^{|Q_1|}$ (and hence the name dimension vector). This is called a polynomial because there exists a unique polynomial $A_{\bm{d}}(t)\in\mathbb{Z}[t]$ such that $A_{\bm{d}}(\mathbb{F}_p)=A_{\bm{d}}(p)$ for any $\mathbb{F}_p$ \cite{kac1980infinite}.

One can then define the doubled quiver $\overline{Q}=(Q_0,Q_1\sqcup Q_1^*)$ where an arrow $X^*$ in opposite direction is added for each arrow $X$ in the quiver $Q$. The preprojective algebra $\Pi_Q$ is defined as the path algebra $\mathbb{C}\overline{Q}$ quotiented by the ideal generated by $\sum\limits_{X\in Q_1}[X,X^*]$. The stack of representations of $\Pi_Q$ is an abelian category denoted as $\text{Rep}\Pi_Q=\bigsqcup\limits_{\bm{d}}\text{Rep}_{\bm{d}}\Pi_Q$. A representation $M$ is called nilpotent if there exists a filtration $\{0\}=M_l\subset\dots\subset M_1\subset M$ such that $\Pi_Q^+(M_i)\subseteq M_{i+1}$, where $\Pi_Q^+\subset\Pi_Q$ is the augmentation ideal \cite{bozec2017number,schiffmann2018kac}. The substack of these nilpotent representations is called the Lusztig nilpotent variety $\Lambda_Q=\bigsqcup\limits_{\bm{d}}\Lambda_{Q,\bm{d}}$. One may also introduce some semi-nilpotent and strongly semi-nilpotent conditions to define the Lagrangian substacks $\Lambda^0_Q$ and $\Lambda^1_Q$ respectively. We shall not expound the details here, and readers are referred to \cite{bozec2017number,schiffmann2017cohomological} for these conditions. As their names suggest, $\Lambda_Q\subseteq\Lambda_Q^1\subseteq\Lambda_Q^0$.

Consider the $T$-equivariant Borel-Moore homology $H^T_*(\text{Rep}\Pi_Q,\mathbb{Q})=\bigoplus\limits_{\bm{d}}H^T_*(\text{Rep}_{\bm{d}}\Pi_Q,\mathbb{Q})$ \cite{borel1960homology}. Its Poincar\'e polynomial\footnote{Since we have infinitely generated homology, this should really be a series, but we shall always refer to it as Poincar\'e polynomial.}, as shown in \cite{davison2016integrality,bozec2017number}, is encoded by the Kac polynomial:
\begin{equation}
    P_Q(t,\bm{z})=\sum_{\bm{d}}P(\text{Rep}_{\bm{d}}\Pi_Q,t)t^{\langle\bm{d},\bm{d}\rangle}\bm{z}^{\bm{d}}=\text{PE}\left[\frac{1}{1-t^{-1}}\sum_{\bm{d}}A_{\bm{d}}(t^{-1})z^{\bm{d}}\right],
\end{equation}
where $P(\text{Rep}_{\bm{d}}\Pi_Q,t)=\sum\limits_{i}\dim H_{2i}(\text{Rep}_{\bm{d}}\Pi_Q)t^i$ and $\langle\bm{d}_1,\bm{d}_2\rangle$ is the Ringel form as defined in Appendix \ref{crystal2bps}. Likewise, for the Borel-Moore homology of $\Lambda_Q^{\flat}$ ($\flat=0,1$), we have\footnote{Similarly, $A_{\bm{d}}^\flat(p)$ gives the number of absolutely indecomposable representations satifying the corresponding nilpotency condition over a finite field $\mathbb{F}_p$ \cite{bozec2017number}.}
\begin{equation}
    P_Q^\flat(t,\bm{z})=\text{PE}\left[\frac{1}{1-t^{-1}}\sum_{\bm{d}}A_{\bm{d}}^\flat(t^{-1})\bm{z}^{\bm{d}}\right].\label{Pflat}
\end{equation}

One can introduce algebra structures on these homology spaces. These ``2d'' COHAs are closely related to the ``3d'' COHAs/quiver Yangians discussed in the previous subsection. For instance, consider the Jordan quiver $Q$, that is, one single node with one loop $X$. Its tripled quiver $\widehat{Q}$ is given by $\overline{Q}$ with a loop $\omega$ added to the node. The (super)potential is then $W=\omega[X,X^*]$. Then the 2d COHAs are the dimensional reductions\footnote{This dimensional reduction is in the sense that the 3d COHAs were defined in the framework of 3-dimensional CY categories in \cite{Kontsevich:2010px} while the 2d ones come from 2-dimensional CY categories in \cite{schiffmann2013cherednik}.} of the corresponding versions of the 3d COHAs associated to quiver Yangian $\mathtt{Y}\left(\widehat{\mathfrak{gl}_1}\right)$ of $\widehat{Q}$ \cite{Behrend:2009dc,Davison:2013nza,Rapcak:2018nsl}.

More generally, given a quiver $Q$, its tripled quiver $\widehat{Q}$ is the doubled quiver $\overline{Q}$ with a loop $\omega_a$ added to each node. The superpotential is then $W=\sum\limits_{a,x}\omega_a[x,x^*]$. For example, the quivers for $\mathbb{C}\times\mathbb{C}^2/\mathbb{Z}_n$ are tripled quivers $\widehat{Q}$ of the affine A-type quivers $Q$. As the expressions here associated to both $Q$ and $\widehat{Q}$ are in the form of PE and the Kac polynomials encode certain graded characters, it would be natural to compare them and expect some relations between them. In general, for other toric CY 3-folds, the quivers are not tripled, but it might be possible that they could also have some interpretations in terms of something similar to Kac polynomials and lead to possible connections between various algebras.

\section{Examples Galore}\label{examples}
We now discuss the BPS partition functions for all toric CY$_3$ without compact 4-cycles and some non-toric examples, along with various relevant aspects. Let us start with the simplest case $\mathbb{C}^3$ which is most well-studied in literature.

\subsection{Plane Partition: $\mathbb{C}^3$}\label{C3}
\begin{figure}[h]
    \centering
    \includegraphics{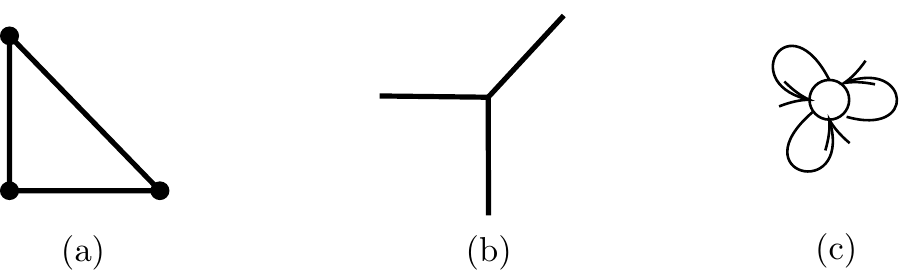}
    \caption{(a) The toric diagram for $\mathbb{C}^3$. (b) Its dual web diagram. (c) The $\mathbb{C}^3$ quiver.}\label{figC3}
\end{figure}

The toric diagram for $\mathbb{C}^3$ is the simplex with vertices $(0,0)$, $(1,0)$ and $(0,1)$. Its dual web is just the trivalent vertex. See Figure \ref{figC3}.
There is no compactly supported D2-branes in this case. The generating function is enumerated by plane partitions, given by the MacMahon function \cite{stanley1997enumerative}:
\begin{equation}
    Z_\text{crystal}=M(q_0)=\prod_{k=1}^\infty\frac{1}{(1-q_0^k)^k}.
\end{equation}
The BPS partition function of D0-branes follows the map $q=-q_0$, that is, $Z_\text{BPS}=M(-q)$. For future convenience\footnote{It seems to be redundant to write $M(q_0)$ (or $M(-q)$) as $M(x)$, but this notation would be easier for our discussions on cases with more variables $q_i$.}, let us also introduce the variable $x=-q$, and then $Z_\text{BPS}=M(x)$. The MacMahon function is precisely the vacuum character of the affine Yangian $\mathtt{Y}\left(\widehat{\mathfrak{gl}_1}\right)$.

It is straightforward to write the generating function as
\begin{equation}
    M(x)=\text{PE}\left[\frac{x}{(1-x)^2}\right].
\end{equation}
It is curious to see that the Hilbert series (HS) for $\mathbb{C}^2$, namely $1/(1-x)^2$, appears inside PE (rather than $\mathbb{C}^3$). Incidentally, $\mathbb{C}^2$ frequently appears in relevant study of instantons and VOAs. The COHA of the $\mathbb{C}^2$ quiver is also isomorphic to the positive part $\mathtt{Y}^+\left(\widehat{\mathfrak{gl}_1}\right)$ of the affine Yangian \cite{Rapcak:2018nsl}. Although similar features are not observed in other cases, the factor $1/(1-x)^2$ is universal in all the examples we consider\footnote{Here, we use $x$ instead of $q$ as it stands for different (but patterned) products of variables for D-branes in different cases.}.

We may now use the method reviewed in Appendix \ref{asymptotic} to get the asymptotics for the generating function. For plane partitions, this is a well-known result \cite{wright1931asymptotic}. At large $n$, the asymptotic expansion of MacMahon function has coefficient
\begin{equation}
    Z_n\sim\frac{\zeta(3)^{7/36}}{\sqrt{12\pi}}\left(\frac{n}{2}\right)^{-25/36}\exp\left(3\zeta(3)^{1/3}\left(\frac{n}{2}\right)^{2/3}+\zeta'(-1)\right).
\end{equation}

Since $\text{PE}[1+f]=\text{PE}[1]\text{PE}[f]=\text{PE}[f]$, we may also write the expression as
\begin{equation}
    M(x)=\text{PE}\left[1+\frac{x}{(1-x)^2}\right]=\text{PE}\left[\frac{1-x+x^2}{(1-x)^2}\right].
\end{equation}
Now the expression inside PE is purely an HS whose Taylor expansion starts from 1. In fact, this is the HS for the complete intersection defined by $\mathcal{X}_1^6+\mathcal{X}_2^3+\mathcal{C}_3^2=0$. By virtue of PE, this gives a one-to-one correspondence between the BPS states labelled by boxes in the plane partition and single-/multi-trace operators generated by $\mathcal{X}_{1,2,3}$. Nevertheless, it is not clear whether this does imply anything non-trivial in physics and mathematics\footnote{It is worth noting that this defining equation could be labelled by $E_{10}$ following \cite{arnol1975critical} though it does not fit in the usual McKay correspondence or belong to the exceptional unimodal singularities. This could probably be in line with the McKay correspdence as equivanlence of derived categories \cite{bridgeland2001mckay,kobayashi2013note}. Moreover, $(1-x+x^2)/(1-x)^2$ was also studied in \cite{He:2010mh} in the context of Hasse-Weil zeta functions and Dirichlet series.}.

\paragraph{Kac polynomials and Poincar\'e polynomials} On the other hand, we find some connections to certain Kac polynomials. Consider the Jordan quiver $Q$ whose doubled quiver $\overline{Q}$ leads to the preprojective algebra $\Pi_Q=\mathbb{C}\overline{Q}/[X,X^*]$. The tripled quiver $\widehat{Q}$ is then the quiver for $\mathbb{C}^3$ having one node with 3 loops $X,X^*,\omega$ and superpotential $W=\omega[X,X^*]$. For the $T$-equivariant Borel-Moore homology $H_*^T(\Lambda^\flat_Q,\mathbb{Q})$, we have \cite{bozec2017number}
\begin{equation}
    P_Q^\flat(t,x)=\text{PE}\left[\frac{tx}{(t-1)(1-x)}\right]=\prod_{d=1}^\infty\prod_{k=0}^\infty\frac{1}{1-t^{-k}x^d}\label{C3Kac}
\end{equation}
with Kac polynomials $A_d^\flat(t)=1$ for both $\flat=0$ and $\flat=1$. In this case, $\Lambda_Q^1=\Lambda_Q^0$. Under the unrefinement $t=x^{-1}$, we find that this agrees with the MacMahon function $M(x)=\text{PE}[x/(1-x)^2]$. This reflects \cite{schiffmann2020cohomological,schiffmann2017cohomological} the fact that the COHA of the moduli stack of coherent sheaves on $\mathbb{C}^2$ with zero-dimensional support is isomorphic to $\mathtt{Y}^+\left(\widehat{\mathfrak{gl}_1}\right)$. One may also check that in this case the Poincar\'e polynomial \cite{schiffmann2013cherednik} of $\mathtt{Y}^+$ is PE$\left[\frac{tx}{(t-1)(1-x)}\right]$. For reference, we also have
\begin{equation}
    P_Q(t,x)=\text{PE}\left[\frac{x}{(t-1)(1-x)}\right]=\prod_{d=1}^\infty\prod_{k=1}^\infty\frac{1}{1-t^{-k}x^d}
\end{equation}
with Kac polynomials $A_d(t)=t$.

\subsection{Conifold}\label{conifold}
\begin{figure}[h]
    \centering
    \includegraphics{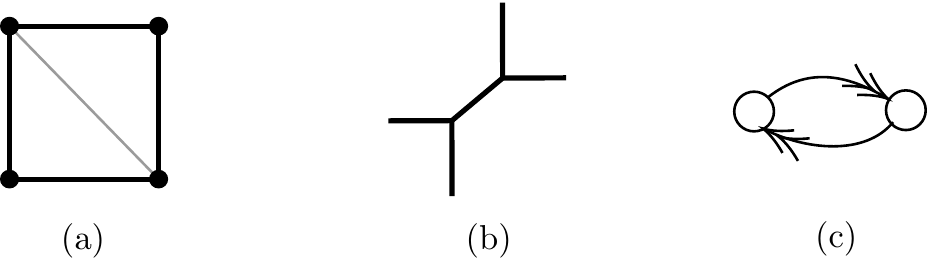}
    \caption{(a) The toric diagram for $\mathcal{C}$. (b) Its dual web diagram. (c) The corresponding quiver.}\label{figconifold}
\end{figure}

Instead of directly move on to $\mathbb{C}^3$ orbifolds, we shall first consider another very well-studied case, that is, the conifold $\mathcal{C}$. The toric diagram is the square enclosed by the four vertices $(p_1,p_2)$ with $p_{1,2}=\{0,1\}$ as shown in Figure \ref{figconifold}, along with its dual web and quiver.
As we can see, the atoms in the crystal (aka pyramid partition) should have two colours $q_{0,1}$. The generating function is well-known from \cite{Szendroi:2007nu,young2009computing}:
\begin{equation}
    Z_\text{crystal}=\frac{M(q_0q_1)^2}{M(-q_1,q_0q_1)M(-q_1^{-1},q_0q_1)}=M(q_0q_1)^2\widetilde{M}(-q_1,q_0q_1)^{-1}.
\end{equation}
We may write this in terms of PE as
\begin{equation}
    \begin{split}
        Z_\text{crystal}&=\text{PE}\left[\sum_{k\in2\mathbb{Z}_{\geq0}+1}kq_1^{k-1}(1+q_1)^2q_0^k\right]\text{PE}\left[\sum_{k\in2\mathbb{Z}_{>0}}\frac{k}{2}q_1^{k-2}(-1+2q_1+4q_1^2+2q_1^3-q_1^4)q_0^k\right]\\
        &=\text{PE}\left[\frac{q_0((1+q_1)^2+q_0^2q_1^2(1+q_1)^2+q_0(-1+2q_1+4q_1^2+2q_1^3-q_1^4))}{(1-q_0q_1)^2}\right].
    \end{split}
\end{equation}
Setting $q_0=q_1=\mathfrak{q}$, we get the pyramid partition without any colouring:
\begin{equation}
    Z=\text{PE}\left[\frac{\mathfrak{q}(1+\mathfrak{q}+3\mathfrak{q}^2+4\mathfrak{q}^3+3\mathfrak{q}^4+\mathfrak{q}^5+\mathfrak{q}^6)}{(1-\mathfrak{q}^4)^2}\right].
\end{equation}
As discussed in Appendix \ref{asymptotic}, this has asymptotic behaviour
\begin{equation}
    Z_n\sim\frac{(7\zeta(3))^{\frac{2}{9}}}{\sqrt{3\pi}}2^{-\frac{25}{36}}n^{-\frac{13}{18}}\exp\left(\frac{2}{3}(7\zeta(3))^{\frac{1}{3}}\left(\frac{n}{2}\right)^{\frac{2}{3}}+2\zeta'(-1)\right).
\end{equation}

We can use the map $q=-q_0q_1$ for D0s and $Q=-q_1$ for D2s to obtain the BPS partition function:
\begin{equation}
    Z_\text{BPS}(q,Q)=M(-q)^2\widetilde{M}(Q,-q)^{-1}.
\end{equation}
In terms of PE, we get
\begin{equation}
    \begin{split}
        Z_\text{BPS}(q,Q)&=\text{PE}\left[\sum_{\substack{k=1\\k\not\in4\mathbb{Z}+2}}^\infty(-1)^{k+1}k\frac{(1-Q)^2}{Q}q^k\right]\text{PE}\left[\sum_{k\in4\mathbb{Z}_{\geq0}+2}-\frac{k}{2}\frac{(1-Q)^2(1+4Q+Q^2)}{Q^2}q^k\right]\\
        &=\text{PE}\left[\frac{q(1-Q)^2(Q-q(1+4Q+Q^2)+3q^2Q-4q^3Q+3q^4Q-q^5(1+4Q+Q^2)+q^6Q)}{Q^2(1-q^4)^2}\right].
    \end{split}
\end{equation}

The expressions in PE are rather tedious in this case. Besides, it is not easy to instantaneously transform between the MacMahon expressions and the PE ones. However, if we change the signs properly, namely getting rid of the minus signs in the arguments of (generalized) MacMahon functions, we can easily get
\begin{equation}
    \widetilde{Z}_c=M(q_0q_1)^2\widetilde{M}(q_1,q_0q_1)^{-1}=\text{PE}\left[-\frac{q_0(1-q_1)^2}{(1-q_0q_1)^2}\right],
\end{equation}
where $\widetilde{Z}_c$ is the sign-changed expression from $Z_\text{crystal}$. As we will see, when writing the generating functions in terms of PE, the patterns are more straightforward for generalized conifolds with the signs properly changed. The coefficients in the expansions of $\widetilde{Z}_c$ and $Z_\text{crystal}$ also agree up to signs. One can simply multiply $(-1)^{n_0+n_1}$ for the terms $q_0^{n_0}q_1^{n_1}$ in $\widetilde{Z}_c$ to recover\footnote{Since the coefficients in the expansion of $Z_\text{crystal}$ are all positive as they simply count the numbers of atoms, this is equivalent to just taking absolute values for the coefficients in the expansion of $\widetilde{Z}_c$.} the correct signs in $Z_\text{crystal}$. Alternatively, one may consider the twisted PE introduced in \cite{Davison:2018zyc}. We find that in general given $\widetilde{Z}_c=\text{PE}[\Tilde{g}]$, the twisted PE of $\Tilde{g}$ is precisely $Z_\text{crystal}$.

Likewise, using $x=-q=q_0q_1$, we have
\begin{equation}
    Z_\text{BPS}(x,Q)=M(x)^2\widetilde{M}(Q,x)^{-1}=\text{PE}\left[-\frac{x(1-Q)^2}{Q(1-x)^2}\right].
\end{equation}
In general given $Z_\text{BPS}(x,Q)=\text{PE}[\Tilde{g}]$, the twisted PE of $\Tilde{g}$ is precisely $Z_\text{BPS}(q,Q)$. Henceforth, we shall always abbreviate $Z_\text{BPS}(x,Q)$ as $Z_\text{BPS}$.

\paragraph{Gluing operators} In \cite{Gaberdiel:2017hcn,Gaberdiel:2018nbs,Li:2019nna}, the vacuum character for the $\mathcal{N}=2$ affine Yangian and its generalization were studied through certain gluing process. Likewise, we may also identify the gluing operators for the affine Yangians discussed in this paper. For the $\mathfrak{u}(1)\oplus\mathcal{W}_\infty^{\mathcal{N}=2}$ algebra, it contains two copies of affine Yangians of $\mathfrak{gl}_1$ as subalgebra. Therefore, in its vacuum character
\begin{equation}
    \chi(x,y)=M(x)^2\widetilde{M}(-yx^{\rho},x)^{-1},
\end{equation}
the factor $M(x)^2$ is identified with the generators contributed from the two $\mathcal{W}_{1+\infty}$ with $\ $ 't Hooft couplings $\lambda_a,\lambda_b$ and central charges $c_a,c_b$. Then the factor
\begin{equation}
    \widetilde{M}(-yx^{\rho},x)^{-1}=\prod_{k=1}^\infty(1+yx^{k+\rho})^k(1+y^{-1}x^{k+\rho})^k
\end{equation}
can be interpreted as gluing operators whose conformal dimensions are controlled by the shifting modulus $\rho$. More precisely, we have $\Delta=1+\rho$. For the $\mathcal{N}=2$ affine Yangian, $\rho=1/2$.

Compared to the vacuum character of affine Yangian of $\mathfrak{gl}_{1|1}$ for the conifold, we find that $M(x)^2$ with $x=q_0q_1=-q$ (and $y=q_1=-Q$) again comes from the two trivalent vertices while their gluing yields the gluing operators with contribution $\widetilde{M}(-y,x)^{-1}$ with no shift, viz, $\rho=0$. Therefore, we may write the character identity
\begin{equation}
    \prod_{k=1}^\infty(1+yx^k)^k=\sum_Ry^{|R|}\chi_R^{\wedge,[\lambda_a]}(x)\chi_{R^*}^{\wedge,[\lambda_b]}(x),
\end{equation}
where the representation $R$ runs over all Young tableaux and $R^*:=\overline{R^\text{T}}$ is the conjugate of $R^\text{T}$. Moreover, $\chi_R^{\wedge,[\lambda]}(x)$ is the wedge part of the character for representation $R$ of $\mathcal{W}_{1+\infty}[\lambda]$, that is \cite{Gaberdiel:2015wpo,Gaberdiel:2017hcn},
\begin{equation}
    \chi_R^{[\lambda]}(x)=\chi_\text{pp}(x)\chi_R^{\wedge,[\lambda]}(x)=M(x)\chi_R^{\wedge,[\lambda]}(x),.
\end{equation}
where $\chi_\text{pp}=\chi_\text{plane partitions}$ is the MacMahon function $M(x)$. Wwith a similar decomposition for the second part in $\widetilde{M}(-yx^{\rho},x)^{-1}$, we arrive at
\begin{equation}
    \begin{split}
        \chi_{\text{vac}, \mathcal{C}}(x,y)&=M(x)^2\widetilde{M}(-y,x)^{-1}\\
        &=\chi_\text{pp}(x)^2\left(\sum_{R_1}y^{|R_1|}\chi_{R_1}^{\wedge,[\lambda_a]}(x)\chi_{R_1^*}^{\wedge,[\lambda_b]}(x)\right)\left(\sum_{R_2}y^{-|R_2|}\chi_{R_2^*}^{\wedge,[\lambda_a]}(x)\chi_{R_2}^{\wedge,[\lambda_b]}(x)\right)\\
        &=\chi_\text{pp}(x)^2+\sum_{R_1}y^{|R_1|}\chi_{R_1}^{[\lambda_a]}(x)\chi_{R_1^*}^{[\lambda_b]}(x)+\sum_{R_2}y^{-|R_2|}\chi_{R_2^*}^{[\lambda_a]}(x)\chi_{R_2}^{[\lambda_b]}(x)+\dots.
    \end{split}
\end{equation}
In particular, the fermionic gluing generators transform as $(R_1,R_1^*)\oplus(R_2^*,R_2)$ under the left and right $\mathcal{W}_{1+\infty}$ algebras\footnote{In \cite{Gaberdiel:2018nbs}, this was denoted as $(R_1\otimes R_2^*,R_1^*\otimes R_2)$, where the notation $R\otimes S^*$ indicates the representation has ``box'' part described by $R$ and ``anti-box'' part described by $S^\text{T}$. In \cite{Li:2019nna}, it was denoted as $(R_1\oplus R_2^*,R_1^*\oplus R_2)$. Here, we shall use the notation which resembles the branching rule.}. This is reflected by the negative power on $\widetilde{M}$ and the minus signs of the arguments therein, as well as the minus signs in the sign-changed $\widetilde{Z}_{c,b}$. The ways of triangulations/gluing simplices in the toric diagrams are also in line with this. It will become more obvious when we discuss those with bosonic generators in the next subsection.

\subsection{Coloured Plane Partitions: $\mathbb{C}\times\mathbb{C}^2/\mathbb{Z}_n$}\label{CC2Zn}
\begin{figure}[h]
    \centering
    \includegraphics{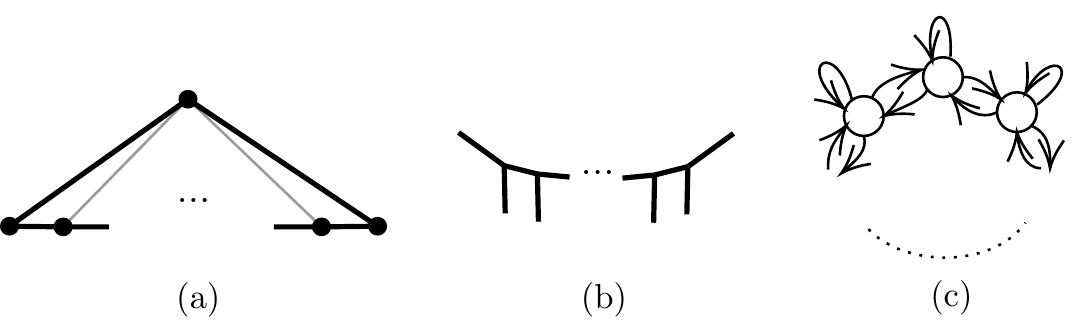}
    \caption{(a) The toric diagram for $\mathbb{C}\times\mathbb{C}^2/\mathbb{Z}_n$ composed of $n$ simplices horizontally arranged in the figure. (b) Its dual web diagram with $n$ vertical lines in the figure. (c) The corresponding quiver with $n$ nodes.}\label{figCC2Zn}
\end{figure}

The toric data for $\mathbb{C}\times\mathbb{C}^2/\mathbb{Z}_n$ is given in Figure \ref{figCC2Zn}.
From the quiver, it is straightforward to see that these are all plane partitions but with multiple colours, one for each node. Therefore, we have $n$ variables $q_{0,1,\dots,n-1}$, and the generating function would reduce to the MacMahon function under $q_0=\dots=q_{n-1}$.

\paragraph{The other bicoloured crystal: $n=2$} Let us start with the simplest case $\mathbb{C}\times\mathbb{C}^2/\mathbb{Z}_2$. When writing the generating functions for the conifold, we observe that they are of form PE$[q_0(1+q_1)^2g_1]$ and PE$\left[\frac{q(1-Q)^2}{Q}g_2\right]$, where $g_{1,2}$ have expansion $1+\dots$. In particular, the two extra factors satisfy $q_0(1+q_1)^2=q(1-Q)^2/Q$ under the matching of variables for conifolds. As one of the only two cases with two colours, it is natural to wonder whether $\mathbb{C}\times\mathbb{C}^2/\mathbb{Z}_2$ would also follow the same pattern with the same extra factor $q_0(1+q_1)^2$ or $q(1-Q)^2/Q$. Recall that we have PE$[xg_1]$ with $g_1=1/(1-x)^2$ for the plane partition with extra factor $x$. Replacing this extra factor with $q_0(1+q_1)^2$, we obtain
\begin{equation}
    Z_\text{crystal}=\text{PE}\left[\frac{q_0(1+q_1)^2}{(1-q_0q_1)^2}\right]=M(q_0q_1)^2\widetilde{M}(q_1,q_0q_1),\label{crystalCC2Z2}
\end{equation}
where we have also substitute $x$ in the denominator with $q_0q_1$ similar to the conifold expression. Indeed, one may check that when taking $q_0=q_1=\mathfrak{q}$, we get $M(\mathfrak{q})=\text{PE}[\mathfrak{q}/(1-\mathfrak{q})^2]$ and recover the plane partition with single colour. As there are no minus signs to be removed in \eqref{crystalCC2Z2}, $\widetilde{Z}_c=Z_\text{crystal}$ in this case.

For $\mathbb{C}\times\mathbb{C}^2/\mathbb{Z}_2$, the D-brane variables follow $q=-q_0q_1$ and $Q=q_1$. Therefore, the extra factor should be $-q(1+Q)^2/Q=q_0(1+q_0q_1)^2$ instead of $q(1-Q)^2/Q$ in this case. Either applying this extra factor to PE$[x/(1-x)^2]$ (with $\mathfrak{q}$ in the denominator changed to $q$) or directly writing \eqref{crystalCC2Z2} in $q,Q$, we can get
\begin{equation}
    \begin{split}
        Z_\text{BPS}(q,Q)=&\text{PE}\left[\frac{-q(1-Q)^2}{Q^2(1-q^4)^2}(Q(1+Q)^2-q(1+2Q+6Q^2+2Q^3+Q^4)+3q^2Q(1+Q)^2\right.\\
        &\left.-4q^3Q(1+Q)^2+3q^4Q(1+Q)^2-q^5(1+2Q+6Q^2+2Q^3+Q^4)+q^6Q(1+Q)^2))\right]\\
        =&\text{PE}\left[\sum_{\substack{k=1\\k\not\in4\mathbb{Z}+2}}^\infty(-1)^{k+1}k\frac{(1+Q)^2}{Q}q^k\right]\text{PE}\left[\sum_{k\in4\mathbb{Z}_{\geq0}+2}\frac{k}{2}\frac{(1-Q)^2(1+2Q+6Q^2+2Q^3+Q^4)}{Q^2}q^k\right]\\
        =&M(-q)^2\widetilde{M}(Q,-q).
    \end{split}\label{BPSCC2Z2}
\end{equation}

In fact, the generating functions for $\mathbb{C}\times\mathbb{C}^2/\mathbb{Z}_n$ were obtained in \cite{Young:2008hn,Cirafici:2010bd}. One can check that \eqref{crystalCC2Z2} and \eqref{BPSCC2Z2} do give the correct expressions.

As before, it is more concise to use $x=-q$:
\begin{equation}
    Z_\text{BPS}=M(x)^2\widetilde{M}(Q,x)=\text{PE}\left[\frac{x(1+Q)^2}{Q(1-x)^2}\right].
\end{equation}

More importantly, comparing $\widetilde{Z}_c$ for $\mathbb{C}\times\mathbb{C}^2/\mathbb{Z}_2$ with the ones for the conifold, or equivalently their $Z_\text{crystal,BPS}$ in (generalized) MacMahon functions, we can see that they only differ by certain minus signs. This is in fact consistent with the analysis of bosonic and fermionic gluing operators. In terms of toric diagrams, they correspond to the two different ways of gluing two simplices. More specifically, here we have
\begin{equation}
    \chi_{\text{vac},\mathbb{C}\times\mathbb{C}^2/\mathbb{Z}_2}=\prod_{k=1}^\infty\frac{1}{(1-x^k)^{2k}(1-yx^k)^k(1-y^{-1}x^k)^k},
\end{equation}
where $x=q_0q_1=-q$ and $y=q_1=Q$. This leads to the bosonic gluing operators with character identity
\begin{equation}
    \prod_{k=1}^\infty(1+yx^k)^{-k}=\sum_Ry^{|R|}\chi_R^{\wedge,[\lambda_a]}(x)\chi_{\overline{R}}^{\wedge,[\lambda_b]}(x).
\end{equation}
As a result, the vacuum character decomposes as
\begin{equation}
    \begin{split}
        \chi_{\text{vac},\mathbb{C}\times\mathbb{C}^2/\mathbb{Z}_2}(x,y)
        &=\chi_\text{pp}(x)^2\left(\sum_{R_1}y^{|R_1|}\chi_{R_1}^{\wedge,[\lambda_a]}(x)\chi_{\overline{R}_1}^{\wedge,[\lambda_b]}(x)\right)\left(\sum_{R_2}y^{-|R_2|}\chi_{\overline{R}_2}^{\wedge,[\lambda_a]}(x)\chi_{R_2}^{\wedge,[\lambda_b]}(x)\right)\\
        &=\chi_\text{pp}(x)^2+\sum_{R_1}y^{|R_1|}\chi_{R_1}^{[\lambda_a]}(x)\chi_{\overline{R}_1}^{[\lambda_b]}(x)+\sum_{R_2}y^{-|R_2|}\chi_{\overline{R}_2}^{[\lambda_a]}(x)\chi_{R_2}^{[\lambda_b]}(x)+\dots.
    \end{split}
\end{equation}
In particular, the bosonic gluing generators transform as $(R_1,\overline{R}_1)\oplus(\overline{R}_2,R_2)$ under the left and right $\mathcal{W}_{1+\infty}$ algebras.

\paragraph{General $n$} We may generalize the above discussion to any $n$. The extra factor now becomes $q_0(1+q_1+q_1q_2+\dots+q_1q_2\dots q_{n-1})(1+q_{n-1}+q_{n-1}q_{n-2}+\dots+q_{n-1}q_{n-2}\dots q_1)$. Therefore,
\begin{equation}
    \begin{split}
        Z_\text{crystal}&=\text{PE}\left[\frac{q_0\left(1+\sum\limits_{i=1}^{n-1}\prod\limits_{j=1}^iq_j\right)\left(1+\sum\limits_{i=1}^{n-1}\prod\limits_{j=1}^iq_{n-j}\right)}{\left(1-\prod\limits_{i=0}^{n-1}q_i\right)^2}\right]\\
        &=M\left(\prod_{i=0}^{n-1}q_i\right)^n\prod_{0<r\leq s<n}\widetilde{M}\left(\prod_{i=r}^sq_i,\prod_{j=0}^{n-1}q_j\right).
    \end{split}\label{crystalCC2Zn}
\end{equation}
As a sanity check, this reduces to the MacMahon function $M(\mathfrak{q})$ under $q_{0,\dots,n-1}=\mathfrak{q}$. More generally, if $m|n$, then $Z_\text{BPS}$ for $n$ can be reduced to the one for $m$ by identifying all $q_i=q_j$ when $i\equiv j\text{ (mod }m)$.

Now that the crystal-to-BPS map reads $q_0\rightarrow-q_0$, $q_{i\neq0}\rightarrow q_i$, we have $q=-\prod\limits_{i=0}^{n-1}q_i$ and $Q_i=q_i$. Thus,
\begin{equation}
    Z_\text{BPS}(q,Q)=M\left(-q\right)^n\prod_{0<r\leq s<n}\widetilde{M}\left(\prod_{i=r}^sQ_i,-q\right).\label{BPSCC2Zn}
\end{equation}
One may check that \eqref{crystalCC2Zn} and \eqref{BPSCC2Zn} agree with the results in \cite{Young:2008hn,Cirafici:2010bd}. By using $x=-q$, we can also get a simpler PE form for $Z_\text{BPS}$:
\begin{equation}
    \begin{split}
        Z_\text{BPS}&=M\left(x\right)^n\prod_{0<r\leq s<n}\widetilde{M}\left(\prod_{i=r}^sQ_i,x\right)\\
        &=\text{PE}\left[\frac{x\left(1+\sum\limits_{i=1}^{n-1}\prod\limits_{j=1}^iQ_j\right)\left(1+\sum\limits_{i=1}^{n-1}\prod\limits_{j=1}^iQ_{n-j}\right)}{(1-x)^2\prod\limits_{i=0}^{n-1}Q_i}\right].
    \end{split}
\end{equation}

Remarkably, it was observed in \cite{Davison:2018zyc} that
\begin{equation}
    Z_\text{crystal}=\text{PE}\left[\frac{x}{(1-x)^2}\left(n+\sum_{\bm{\alpha}\in\Psi}\bm{q}_*^{\bm{\alpha}}\right)\right],\label{adjrep}
\end{equation}
where $x=\prod\limits_{i=0}^{n-1}q_i$ and $\bm{q}_*^{\bm{\alpha}}=\prod\limits_{i=1}^{n-1}q_i^{\alpha_i}$ while $\Psi$ is the root system of the Lie algebra of type $A_{n-1}$. In particular, $\left(n+\sum\limits_{\bm{\alpha}\in\Phi}\bm{q}_*^{\bm{\alpha}}\right)$ is the character of the adjoint representation. This reflects the enhanced gauge symmetry when the target spaces of type IIA strings have $A_{n-1}$ singularities \cite{Katz:1996ht}.

\paragraph{General gluings} Given the vacuum characters for affine Yangians $\mathtt{Y}\left(\widehat{\mathfrak{gl}_n}\right)$, we are now able to generalize the gluing process to $n$ trivalent vertices. In \eqref{crystalCC2Zn}, the factor $M(x)^n$ arises from $n$ disjoint trivalent vertices. This corresponds to the subalgebra of $n$ copies of $\mathcal{W}_{1+\infty}$. Hence, the remaining product of generalized MacMahon functions are contributions from the gluing operators.

Suppose we only have the first two vertices and glue them following the pattern in Figure \ref{figCC2Zn}(b). Then we obtain
\begin{equation}
    M(x)^2{\color{blue}\widetilde{M}(q_1,x)}=\chi_\text{pp}(x)^2{\color{blue}\left(\sum_{R_1}q_1^{|R_1|}\chi_{R_1}^{\wedge,[\lambda_a]}(x)\chi_{\overline{R}_1}^{\wedge,[\lambda_b]}(x)\right)\left(\sum_{R_2}q_1^{-|R_2|}\chi_{\overline{R}_2}^{\wedge,[\lambda_a]}(x)\chi_{R_2}^{\wedge,[\lambda_b]}(x)\right)}
\end{equation}
as in the $\mathbb{C}\times\mathbb{C}^2/\mathbb{Z}_{n=2}$ case, where the blue part corresponds to the two bosonic gluing operators.

Now let us glue a third vertex following Figure \ref{figCC2Zn}(b). We should expect different non-trivial factors as this is not a gluing of two trivalent vertices any more. According to the vacuum character in the $n=3$ case, we should get
\begin{equation}
    \begin{split}
        M(x)^3{\color{blue}\widetilde{M}(q_1,x)}{\color{red}\widetilde{M}(q_2,x)}{\color{purple}\widetilde{M}(q_1q_2,x)}=&\chi_\text{pp}^3{\color{blue}\left(\sum_{R_1}q_1^{|R_1|}\chi_{R_1}^{\wedge}\chi_{\overline{R}_1}^{\wedge}\right)\left(\sum_{R_2}q_1^{-|R_2|}\chi_{\overline{R}_2}^{\wedge}\chi_{R_2}^{\wedge}\right)}\\
        &\times{\color{red}\left(\sum_{R_3}q_2^{|R_3|}\chi_{R_3}^{\wedge}\chi_{\overline{R}_3}^{\wedge}\right)\left(\sum_{R_4}q_2^{-|R_4|}\chi_{\overline{R}_4}^{\wedge}\chi_{R_4}^{\wedge}\right)}\\
        &\times{\color{purple}\left(\sum_{R_5}(q_1q_2)^{|R_5|}\chi_{R_5}^{\wedge}\chi_{\overline{R}_5}^{\wedge}\right)\left(\sum_{R_6}(q_1q_2)^{-|R_6|}\chi_{\overline{R}_6}^{\wedge}\chi_{R_6}^{\wedge}\right)},
    \end{split}
\end{equation}
where we have omitted the superscripts coming from the three copies $\mathcal{W}_{1+\infty}[\lambda_{a,b,c}]$ in $\chi^\wedge$ for brevity. In particular, the red part corresponds to the bosonic operators when the second and third vertices are glued together (ignoring the first vertex). On the other hand, the purple part indicates that there are new bosonic generators arising from blue and red ones. For convenience, we shall refer to the generators like those in blue and red as ``basic'' gluing operators while the ones like those in purple as ``derived'' gluing operators. The vacuum character can be decomposed as
\begin{equation}
    \begin{split}
        \chi_{\text{pp}_3}=&\chi_\text{pp}^3+{\color{blue}\sum_{R_1}q_1^{|R_1|}\chi_{R_1}\chi_{\overline{R}_1}\chi_\text{pp}}+{\color{blue}\sum_{R_2}q_1^{-|R_2|}\chi_{\overline{R}_2}\chi_{R_2}\chi_\text{pp}}\\
        &+{\color{red}\sum_{R_3}q_2^{|R_3|}\chi_\text{pp}\chi_{R_3}\chi_{\overline{R}_3}}+{\color{red}\sum_{R_4}q_2^{-|R_4|}\chi_\text{pp}\chi_{\overline{R}_4}\chi_{R_4}}\\
        &+{\color{purple}\sum_{R_5}(q_1q_2)^{|R_5|}\chi_{R_5}\chi_{\overline{R}_5}\chi_\text{pp}}+{\color{purple}\sum_{R_4}(q_1q_2)^{-|R_6|}\chi_{\overline{R}_6}\chi_{R_6}\chi_\text{pp}}+\dots,
    \end{split}
\end{equation}
where $\text{pp}_n$ denotes the $n$-coloured plane partitions. Here, some generators transform as $(R_1,\overline{R}_1,1)\oplus(\overline{R}_2, R_2,1)$ and $(1,R_3,\overline{R}_3)\oplus(1,\overline{R}_4,R_4)$. The remaining ones transform as $(R_5,\overline{R}_5,1)\oplus(\overline{R}_6, R_6,1)$ under a subalgebra composed of three different copies of $\mathcal{W}'_{1+\infty}$ (which can be thought of as a mixing of $\mathcal{W}_{1+\infty}[\lambda_{a,b,c}]$). We shall illustrate this gluing in the shorthand notation
\begin{equation}
    \includegraphics{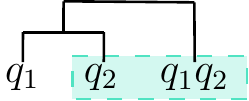},
\end{equation}
where those in the dashed box correspond to the new bosonic gluing operators.

Moving on to $n=4$, we further glue another vertex following Figure \ref{figCC2Zn}(b). According to \eqref{crystalCC2Zn},
\begin{equation}
    \chi_{\text{pp}_4}=M(x)^4\widetilde{M}(q_1,x){\color{blue}\widetilde{M}(q_2,x)}{\color{red}\widetilde{M}(q_1q_2,x)}{\color{green}\widetilde{M}(q_3,x)}{\color{cyan}\widetilde{M}(q_2q_3,x)}{\color{bananayellow}\widetilde{M}(q_1q_2q_3,x)}.
\end{equation}
As we can see, gluing the third and fourth vertices (while ignoring the other two) leads to the bosonic operators of the green part. Then the blue and green operators give rise to the new cyan bosonic gluing operators while the red and green parts yield the new yellow ones. The character decomposition can be obtained likewise as before. In the above shorthand notation,
\begin{equation}
    \includegraphics{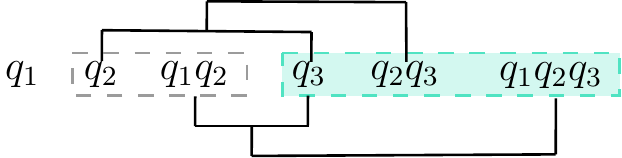}.
\end{equation}
As we can see, we have $q_{1,2,3}$ corresponding to basic operators while $q_1q_2$ and $q_2q_3$ corresponds to derived operators arising from basic ones. Furthermore, we also have derived ones that are derived from both basic and derived generators.

We can thence get the gluing operators for any $n$. For instance, at the next level, in the shorthand notation we have
\begin{equation}
    \includegraphics{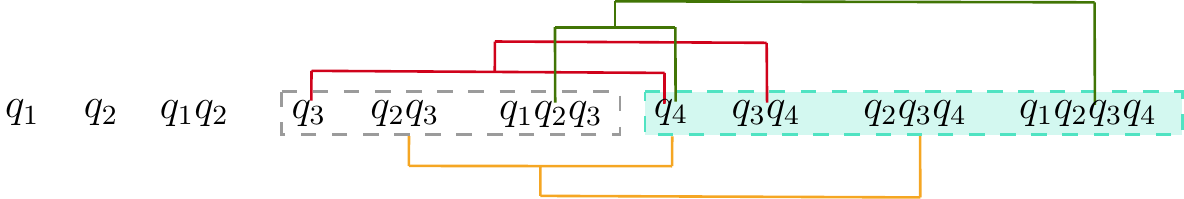}.
\end{equation}
Here, we only have bosonic gluing operators, so we do not need to worry about their $\mathbb{Z}_2$-gradings. When considering any generalized conifolds, we will also have fermionic gluing operators. Although the process is the same, we will discuss the way to determine their $\mathbb{Z}_2$-gradings for multiple vertices.

\paragraph{Kac polynomials and Poincar\'e polynomials} As in the $\mathbb{C}^3$ case, let us view the quiver in Figure \ref{figCC2Zn}(c) as the tripled quiver $\widetilde{Q}$ of some quiver $Q$. Then the quiver $Q$ is simply the cyclic affine $\hat{A}_{n-1}$ quiver with arrows in the same orientation. From \cite{bozec2017number}, we know that
\begin{equation}
    P_Q^0(t,\bm{q})=\text{PE}\left[\sum_{\bm{d}\in\Phi_0^+}\frac{t\bm{q}^{\bm{d}}}{(t-1)(1-\bm{q}^{\bm{\delta}})}\right]\text{PE}\left[\frac{nt\bm{q}^{\bm{\delta}}}{(t-1)(1-\bm{q}^{\bm{\delta}})}\right]\text{PE}\left[\sum_{\bm{d}\in\Phi_0^-}\frac{t\bm{q}^{\bm{d}+\bm{\delta}}}{(t-1)(1-\bm{q}^{\bm{\delta}})}\right],\label{P0}
\end{equation}
where $\bm{q}^{\bm{d}}=\prod\limits_{i=0}^{n-1}q_i^{d_i}$. Here, let $\Phi^+=\Phi_\text{Re}^+\sqcup\Phi_\text{Im}^+$ denote the set of positive roots with real and imaginary roots $\Phi_\text{Re}^+=\{\Phi_0^++\bm{\delta}\mathbb{Z}_{\geq0}\}\sqcup\{\Phi_0^-+\bm{\delta}\mathbb{Z}_{>0}\}$ and $\Phi_\text{Im}^+=\bm{\delta}\mathbb{Z}_{>0}$ respectively, where $\bm{\delta}$ is the minimal positive imaginary root. For affine $A$-type, we simply have $\bm{\delta}=(1,\dots,1)=\bm{1}_n$. Then $\Phi_0$ is the root system of the underlying finite type quiver $Q_0\subset Q$. For reference, we also have
\begin{equation}
    P_Q(t,\bm{q})=P_Q^1(t,\bm{q})=\text{PE}\left[\sum_{\bm{d}\in\Phi_0^+}\frac{t\bm{q}^{\bm{d}}}{(t-1)(1-\bm{q}^{\bm{\delta}})}\right]\text{PE}\left[\frac{(1+(n-1)t)\bm{q}^{\bm{\delta}}}{(t-1)(1-\bm{q}^{\bm{\delta}})}\right]\text{PE}\left[\sum_{\bm{d}\in\Phi_0^-}\frac{t\bm{q}^{\bm{d}+\bm{\delta}}}{(t-1)(1-\bm{q}^{\bm{\delta}})}\right].
\end{equation}
The Kac polynomials are
\begin{equation}
    \begin{cases}
    A_{\bm{d}}(t)=A_{\bm{d}}^\flat(t)=1,&\qquad\bm{d}\in\Phi_\text{Re}^+\\
    A_{\bm{d}}(t)=A_{\bm{d}}^1(t)=t+n-1,~A_{\bm{d}}^0(t)=n,&\qquad\bm{d}\in\Phi_\text{Im}^+
    \end{cases}.
\end{equation}

To compare this with the character of the affine Yangian, let us further introduce a ``negative'' counterpart of the COHA associated to $\Lambda^0_Q$ such that the Poincar\'e polynomial takes the sum over $\Phi^-=\Phi_\text{Re}^-\sqcup\Phi_\text{Im}^-$ with $\Phi_\text{Re}^-=\{\Phi_0^--\bm{\delta}\mathbb{Z}_{\geq0}\}\sqcup\{\Phi_0^+-\bm{\delta}\mathbb{Z}_{>0}\}$ and $\Phi_\text{Im}^-=-\bm{\delta}\mathbb{Z}_{>0}$. This simply takes $q_i\rightarrow q_i^{-1}$ in \eqref{P0}. Notice that $A_{\bm{d}}^0$ is independent of $t$, and the $t$ dependence in $P_Q^0$ only comes from the factor $1/(1-t^{-1})$ in \eqref{Pflat}. Therefore, we also treat $t$ as a formal variable and take $t\rightarrow t^{-1}$. Then
\begin{equation}
    \begin{split}
        \widetilde{P}_Q^0(t,\bm{q})&=\text{PE}\left[\sum_{\bm{d}\in\Phi_0^+}\frac{t^{-1}\bm{q}^{-\bm{d}}}{(t^{-1}-1)(1-\bm{q}^{-\bm{\delta}})}\right]\text{PE}\left[\frac{nt^{-1}\bm{q}^{-\bm{\delta}}}{(t^{-1}-1)(1-\bm{q}^{-\bm{\delta}})}\right]\text{PE}\left[\sum_{\bm{d}\in\Phi_0^-}\frac{t^{-1}\bm{q}^{-\bm{d}-\bm{\delta}}}{(t^{-1}-1)(1-\bm{q}^{-\bm{\delta}})}\right]\\
        &=\text{PE}\left[\sum_{\bm{d}\in\Phi_0^-}\frac{\bm{q}^{\bm{d}}}{(1-t)(1-\bm{q}^{-\bm{\delta}})}\right]\text{PE}\left[\frac{n\bm{q}^{-\bm{\delta}}}{(1-t)(1-\bm{q}^{-\bm{\delta}})}\right]\text{PE}\left[\sum_{\bm{d}\in\Phi_0^+}\frac{\bm{q}^{\bm{d}-\bm{\delta}}}{(1-t)(1-\bm{q}^{-\bm{\delta}})}\right]\\
        &=\text{PE}\left[\sum_{\bm{d}\in\Phi_0^-}\frac{\bm{q}^{\bm{d}+\bm{\delta}}}{(t-1)(1-\bm{q}^{\bm{\delta}})}\right]\text{PE}\left[\frac{n}{(t-1)(1-\bm{q}^{\bm{\delta}})}\right]\text{PE}\left[\sum_{\bm{d}\in\Phi_0^+}\frac{\bm{q}^{\bm{d}}}{(t-1)(1-\bm{q}^{\bm{\delta}})}\right].
    \end{split}
\end{equation}

Consider the product
\begin{equation}
    \begin{split}
        P_Q^0(t,\bm{q})\widetilde{P}_Q^0(t,\bm{q})=&\text{PE}\left[\sum_{\bm{d}\in\Phi_0^+}\frac{\bm{q}^{\bm{d}}}{(t-1)(1-\bm{q}^{\bm{\delta}})}\right]\text{PE}\left[\frac{nt\bm{q}^{\bm{\delta}}}{(t-1)(1-\bm{q}^{\bm{\delta}})}\right]\text{PE}\left[\sum_{\bm{d}\in\Phi_0^-}\frac{t\bm{q}^{\bm{d}+\bm{\delta}}}{(t-1)(1-\bm{q}^{\bm{\delta}})}\right]\\
        &\times\text{PE}\left[\sum_{\bm{d}\in\Phi_0^+}\frac{t\bm{q}^{\bm{d}}}{(t-1)(1-\bm{q}^{\bm{\delta}})}\right]\text{PE}\left[\sum_{\bm{d}\in\Phi_0^-}\frac{\bm{q}^{\bm{d}+\bm{\delta}}}{(t-1)(1-\bm{q}^{\bm{\delta}})}\right]\text{PE}\left[\frac{n}{(t-1)(1-\bm{q}^{\bm{\delta}})}\right]\\
        =&\text{PE}\left[\left(\sum_{\bm{d}\in\Phi_0^+}\frac{\bm{q}^{\bm{d}}}{(t-1)(1-x)}\right)+\frac{ntx}{(t-1)(1-x)}+\left(\sum_{\bm{d}\in\Phi_0^-}\frac{tx\bm{q}^{\bm{d}}}{(t-1)(1-x)}\right)\right]\\
        &\times\text{PE}\left[\left(\sum_{\bm{d}\in\Phi_0^+}\frac{t\bm{q}^{\bm{d}}}{(t-1)(1-x)}\right)+\frac{n}{(t-1)(1-x)}+\left(\sum_{\bm{d}\in\Phi_0^-}\frac{x\bm{q}^{\bm{d}}}{(t-1)(1-x)}\right)\right],
    \end{split}
\end{equation}
where we have again used $x=\prod\limits_{i=0}^{n-1}q_i$. Henceforth, we shall abbreviate the second PE in the last equality as an ellipsis. As before, taking $t=x^{-1}$, we get
\begin{equation}
    P_Q^0(1/x,\bm{q})\widetilde{P}_Q^0(1/x,\bm{q})=\text{PE}\left[\frac{x}{(1-x)^2}\left(n+\sum_{\bm{d}\in\Phi_0}\bm{q}^{\bm{d}}\right)\right]\times\dots.
\end{equation}

Recall the character of the affine Yangian $\mathtt{Y}\left(\widehat{\mathfrak{gl}_n}\right)$ in \eqref{crystalCC2Zn} and especially in \eqref{adjrep}. Inside PE, we have the root system $\Psi$ of $A_{n-1}$ while $\Phi_0$ here is the root system of $A_n$. Hence, $\Psi$ is the subset of $\Phi_0$ with $d_0=0$. As a result, we obtain
\begin{equation}
    \begin{split}
        P_Q^0(1/x,\bm{q})\widetilde{P}_Q^0(1/x,\bm{q})&=\text{PE}\left[\frac{x}{(1-x)^2}\left(n+\sum_{\substack{\bm{d}\in\Phi_0\\d_0=0}}\bm{q}^{\bm{d}}\right)\right]\text{PE}\left[\frac{x}{(1-x)^2}\left(\sum_{\substack{\bm{d}\in\Phi_0\\d_0\neq0}}\bm{q}^{\bm{d}}\right)\right]\times\dots\\
        &=\chi_{\text{pp}_n}\text{PE}\left[\frac{x}{(1-x)^2}\left(\sum_{\substack{\bm{d}\in\Phi_0\\d_0\neq0}}\bm{q}^{\bm{d}}\right)\right]\times\dots.
    \end{split}
\end{equation}
Therefore, it is tempting to conjecture that the double copy of the COHA associated to $\Lambda_Q^0$ contains (the positive part of) the affine Yangian as a subalgebra. In \S\ref{refined}, we will check this with the refined partition functions.

Let us illustrate this with a concrete example. Consider $n=2$, then we have
\begin{equation}
    \chi_{\text{pp}_2}=\text{PE}\left[\frac{q_0(1+q_1)^2}{(1-q_0q_1)^2}\right]=\text{PE}\left[\frac{q_0q_1({\color{orange}q_1+2+q_1^{-1}})}{(1-q_0q_1)^2}\right]
\end{equation}
while
\begin{equation}
    P_Q^0(1/x,\bm{q})\widetilde{P}_Q^0(1/x,\bm{q})=\text{PE}\left[\frac{q_0q_1}{(1-q_0q_1)^2}({\color{orange}q_1+2+q_1^{-1}}+q_0+q_0^{-1}+q_0q_1+q_0^{-1}q_1^{-1})\right]\times\dots.
\end{equation}

\subsection{Generalized Conifolds}\label{genC}
\begin{figure}[h]
    \centering
    \includegraphics{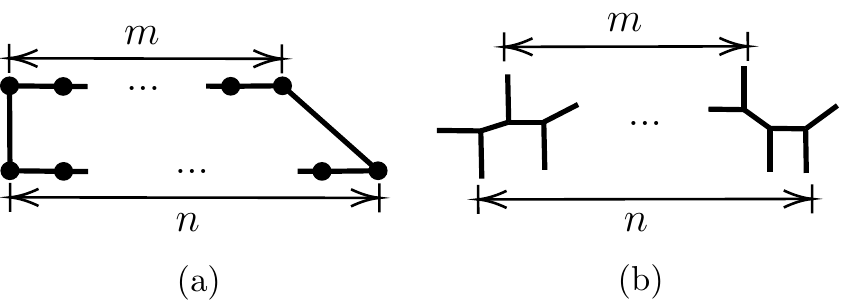}
    \caption{The toric diagram for the generalized conifold and its dual web.}\label{figgenC}
\end{figure}

The generalized conifold is defined by $xy=z^mw^n$. Its crystal melting partition function will give the vacuum character of the affine Yangian $\mathtt{Y}\left(\widehat{\mathfrak{gl}_{m|n}}\right)$. The toric diagram and its dual web are depicted in Figure \ref{figgenC}.

\begin{figure}[h]
    \centering
    \includegraphics{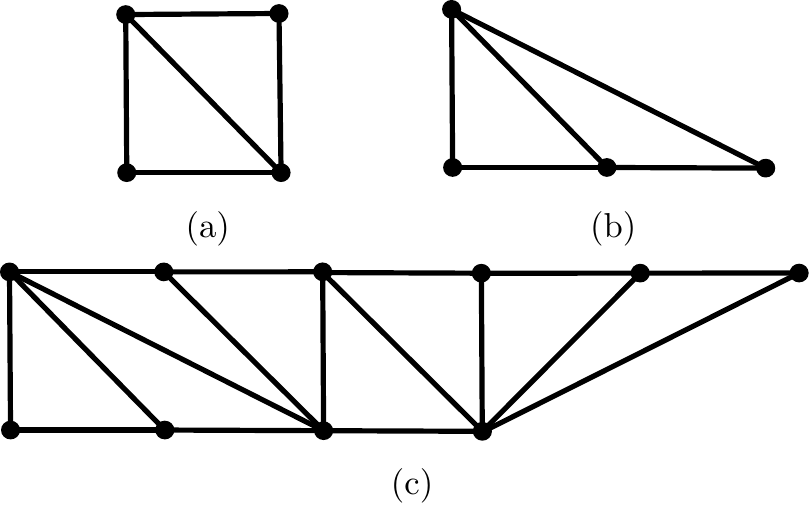}
    \caption{In these examples, we have (a) $\sigma=\{+1,-1\}$, (b) $\sigma=\{+1,+1\}$, (c) $\sigma=\{+1,+1,-1,-1,+1,-1,-1,-1\}$.}\label{sigma}
\end{figure}

Given a toric diagram of the generalized conifold, it can have multiple distinct ways of triangulations. These triangulations correspond to quiver theories in different phases and are related by Seiberg duality \cite{Franco:2005rj}. Their crystal/BPS partition functions are related by ``wall crossing of the second kind'' according to \cite{Aganagic:2010qr}. In the dual web, these phases are connected by flop transitions. The triangulations can be concisely encoded by a sequence of signs $\sigma=\{\sigma_a\}$ ($a\in\mathbb{Z}_{m+n}$) consisting of $m$ plus ones and $n$ minus ones \cite{nagao2008derived,Nagao:2009rq}. When two adjacent simplices are glued side by side, they have the same signs. When they are glued in an alternative way, they have opposite signs. This is illustrated in Figure \ref{sigma}.

We can then use this information to construct the quiver as follows. First, the quiver has $(m+n)$ nodes forming a closed cycle. There is always one arrow from node $a$ to $(a+1)$ and one arrow from $(a+1)$ to $a$. Next, the adjoint loops can be added to the nodes based on $\sigma$. If $\sigma_a=\sigma_{a+1}$, then node $a$ has an adjoint loop. If $\sigma_a=-\sigma_{a+1}$, then node $a$ does not have such loop. The superpotential can also be read off from $\sigma$. See for example (8.83) in \cite{Li:2020rij}.

From this, we can deduce that the crystal-to-BPS map reads $q_0\rightarrow(-1)^{\frac{\sigma_0+\sigma_1}{2}}q_0$ and $q_{a}\rightarrow(-1)^{\frac{\sigma_a-\sigma_{a+1}}{2}}q_a$ for $a\neq 0$. Now we can write our ansantz for $\widetilde{Z}_c$ which recovers $Z_\text{crystal}$ under the crystal-to-BPS map. The sign-changed expression is $\widetilde{Z}_c=\text{PE}\left[\frac{g}{(1-x)^2}\right]$, where $x=\prod\limits_{i=0}^{m+n-1}q_i$. The extra factor $g$ is
\begin{equation}
    g=(-1)^{\frac{\sigma_0+\sigma_1}{2}}q_0\left(1+\sum_{i=1}^{m+n-1}(-1)^{\frac{\sigma_1-\sigma_{i+1}}{2}}\prod_{j=1}^iq_j\right)\left(1+\sum_{i=1}^{m+n-1}(-1)^{\frac{\sigma_{m+n-i}-\sigma_0}{2}}\prod_{j=1}^iq_{m+n-j}\right).
\end{equation}
In the expansion of $\widetilde{Z}_c$, the coefficients are equal to the numbers of atoms given by $Z_\text{crystal}$ up to signs. As $Z_\text{crystal}$ always has positive coefficients in its expansion, the correct signs are recovered simply by taking absolute values.

Write $\widetilde{Z}_c$ using (generalized) MacMahon functions and apply the crystal-to-BPS map, we find
\begin{equation}
    Z_\text{crystal}=M\left(\prod_{i=0}^{m+n-1}q_i\right)^{m+n}\prod_{0<r\leq s<m+n}\widetilde{M}\left((-1)^{\frac{\sigma_r-\sigma_{s+1}}{2}}\prod_{j=r}^sq_j,\prod_{i=0}^{m+n-1}q_i\right)^{(-1)^{\frac{\sigma_r-\sigma_{s+1}}{2}}}.
\end{equation}
As we can see, such expression in terms of (generalized) MacMahon functions also follows a nice pattern. One may check that all the cases discussed before obey this expression.

Now from the crystal-to-BPS map, we obtain $q=-\prod\limits_{i=0}^{m+n-1}q_i$, $Q_j=(-1)^{\frac{\sigma_j-\sigma_{j+1}}{2}}q_j$. Therefore,
\begin{equation}
    Z_\text{BPS}(q,Q)=M(-q)^{m+n}\prod_{0<r\leq s<m+n}\widetilde{M}\left(\prod_{i=r}^sQ_i,-q\right)^{(-1)^{\frac{\sigma_r-\sigma_{s+1}}{2}}}.
\end{equation}
In terms of $x=-q$, $Z_\text{BPS}=\text{PE}\left[\frac{\Tilde{g}}{(1-x)^2}\right]$, where the extra factor $\Tilde{g}$ reads
\begin{equation}
    \Tilde{g}=\frac{x\left(1+\sum\limits_{i=1}^{m+n-1}(-1)^{\frac{\sigma_1-\sigma_{i+1}}{2}}\prod\limits_{j=1}^iQ_j\right)\left(1+\sum\limits_{i=0}^{m+n-1}(-1)^{\frac{\sigma_{m+n-i}-\sigma_0}{2}}\prod\limits_{j=1}^iQ_{m+n-j}\right)}{\prod\limits_{i=1}^{m+n-1}Q_i}.
\end{equation}

One may expect that these expressions agree with the topological vertex formalism in \cite{Aganagic:2003db,Iqbal:2004ne} as well as the results in \cite{Mozgovoy:2020has} from a more mathematical approach. They should also satisfy the following properties:
\begin{itemize}
    \item The perturbative expansion would recover the number of configurations at each level in the crystal in light of the melting rule.
    \item As a self-consistency check, we can make identifications among the variables $q_{0,\dots,m+n-1}$. This should reduce to $Z_\text{crystal}$ with fewer colours of the same crystal configuration.
    \item The general gluing operators should be consistent with the factors in the character.
\end{itemize}

\paragraph{The gluing process} Let us explain the gluings in more detail. When gluing two ``free'' vertices, there will be fermionic or bosonic generators depending on the way of gluing them. For Figure \ref{sigma}(a), this gives rise to fermionic generators. For Figure \ref{sigma}(b), we get bosonic generators. More generally, when there are multiple vertices glued together, the $\mathbb{Z}_2$-gradings of the basic generators are determined via $\sigma$. In other words, if $\sigma_a=\sigma_{a+1}$, the basic gluing operators are bosonic for $q_a$. If $\sigma_a=-\sigma_{a+1}$, the basic gluing operators are fermionic for $q_a$. As a result, we cannot separate the two triangles/trivalent vertices and treat them as two ``free'' building blocks to determine the $\mathbb{Z}_2$-grading of the basic generators. Therefore, for the conifold $\mathcal{C}$, we have fermionic gluing operators since $\sigma_1=-\sigma_{2\equiv0}$. On the other hand, we only have bosonic ones for $\mathbb{C}\times\mathbb{C}^2/\mathbb{Z}_n$ since $\sigma_1=\sigma_2=\dots=\sigma_{n-1}=\sigma_{n\equiv0}$.

Recall the criterion of adding adjoint loops to quiver nodes. We find that $a$ yields bosonic gluing operators when it has an odd number of adjoint loops while it gives fermionic ones when it has no adjoint loop\footnote{Here, the ``odd number'' is used to include the $\mathbb{C}^3$ case. We can likewise extend the fermionic case to even number of adjoints. Of course, for generalized conifolds, this even number can only be zero. It seems that a non-zero even number of adjoints does not exist for physical quiver theories \cite{Li:2020rij}.}. This is exactly the same as the grading rule in \cite{Li:2020rij} for determining whether $e_n^{(a)}$ and $f_n^{(a)}$ are bosonic or fermionic generators.

Moreover, there will also be derived gluing operators as discussed before. These extra generators can be simply determined by the usual $\mathbb{Z}_2$-grading, namely, $\text{b}\times\text{b}=\text{f}\times\text{f}=\text{b}$ and $\text{b}\times\text{f}=\text{f}$. One may check that the generalized MacMahon functions in the characters do follow the discussions here.

\begin{figure}[h]
    \centering
    \includegraphics{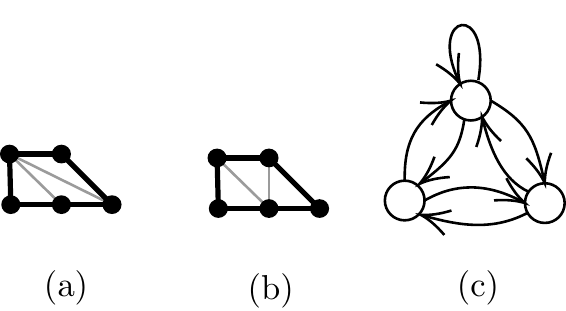}
    \caption{The toric diagram with two different triangulations is shown in (a), (b). They give the same quiver as in (c). In the crystals, (a) has the initial atom corresponding to the node with an adjoint while (b) has the initial atom corresponding to one of the nodes without adjoints.}\label{figSPP}
\end{figure}

\paragraph{Example: SPP} As an example, let us consider the suspended pinched point (SPP) as in Figure \ref{figSPP}; this corresponds to $m=1, n=2$ from the above.

For the crystal from Figure \ref{figSPP}(a), the crystal partition function reads
\begin{equation}
    \begin{split}
        Z_\text{crystal}=&M(q_0q_1q_2)\widetilde{M}(-q_1,q_0q_1q_2)^{-1}\widetilde{M}(-q_2,q_0q_1q_2)^{-1}\widetilde{M}(q_1q_2,q_0q_1q_2)\\
        =&\text{PE}\left[\frac{q_0}{\left(1-q_0^2 q_1^2 q_2^2\right)^2}\left(q_0 q_2^2 \left(q_0 q_2 \left(q_2+1\right)-1\right) q_1^4+q_0 q_2^2 \left(q_0 q_2^2+3 q_0 q_2+2 q_2+q_0+2\right) q_1^3\right.\right.\\
        &+\left(q_2+1\right) \left(q_0^2 q_2^2+q_2-q_0 \left(q_2^3-3 q_2^2-3 q_2+1\right)\right) q_1^2+\left(\left(2 q_0+1\right)q_2^2+\left(2 q_0+3\right) q_2+1\right) q_1\\
        &\left.-q_0 q_2^2+q_2+1\right)\Bigg].
    \end{split}\label{crystalSPP}
\end{equation}
The sign-changed expression is
\begin{equation}
    \begin{split}
        \widetilde{Z}_c=&M(q_0q_1q_2)\widetilde{M}(q_1,q_0q_1q_2)^{-1}\widetilde{M}(q_2,q_0q_1q_2)^{-1}\widetilde{M}(q_1q_2,q_0q_1q_2)\\
        =&\text{PE}\left[\frac{q_0 \left(q_2^2 q_1^2-q_2 q_1^2-q_2^2 q_1+3 q_2 q_1-q_1-q_2+1\right)}{\left(1-q_0 q_1 q_2\right)^2}\right].
    \end{split}
\end{equation}
They have perturbative expansions
\begin{equation}
    Z_\text{crystal}=1+q_0+(q_0^2+q_0q_1+q_0q_2)+(q_0^3+q_0^2q_1+q_0^2q_2+3q_0q_1q_2)+\dots
\end{equation}
and
\begin{equation}
    \widetilde{Z}_c=1+q_0+(q_0^2-q_0q_1-q_0q_2)+(q_0^3-q_0^2q_1-q_0^2q_2+3q_0q_1q_3)+\dots.
\end{equation}
Indeed, the terms only differ by signs. We may take $\mathfrak{q}=q_0=q_1=q_2$ to get the monochrome crystal\footnote{Of course, $\widetilde{Z}_c(\mathfrak{q})=1+\mathfrak{q}-\mathfrak{q}^2+2\mathfrak{q}^3+\dots$ would have different coefficients.}:
\begin{equation}
    \begin{split}
        Z&=\text{PE}\left[\frac{\mathfrak{q}(1+2\mathfrak{q}+3\mathfrak{q}^2+2\mathfrak{q}^3+5\mathfrak{q}^4+6\mathfrak{q}^5+5\mathfrak{q}^6+2\mathfrak{q}^7+3\mathfrak{q}^8+2\mathfrak{q}^9+\mathfrak{q}^{10})}{(1-\mathfrak{q}^6)^2}\right]\\
        &=1+\mathfrak{q}+3\mathfrak{q}^2+6\mathfrak{q}^3+\dots.
    \end{split}\label{crystalSPPq}
\end{equation}
As a byproduct, its asymptotic behaviour is
\begin{equation}
    Z_n\sim\frac{\text{e}^{\frac{7}{3}\zeta'(-1)}\Gamma\left(\frac{1}{6}\right)^{\frac{2}{3}}\zeta(3)^{\frac{17}{108}}}{2^{\frac{23}{54}}3^{\frac{47}{54}}\pi^{\frac{5}{6}}}n^{-\frac{71}{108}}\exp\left(6^{\frac{1}{3}}\zeta(3)^{\frac{1}{3}}n^{\frac{2}{3}}\right).
\end{equation}

Under $q=-q_0q_1q_2$ and $Q_{1,2}=-q_{1,2}$, we have
\begin{equation}
    \begin{split}
        Z_\text{BPS}(q,Q)=&M(-q)^3\widetilde{M}(Q_1,-q)^{-1}\widetilde{M}(Q_2,-q)^{-1}\widetilde{M}(Q_1Q_2,-q)\\
        =&\text{PE}\left[\frac{q}{Q_1^2 Q_2^2\left(1-q^4\right)^2}(-q^6 Q_1^3Q_2^3+q^6 Q_1^2 Q_2^3+q^6 Q_1^3 Q_2^2-3 q^6 Q_1^2 Q_2^2\right.\\
        &+q^6 Q_1 Q_2^2+q^6 Q_1^2 Q_2-q^6 Q_1 Q_2+q^5 Q_1^4 Q_2^4-q^5 Q_1^2 Q_2^4+2 q^5 Q_1^3 Q_2^3-2 q^5 Q_1^2 Q_2^3\\
        &-q^5 Q_1^2-q^5 Q_1^4 Q_2^2-2 q^5 Q_1^3 Q_2^2+9 q^5 Q_1^2 Q_2^2-2 q^5 Q_1
   Q_2^2-q^5 Q_2^2-2 q^5 Q_1^2 Q_2\\
   &+2 q^5 Q_1 Q_2+q^5-3 q^4 Q_1^3 Q_2^3+3 q^4 Q_1^2 Q_2^3+3 q^4 Q_1^3 Q_2^2-9 q^4 Q_1^2 Q_2^2+3 q^4 Q_1 Q_2^2\\
   &+3 q^4 Q_1^2 Q_2-3 q^4 Q_1 Q_2+4 q^3 Q_1^3 Q_2^3-4 q^3 Q_1^2 Q_2^3-4 q^3 Q_1^3 Q_2^2+12 q^3 Q_1^2 Q_2^2-4 q^3 Q_1 Q_2^2\\
   &-4 q^3Q_1^2 Q_2+4 q^3 Q_1 Q_2-3 q^2 Q_1^3 Q_2^3+3 q^2 Q_1^2 Q_2^3+3 q^2 Q_1^3 Q_2^2-9 q^2 Q_1^2 Q_2^2+3 q^2 Q_1 Q_2^2\\
   &+3 q^2 Q_1^2 Q_2-3 q^2 Q_1 Q_2+q Q_1^4 Q_2^4-q Q_1^2 Q_2^4+2 q Q_1^3 Q_2^3-2 q Q_1^2 Q_2^3-q Q_1^2-q Q_1^4 Q_2^2\\
   &-2 q Q_1^3 Q_2^2+9 q Q_1^2 Q_2^2-2qQ_1 Q_2^2-q Q_2^2-2 q Q_1^2 Q_2+2 q Q_1 Q_2+q-Q_1^3 Q_2^3+Q_1^2 Q_2^3\\
   &+Q_1^3 Q_2^2-3 Q_1^2 Q_2^2+Q_1 Q_2^2+Q_1^2 Q_2-Q_1 Q_2)\Bigg]\\
   =&1+\left(-3+Q_1+Q_2+\frac{1}{Q_1}+\frac{1}{Q_2}-Q_1Q_2-\frac{1}{Q_1Q_2}\right)q+\dots.
    \end{split}
\end{equation}
More concisely, with $x=-q$, we have
\begin{equation}
    Z_\text{BPS}=M(x)^3\widetilde{M}(Q_1,x)^{-1}\widetilde{M}(Q_2,x)^{-1}\widetilde{M}(Q_1Q_2,x)=\text{PE}\left[\frac{x(1-Q_1+Q_1Q_2)(1-Q_2+Q_1Q_2)}{Q_1Q_2(1-x)^2}\right].
\end{equation}

From the generalized MacMahon functions, it is straightforward to find out the gluing operators. In particular, the basic generators for $\widetilde{M}(q_1,q_0q_1q_2)^{-1}$ and $\widetilde{M}(q_2,q_0q_1q_2)^{-1}$ are both fermionic. This is consistent with $\sigma_1=-\sigma_2$ and $\sigma_2=-\sigma_{3\equiv0}$. Their derived gluing operators $\widetilde{M}(q_1q_2,q_0q_1q_2)$ are thus bosonic as expected. The shorthand notation is simply
\begin{equation}
    \includegraphics{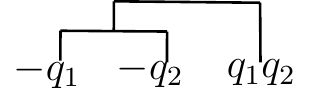},
\end{equation}
where the minus signs indicate the fermionic generators.

Likewise, for Figure \ref{figSPP}(b), we have
\begin{equation}
    \begin{split}
        Z_\text{crystal}&=M(q_0q_1q_3)^3\widetilde{M}(-q_1,q_0q_1q_2)^{-1}\widetilde{M}(q_2,q_0q_1q_2)\widetilde{M}(-q_1q_2,q_0q_1q_2)^{-1}\\
        &=1+q_0+(q_0q_1+q_0q_2)+(3q_0q_1q_2+q_0^2q_1+q_0^2q_2)+\dots
    \end{split}
\end{equation}
and
\begin{equation}
    \begin{split}
        Z_\text{BPS}(q,Q)&=M(-q)^3\widetilde{M}(Q_1,-q)^{-1}\widetilde{M}(Q_2,-q)\widetilde{M}(Q_1Q_2,-q)^{-1}\\
        &=1+\left(-3+Q_1-Q_2+\frac{1}{Q_1}-\frac{1}{Q_2}+Q_1Q_2+\frac{1}{Q_1Q_2}\right)q+\dots.
    \end{split}
\end{equation}
One may also check that the gluing operators follow our discussions above.

As another check, let us consider for instance two copies of the (triangulated) trapezia in Figure \ref{figSPP}(a) glued together. This is SPP$/\mathbb{Z}_2$ with action $(1,0,0,1)$. Its defining equation is $xy=z^2w^4$. Its crystal has four colours with generating function
\begin{equation}
    \begin{split}
        Z_\text{BPS}=&M(x)^6\widetilde{M}(-q_1,x)^{-1}\widetilde{M}(-q_2,x)^{-1}\widetilde{M}(q_3,x)\widetilde{M}(-q_4,x)^{-1}\widetilde{M}(-q_5,x)^{-1}\widetilde{M}(q_1q_2,x)\\
        &\times\widetilde{M}(-q_2q_3,x)^{-1}\widetilde{M}(-q_3q_4,x)^{-1}\widetilde{M}(q_4q_5,x)\widetilde{M}(q_1q_2q_3,x)\widetilde{M}(q_2q_3q_4,x)\widetilde{M}(q_3q_4q_5,x)\\
        &\times\widetilde{M}(-q_1q_2q_3q_4,x)^{-1}\widetilde{M}(-q_2q_3q_4q_5,x)^{-1}\widetilde{M}(q_1q_2q_3q_4q_5,x),
    \end{split}
\end{equation}
where $x=\prod\limits_{i=0}^5q_i$. One may check that under $q_0=\dots=q_5=\mathfrak{q}$, this reduces to the SPP partition without colouring as in \eqref{crystalSPPq}. Moreover, taking $q_0=q_3$, $q_1=q_4$ and $q_2=q_5$, we get the crystal partition function \eqref{crystalSPP} for SPP, that is, the SPP partition with three colours.

\subsection{The Remaining Case: $\mathbb{C}^3/(\mathbb{Z}_2\times\mathbb{Z}_2)$}\label{C3Z2Z2}
\begin{figure}[h]
    \centering
    \includegraphics{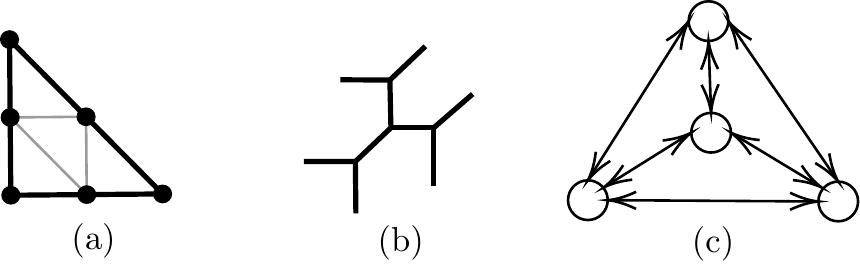}
    \caption{(a) The toric diagram for $\mathbb{C}^3/(\mathbb{Z}_2\times\mathbb{Z}_2)$. (b) Its dual web diagram. (c) The corresponding quiver (the Mercedes-Benz quiver).}\label{figC3Z2Z2}
\end{figure}

Besides generalized conifolds, there is another one which does not have compact four cycles, that is, $\mathbb{C}^3/(\mathbb{Z}_2\times\mathbb{Z}_2)$ as shown in Figure \ref{figC3Z2Z2}.

The generating functions were already obtained in \cite{Young:2008hn,Cirafici:2010bd,Mozgovoy:2020has,mozgovoy2021donaldson}. We have
\begin{equation}
    \begin{split}
        Z_\text{crystal}=&M(q_0q_1q_2q_3)^4\widetilde{M}(-q_1,q_0q_1q_2q_3)^{-1}\widetilde{M}(-q_2,q_0q_1q_2q_3)^{-1}\widetilde{M}(-q_3,q_0q_1q_2q_3)^{-1}\\
        &\times\widetilde{M}(q_1q_2,q_0q_1q_2q_3)\widetilde{M}(q_1q_3,q_0q_1q_2q_3)\widetilde{M}(q_2q_3,q_0q_1q_2q_3)\widetilde{M}(-q_1q_2q_3,q_0q_1q_2q_3)^{-1}
    \end{split}\label{crystalC3Z2Z2}
\end{equation}
and
\begin{equation}
    \begin{split}
        Z_\text{BPS}(q,Q)=&M(-q)^4\widetilde{M}(Q_1,-q)^{-1}\widetilde{M}(Q_2,-q)^{-1}\widetilde{M}(Q_3,-q)^{-1}\widetilde{M}(Q_1Q_2,-q)\widetilde{M}(Q_1Q_3,-q)\\
        &\times\widetilde{M}(Q_2Q_3,-q)\widetilde{M}(Q_1Q_2Q_3,-q)^{-1}.
    \end{split}
\end{equation}
The expressions in terms of PE are rather tedious. Hence, we shall not list them here. Instead, by removing the minus signs, the sign-changed expression $\widetilde{Z}_c$ is more concise:
\begin{equation}
    \begin{split}
        \widetilde{Z}_c=&\text{PE}\left[\frac{q_0}{\left(1-q_0 q_1 q_2 q_3\right)^2}(-q_2^2 q_3^2 q_1^2+q_2 q_3^2 q_1^2+q_2^2 q_3 q_1^2-q_2 q_3 q_1^2+q_2^2 q_3^2 q_1-q_2 q_3^2 q_1\right.\\
    &-q_2 q_1-q_2^2 q_3 q_1+4 q_2 q_3 q_1-q_3 q_1+q_1+q_2-q_2 q_3+q_3-1)\Bigg]
    \end{split}.
\end{equation}
Likewise, again with $x=-q$,
\begin{equation}
    \begin{split}
        Z_\text{BPS}=&\text{PE}\left[\frac{x}{Q_1Q_2Q_3\left(1-x\right)^2}(Q_2^2 Q_3^2 Q_1^2+Q_2 Q_3^2 Q_1^2+Q_2^2 Q_3 Q_1^2+Q_2 Q_3 Q_1^2+Q_2^2 Q_3^2 Q_1\right.\\
    &+Q_2 Q_3^2 Q_1+Q_2 Q_1+Q_2^2 Q_3 Q_1+4 Q_2 Q_3 Q_1+Q_3 Q_1+Q_1+Q_2+Q_2 Q_3+Q_3+1)\Bigg].
    \end{split}
\end{equation}
One may check that $Z_\text{crystal}$ reduces to PE$\left[\frac{\mathfrak{q}}{(1-\mathfrak{q})^2}\right]$, namely the (monochrome) crystal for $\mathbb{C}^3$, when taking $q_{0,1,2,3}=\mathfrak{q}$.

Moreover,
\begin{equation}
    Z_\text{BPS}=\text{PE}\left[\frac{x}{(1-x)^2}\left(2+\prod_{i=1}^4\left(Q_i^{1/2}+Q_i^{-1/2}\right)\right)\right],
\end{equation}
where $Q_4:=Q_1Q_2Q_3$. In particular, it contains the fundamental representation of SU(2)$^4$. Physically, the web diagram decribes the $T[A_{N-1}]$ theory where $N$ M5-branes wrap a sphere with three full punctures when $N=2$ \cite{Benini:2009gi}. Therefore, it should have SU$(2)^3$ flavour symmetry \cite{Gaiotto:2009we}, which is reflected by the factors with $Q_{1,2,3}$. On the other hand, the $Q_4$ part should indicate the $\mathbb{Z}_3$ action on the brane web which reduces the above SU$(2)^3$ to a single SU$(2)$ as discussed in \cite{Acharya:2021jsp}\footnote{Notice that the full flavour symmetry under gauging this $\mathbb{Z}_3$ discrete symmetry would further have an extra SU(3) factor.}.

\paragraph{The gluing process} As shown in Figure \ref{figC3Z2Z2}, there is one trivalent vertex glued to each leg of the centre one. As a result, the gluing operators in this picture would also be different\footnote{However, we should emphasize that the gluing process here is essentially in line with the ones for generalized conifolds. The algebraic gluing rules we have still consist of the corresponding holomorphic curves on the geometric side for topological string amplitudes.}. This is again indicated by the vacuum character. From \eqref{crystalC3Z2Z2}, we see that the basic gluing operators are all fermionic. Furthermore, we have gluing operators associated to $q_iq_j$ for all pairs $(i,j)$ with $i<j$ and $q_1q_2q_3$ all derived from the basic operators. In our shorthand notation, we have
\begin{equation}
    \includegraphics{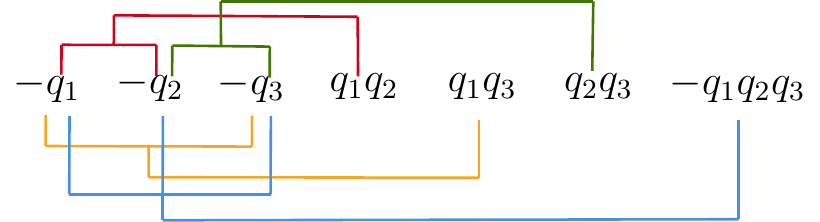}.
\end{equation}

\subsection{Some Non-Toric Examples}\label{nontoric}
Based on the discussions on A-type singularities in \S\ref{CC2Zn}, we may try to generalize to D- and E-type singularities. Now, $\mathbb{C}\times\mathbb{C}^2/\Gamma$ are not toric, where $\Gamma\in\{\text{Dic}_r,\text{BT},\text{BO},\text{BI}\}$ is the binary dihedral/tetrahedral/octaheral/icosahedral group, i.e., the $D_r$ and $E_{6,7,8}$ subgroups of SU(2). Nevertheless, they should still admit quiver descriptions which are the tripled quivers $\widehat{Q}$ of the affine D-/E-type quivers $Q$ \cite{Lawrence:1998ja}.

Similar to \eqref{crystalCC2Zn} and \eqref{adjrep}, we may conjecture that the parititon function in such case is
\begin{equation}
    \chi_r=\text{PE}\left[\frac{x}{(1-x)^2}\left(r+\sum_{\bm{\alpha}\in\Psi}\bm{q}_*^{\bm{\alpha}}\right)\right],\label{ADEpartitionfunction}
\end{equation}
where $x:=\prod\limits_{i=0}^{r-1}q_i^{\delta_i}$ and $\bm{q}_*^{\bm{\alpha}}=\prod\limits_{i=1}^{r-1}q_i^{\alpha_i}$ while $\Psi$ is the root system of the Lie algebra of type $D_{r}$ or $E_{6,7,8}$. In particular, $\left(r+\sum\limits_{\bm{\alpha}\in\Phi}\bm{q}_*^{\bm{\alpha}}\right)$ is the character of the adjoint representation.

Notice that here we let the convention to be $x=\prod\limits_{i=0}^{r-1}q_i^{\delta_i}$ due to the non-trivial minimal positive imaginary root $\bm{\delta}$. For the affine ADE types, $\bm{\delta}$ are the Dynkin labels (dual Coxeter numbers) \cite{kac1990infinite}:
\begin{equation}
    \includegraphics[width=15cm]{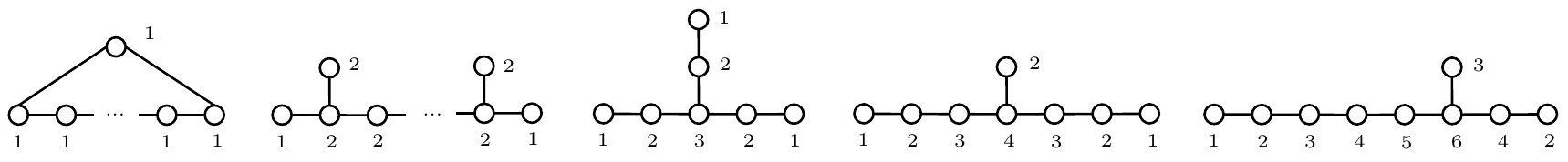}.
\end{equation}
Then $\delta_{0,\dots,r-1}$ should take the values associated to the nodes of the underlying finite quiver.

This is in line with the discussions on Kac polynomials. For affine DE's, the Kac polynomials are \cite{bozec2017number}
\begin{equation}
    \begin{cases}
    A_{\bm{d}}(t)=A_{\bm{d}}^\flat(t)=1,&\qquad\bm{d}\in\Phi_\text{Re}^+\\
    A_{\bm{d}}(t)=A_{\bm{d}}^1(t)=t+r,&\qquad\bm{d}\in\Phi_\text{Im}^+
    \end{cases}.\label{KacADE}
\end{equation}
Here, the notations are the same as in \S\ref{CC2Zn}. The Poincar\'e polynomials are
\begin{equation}
    P_Q^\flat(t,\bm{q})=P_Q(t,\bm{q})=\text{PE}\left[\sum_{\bm{d}\in\Phi_0^+}\frac{t\bm{q}^{\bm{d}}}{(t-1)(1-\bm{q}^{\bm{\delta}})}\right]\text{PE}\left[\frac{(1+rt)\bm{q}^{\bm{\delta}}}{(t-1)(1-\bm{q}^{\bm{\delta}})}\right]\text{PE}\left[\sum_{\bm{d}\in\Phi_0^-}\frac{t\bm{q}^{\bm{d}+\bm{\delta}}}{(t-1)(1-\bm{q}^{\bm{\delta}})}\right].\label{PoincareADE1}
\end{equation}
Again, the double copy $P_Q^0(t,\bm{q})\widetilde{P}_Q^0(t,\bm{q})$ contains $\chi_r$ as a factor:
\begin{equation}
    P_Q^0(t,\bm{q})\widetilde{P}_Q^0(t,\bm{q})=\text{PE}\left[\frac{x}{(1-x)^2}\left(r+\sum_{\substack{\bm{d}\in\Phi_0\\d_0=0}}\bm{q}^{\bm{d}}\right)\right]\times\dots=\chi_r\times\dots\label{PoincareADE2}
\end{equation}
under the unrefinement $t^{-1}=x$. This seems to indicate some subalgebra struture. In \S\ref{refined}, we will check this with the refined partition functions.

It is worth noting that the partition functions $Z_\text{DT}$ and $Z_\text{PT}$ for DT and Pandharipande-Thomas (PT) invariants were obtained in \cite{gholampour2009counting,Mozgovoy:2011ps} for ADE singularities $\mathbb{C}\times\mathbb{C}^2/\Gamma$ with $\Gamma\subset\text{SU}(2)$ finite. One may then verify that \eqref{ADEpartitionfunction} agrees with these results under wall crossings discussed in the next section. More generally, one may also consider all the other affine quivers as classified in \cite[Table Aff 1-3]{kac1990infinite}. Although the 3-fold geometry may not be clear, it would be natural to conjecture that \eqref{ADEpartitionfunction} would still give the partition functions for the tripled quivers of these affine quivers. Moreover, the Kac polynomials and Poincar\'e polynomials would again follow \eqref{KacADE}-\eqref{PoincareADE2}.

\section{Wall Crossings}\label{wallcrossing}
Having presented in detail, in the previous section, explicit expressions for $Z_\text{crystal}$ and $Z_\text{BPS}$ for a variety of examples, let us now move on to discuss the wall-crossing phenomena which have been intensively studied for such partition functions. In this section, we first rapidly summarize some of the standard results in the literature. Then we will make an attempt to generalize the crystal descriptions for arbitrary chambers.

It is well-known that there are walls of marginal stability of codimension 1 in the moduli space of the quiver theory. When BPS particles cross a wall from one chamber to another, they might decay due to the stability conditions. So far, we have only focused on the BPS states in the non-commutative Donaldson-Thomas (NCDT) chamber \cite{Szendroi:2007nu}. It is related to the toplogical string amplitudes by $Z_\text{BPS}=Z_\text{top}(x,\bm{Q})Z_\text{top}(x,\bm{Q}^{-1})$. On the other hand, the BPS parition function in the core chamber is trivially $Z_\text{BPS}=1$. There are many other chambers between these two where the (anti-)D2s on different 2-cycles form stable states with various numbers of D0s.

For example, the most well-studied conifold case has the chamber structure which can be depicted as\footnote{This can be understood as follows. Starting from the region where only the D6 itself is stable (which is known as the core chamber), every time one crosses a wall labeled by $\text{D}2+N\text{D}0$, an arbitrary number of $\text{D}2+N\text{D}0$ can bind to the D6. Then one encounters the D0 wall where any number of D0s can bind to the D6. After that, $\overline{\text{D}2}+N\text{D}0$ particles start to bind to the D6 every time one crosses a $\overline{\text{D}2}+N\text{D}0$ wall.} \cite{Yamazaki:2010fz,Dimofte:2010wxa}
\begin{equation}
    \includegraphics[width=15cm]{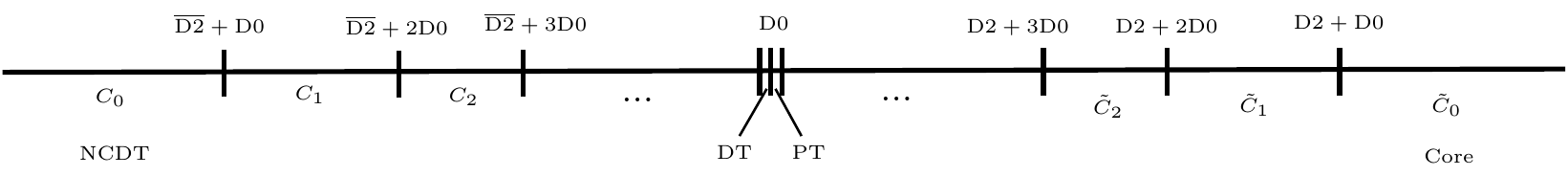}.
\end{equation}
The BPS partition function $Z_\text{BPS}$ in the NCDT/Szendr\"oi chamber is the one discussed in \S\ref{conifold} while $Z_\text{BPS}=Z_\text{DT}=Z_\text{top}(x,Q)$ in the DT chamber. Therefore, one loses a factor $(1-x^kQ^{-1})^k$ when crossing the wall from the chamber $C_{k-1}$ to $C_k$. In the other half, if we start from the core chamber, one obtains a factor $(1-x^kQ)^k$ when crossing the wall from the chamber $\widetilde{C}_{k-1}$ to $\widetilde{C}_k$, and in the PT chamber, we have $Z_\text{BPS}=M(Q,x)^{-1}$ such that the BPS invariants are identified with PT invariants. Let $R$ denote the inverse D0-brane central charge (up to some complex constant)\footnote{This notation $R$ comes from the Taub-NUT circle in the M-theory uplift \cite{Aganagic:2009kf}.}, and let $B$ denote the NS-NS B-field through the 2-cycles wrapped by the D2s in the CY manifold. Then $R>0$ from NCDT to DT chambers while $R<0$ from PT to core chambers. The B-fields satisfy $k-1<B<k$ and $-k-1<B<-k$ respectively. In fact, there is another half with flopped geometry going from NCDT to core chambers. Together the two pieces form a closed circle.

More generally, for any toric CYs without compact 4-cycles, one would obtain/lose a factor $\left(1-x^k\mathtt{Q}^{\pm1}\right)^{\pm k}$ every time we cross a wall of marginal stability similar to the conifold case. Here, $\mathtt{Q}=\left(\prod\limits_{i\in J}Q_i\right)$ where $J$ refers to the set of indices for any possible combination of $Q_i$'s that would appear in $Z_\text{NCDT}$.

For generalized conifolds $xy=z^mw^n$, the BPS partition function in any chamber can be written as \cite{Aganagic:2009kf,nagao2008derived}
\begin{equation}
    Z_\text{BPS}=\prod_{(k,\bm{\beta}):\mathcal{Z}(k,\bm{\beta})>0}\left(1-x^k\bm{Q}^{\bm{\beta}}\right)^{kN_{\bm{\beta}}^0},
\end{equation}
where $\mathcal{Z}$ denotes the central charge and $N_{\bm{\beta}}^0$ is the genus-0 Gopakumar-Vafa invariant specified by the 2-cycle $\bm{\beta}=\sum\limits_{i\leq l\leq j}\alpha_l$ with $\alpha_l$ the basis of 2-cycles. Therefore, $N_{0}^0=|\mathcal{Q}_0|-1=m+n$ and $N_{-\bm{\beta}}^0=N_{\bm{\beta}}^0$. The central charge is $\mathcal{Z}=(k+B(\bm{\beta}))/R$ where $B(\bm{\beta})$ is the B-field flux through the 2-cycle $\bm{\beta}$. Recall that $\sigma=\{\sigma_l\}$ denotes the signs of simplices in the triangulation in \S\ref{genC}. Then \cite{Nagao:2009rq}
\begin{equation}
    N_{\bm{\beta}}^0=(-1)^{1+\#\{l\in[i,j]:~\alpha_l\text{ is an }\mathcal{O}(-1,-1)\text{-curve}\}}=(-1)^{1+\#\{l\in[i,j]:~\sigma_l\neq\sigma_{l+1}\}}.
\end{equation}
The BPS partition function is therefore
\begin{equation}
    Z_\text{BPS}=M(x)^{m+n}\prod_{0<r\leq s<m+n}\left(M\left(\prod_{i=r}^sQ_i,x\right)M_\wedge\left(\prod_{i=r}^sQ_i^{-1},x;B_{r,\dots,s}\right)\right)^{(-1)^{\frac{\sigma_r-\sigma_{s+1}}{2}}}\label{BPSchamber1}
\end{equation}
or
\begin{equation}
    Z_\text{BPS}=\prod_{0<r\leq s<m+n}M^\wedge\left(\prod_{i=r}^sQ_i,x;B_{r,\dots,s}\right)^{(-1)^{\frac{\sigma_r-\sigma_{s+1}}{2}}}\label{BPSchamber2}
\end{equation}
based on the chamber, where $B_{r,\dots,s}:=[B(\alpha_r+\dots+\alpha_s)]$, which labels the chamber, is the integer part of the value of the B-field through the 2-cycle $\bm{\beta}=\alpha_r+\dots+\alpha_s$.

The remarkable result in \cite{Nagao:2009rq} says that $[B]$ are not completely independent and can be determined by the map $\theta:\frac{1}{2}\mathbb{Z}_\text{odd}\rightarrow\frac{1}{2}\mathbb{Z}_\text{odd}$ such that $\theta(h+m+n)=\theta(h)+m+n$ for any half-integer $h$ and $\sum\limits_{i=1}^{m+n}\theta(i-\frac{1}{2})=\sum\limits_{i=1}^{m+n}(i-\frac{1}{2})$. If $\theta(1/2)<\theta(3/2)<\dots<\theta(m+n-1/2)$, then
\begin{equation}
    [B_{\theta}(\alpha_r+\dots+\alpha_s)]=\#\{k\in\mathbb{Z}|\theta(r-1/2)<k(m+n)<\theta(s+1/2)\}.
\end{equation}
If $\theta$ is not increasing, then we can choose a permutation $\tau\in\mathfrak{S}_{m+n}$ such that $\theta(\tau(1/2))<\theta(\tau(3/2))<\dots<\theta(\tau(m+n+1/2))$ and replace $\theta$ by $\theta\circ\tau$. For instance, in the SPP example in Figure \ref{figSPP}, if $\theta(1/2)=11/2,\theta(3/2)=3/2,\theta(5/2)=-5/2$, then $[B_{\theta\circ\tau}(\alpha_1)]=[B_{\theta\circ\tau}(\alpha_2)]=1,[B_{\theta\circ\tau}(\alpha_1+\alpha_2)]=2$ where $\tau=(132)$. This specifies the truncations of MacMahon functions in \eqref{BPSchamber1} and \eqref{BPSchamber2}. Notice that $\theta(1/2)=-5/2,\theta(3/2)=3/2,\theta(5/2)=11/2$ gives $[B_{\theta}]$ of the same values, but generically they parametrize different chambers \cite{Nagao:2009rq}.

It is also straightforward to write $Z_\text{BPS}$ in different chambers using PE. This simply follows from
\begin{equation}
    M_\wedge(p,q;k_0)=\text{PE}\left[\sum_{k=k_0}^\infty kpq^k\right],\quad M^\wedge(p,q;k_0)=\text{PE}\left[\sum_{k=1}^{k_0} kpq^k\right],
\end{equation}
along with $\text{PE}[f]\text{PE}[g]=\text{PE}[f+g]$ and $\text{PE}[f]^{-1}=\text{PE}[-f]$.

\subsection{Towards a Crystal Description}\label{crystaldescription}
It could be possible that there are certain crystal models describing other chambers. It is well-known that the crystal in the chamber $C_N$ for the conifold is the pyramid partition with a ridge of $(N+1)$ atoms on the top row. The crystal partition function reads \cite{Szendroi:2007nu,young2009computing}
\begin{equation}
    Z_\text{crystal}\left(q_{0,N},q_{1,N}\right)=M(q_{0,N}q_{1,N})^2M\left(-q_{0,N}^{N-1}q_{1,N}^N,q_{0,N}q_{1,N}\right)^{-1}M_\wedge\left(-q_{0,N}^{-(N-1)}q_{1,N}^{-N},q_{0,N}q_{1,N};N\right)^{-1},
\end{equation}
where $q_{i,N}$ is the variable for the atom of $i^\text{th}$ colour in the crystal for $C_N$. Under $q_{0,N}=q_0^Nq_1^{N-1}$ and $q_{1,N}=q_0^{-N+1}q_1^{-N+2}$, we obtain
\begin{equation}
    Z_\text{crystal}(q_0,q_1)=M(q_0q_1)^2M(-q_1,q_0q_1)^{-1}M_\wedge(-q_1,q_0q_1;N)^{-1}.
\end{equation}
This is similar in the $\widetilde{C}_N$ chamber, where the crystal is finite and
\begin{equation}
    Z_\text{crystal}\left(q_{0,N},q_{1,N}\right)=M^\wedge\left(-q_{0,N}^{N+1}q_{1,N}^N,q_{0,N}^{-1}q_{1,N}^{-1};N\right)^{-1}
\end{equation}
with $q_{0,N}=q_0^{N}q_1^{N+1}$, $q_{1,N}=q_0^{-N-1}q_1^{-N-2}$ \cite{Chuang:2009crq}.

\begin{figure}[h]
    \centering
    \includegraphics{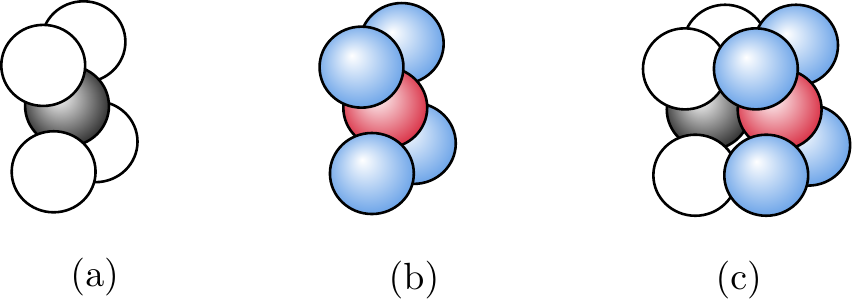}
    \caption{(a) The crystal for $\widetilde{C}_2$ for the conifold $\mathcal{C}$. (b) The same copy with different colours. (c) The crystal which is a disjoint union of (a) and (b).}\label{figtC2tC2}
\end{figure}

For a general CY, it is still not clear whether there is a crystal for every chamber. We conjecture that such crystal should exist, at least upon ``artificial'' constructions. For instance, the crystal for $\widetilde{C}_2$ for the conifold is shown in Figure \ref{figtC2tC2}(a).
Together with another copy with different colours in Figure \ref{figtC2tC2}(b), we have a crystal with two disjoint parts as in Figure \ref{figtC2tC2}(c). In other words, they form a crystal whose white-black part and blue-red part have no chemical bonds between each other. More generally, for two copies for the chamber $\widetilde{C}_N$ of the conifold, this gives the partition function
\begin{equation}
    \begin{split}
        Z_\text{crystal}&=M^\wedge\left(-q_\text{white}^{N+1}q_\text{black}^N,q_\text{white}^{-1}q_\text{black}^{-1};N\right)^{-1}M^\wedge\left(-q_\text{blue}^{N+1}q_\text{red}^N,q_\text{blue}^{-1}q_\text{red}^{-1};N\right)^{-1}\\
        &=M^\wedge\left(-q_1,x;N\right)^{-1}M^\wedge\left(-q_3,x;N\right)^{-1},
    \end{split}
\end{equation}
where the second line is obtained under the substitutions $q_\text{white}=q_0^Nq_1^{N+1}q_2^Nq_3^N$, $q_\text{black}=q_0^{-N-1}q_1^{-N-2}q_2^{-N-1}q_3^{-N-1}$, $q_\text{blue}=q_0^Nq_1^Nq_2^Nq_3^{N+1}$, $q_\text{red}=q_0^{-N-1}q_1^{-N-1}q_2^{-N-1}q_3^{-N-2}$ and $x=q_0q_1q_2q_3$. Notice that however, this partition function does not correspond to any chamber for $\mathcal{C}/\mathbb{Z}_2$ due to the constraints on $[B]$ discussed above. To recover the partition function of certain chamber, one should not only consider such crystal associated to $M^\wedge(\bullet,\bullet;N)^{-1}$, but also consider the crystal for $\widetilde{C}_N$ of $\mathbb{C}\times\mathbb{C}^2/\mathbb{Z}_2$. This is because in general we would also have $M^\wedge(\bullet,\bullet;N)$ in the partition function.

\begin{figure}[h]
    \centering
    \includegraphics[width=10cm]{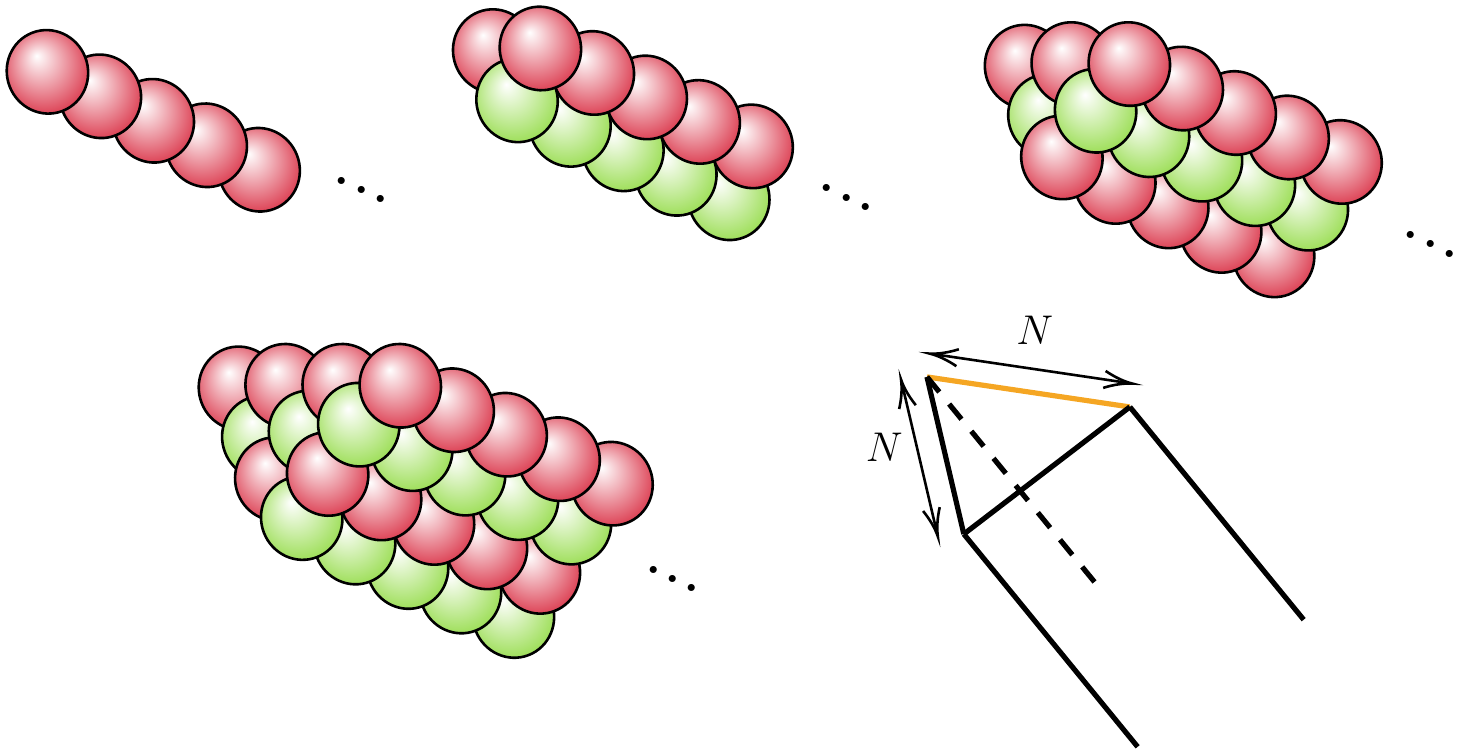}
    \caption{The crystal for chamber $\widetilde{C}_N$ for $\mathbb{C}\times\mathbb{C}^2/\mathbb{Z}_2$. Here we show the cases for $N=1,2,3,4$. We also give a sketch of general $N$ with the orange line as the top row. The crystal is infinitely long.}\label{figtriangularlog}
\end{figure}

Here, we propose that the (natural) crystal for the chamber $\widetilde{C}_N$ for $\mathbb{C}\times\mathbb{C}^2/\mathbb{Z}_2$ has the shape of a tilted (semi-)infinite ``triangular log store'' as shown in Figure \ref{figtriangularlog}.
The crystal partition function is
\begin{equation}
    Z_\text{crystal}\left(q_{0,N},q_{1,N}\right)=M^\wedge\left(q_{0,N}^{N+1}q_{1,N}^N,q_{0,N}^{-1}q_{1,N}^{-1};N\right).
\end{equation}
Under $q_{0,N}=q_0^{N}q_1^{N+1}$, $q_{1,N}=q_0^{-N-1}q_1^{-N-2}$, we recover $M^\wedge(q_1,q_0q_1;N)$ as expected. As an illustration, we list the perturbative expansion of $Z_\text{crystal}\left(q_{0,N},q_{1,N}\right)$ for some small $N$:
\begin{equation}
    \begin{split}
        N=1:~&1+q_{0,1}+q_{0,1}^2+q_{0,1}^3+\dots;\\
        N=2:~&1+2q_{0,2}+(3+q_{1,2})q_{0,2}^2+(4+2q_{1,2})q_{0,2}^3+\dots;\\
        N=3:~&1+3q_{0,3}+(6+2q_{1,3})q_{0,3}^2+(10+6q_{1,3}+q_{1,3}^2)q_{0,3}^3+\dots;\\
        N=4:~&1+4q_{0,4}+(10+3q_{1,3})q_{0,4}^2+(20+12q_{1,4}+2q_{1,4}^2)q_{0,4}^3+\dots.
    \end{split}
\end{equation}

Now, any chamber $\widetilde{C}$ for any toric CY without compact 4-cycles could be represented by a disjoint union of the crystals in Figure \ref{figtC2tC2} and Figure \ref{figtriangularlog}. For instance, three copies of Figure \ref{figtC2tC2} and two copies of Figure \ref{figtriangularlog} (all with distinct colours) yield the chamber with
\begin{equation}
    Z_\text{crystal}=M^\wedge\left(-q_1,x;1\right)^{-1}M^\wedge\left(-q_3,x;1\right)^{-1}M^\wedge\left(q_1q_2,x;1\right)M^\wedge\left(q_2q_3,x;1\right)M^\wedge\left(-q_1q_2q_3,x;2\right)^{-1}
\end{equation}
for $\mathcal{C}/\mathbb{Z}_2$. The maps from $q_{i,N}$ to $q_j$ should be straightforward from the above discussions\footnote{One may check that this indeed corresponds to some chamber. For example, the $\theta$ map can be chosen as $\theta(1/2)=-7/2$, $\theta(3/2)=3/2$, $\theta(5/2)=7/2$ and $\theta(7/2)=9/2$.}.

In the case of $\mathbb{C}\times\mathbb{C}^2/\mathbb{Z}_n$ or $\mathcal{C}/\mathbb{Z}_n$, a more natural crystal could be the same as the one for $\mathbb{C}\times\mathbb{C}^2/\mathbb{Z}_2$ or $\mathcal{C}$ but with more colours. Nevertheless, this artificial method allows us to construct the crystal for arbitrary chamber $\widetilde{C}$ for any toric CY without compact 4-cycles.

\begin{figure}[h]
    \centering
    \includegraphics{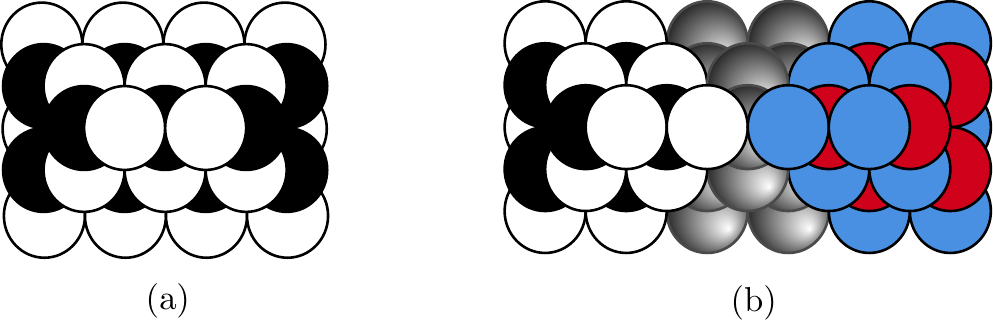}
    \caption{(a) The crystal for $C_1$ for the conifold $\mathcal{C}$. (b) ``Merging'' two such copies. Here, we still stay in the case $\mathcal{C}$, but with a different chamber. Each copy gives $M(x)^2\widetilde{M}_\wedge(2)^{-1}$ where $\widetilde{M}_\wedge(2):=M(\prod q_i,x)M_\wedge(\prod q_i,x;2)$. After merging the shaded pyramid (where the colours are ignored), we reach the chamber $C_3$ with partition function $M^2\widetilde{M}_\wedge(4)^{-1}$ with the blue (red) colour identified with the white (black) colour.}\label{figmerge}
\end{figure}

One may consider a similar construction for a chamber $C$ such that there is a crystal model for each $(M(\prod q_i,x)M_\wedge(\prod q_i,x;N))^{\pm1}$ where $\pm1$ determines the crystal being either pyramid partition or bicoloured plane partition. Then the union of the crystals would give all the factors in the partition function. For such constructions, we need to point out the followings:
\begin{itemize}
    \item There could be more colours $q_{i,\cup}$ of this union than the actual number of variables $q_i$. Therefore, the map from $\{q_{i,\cup}\}$ to $\{q_i\}$ should reduce such number. This is similar to the case for $\widetilde{C}$.
    \item Every (sub-)crystal in the union would introduce a factor of $M(x)^2$ in the product. To remove these extra factors, we need to make identifications of some atoms when gluing the crystals together. For each factor of $M(x)$, a pair of $\mathbb{C}^3$ sub-crystal in the union should ``merge'' into one. For some special/simpler cases, one may also consider merging a different sub-crystal. This is illustrated in Figure \ref{figmerge} where the CY geometry is not even changed but we have a different chamber\footnote{Of course, for general CY it would be easier to consider merging its own crystals rather than combining copies of bicoloured pyramid or plane partition and identifying $\mathbb{C}^3$ sub-crystals. However, the premise is to know the crystals for this general CY in different (or at least a few) chambers.}.
    \item After merging, the truncations $N$ in the (remaining) factors $M_\wedge(\bullet,\bullet;N)$ could change. Again, Figure \ref{figmerge} provides an example. It could be possible that cancelling the surplus colours in $\{q_{i,\cup}\}$ would simultaneously correct $N$ in the remaining $M_\wedge(\bullet,\bullet;N)$.
\end{itemize}

It is not clear whether such construction would give a ``natural'' crystal description of the BPS states in different chambers. Nevertheless, if there does exist a natural crystal description, the 2d projection of the crystal shape should coincide with the web diagram of the toric CY. This is because the thickening of the web would give the 2d projection of the crystal melting in the thermodynamic limit\footnote{The thickening of the web is known as the amoeba \cite{Kenyon:2003uj,Feng:2005gw}. As the (thermodynamic) limit shape of the crystal and the amoeba are general features for any CY, we expect the discussion here would also work for CYs with compact 4-cycles.}. Then the tops of the crystals would be the finite ridges in the webs with different numbers of coloured atoms for different chambers.

\begin{figure}[h]
    \centering
    \includegraphics[width=15cm]{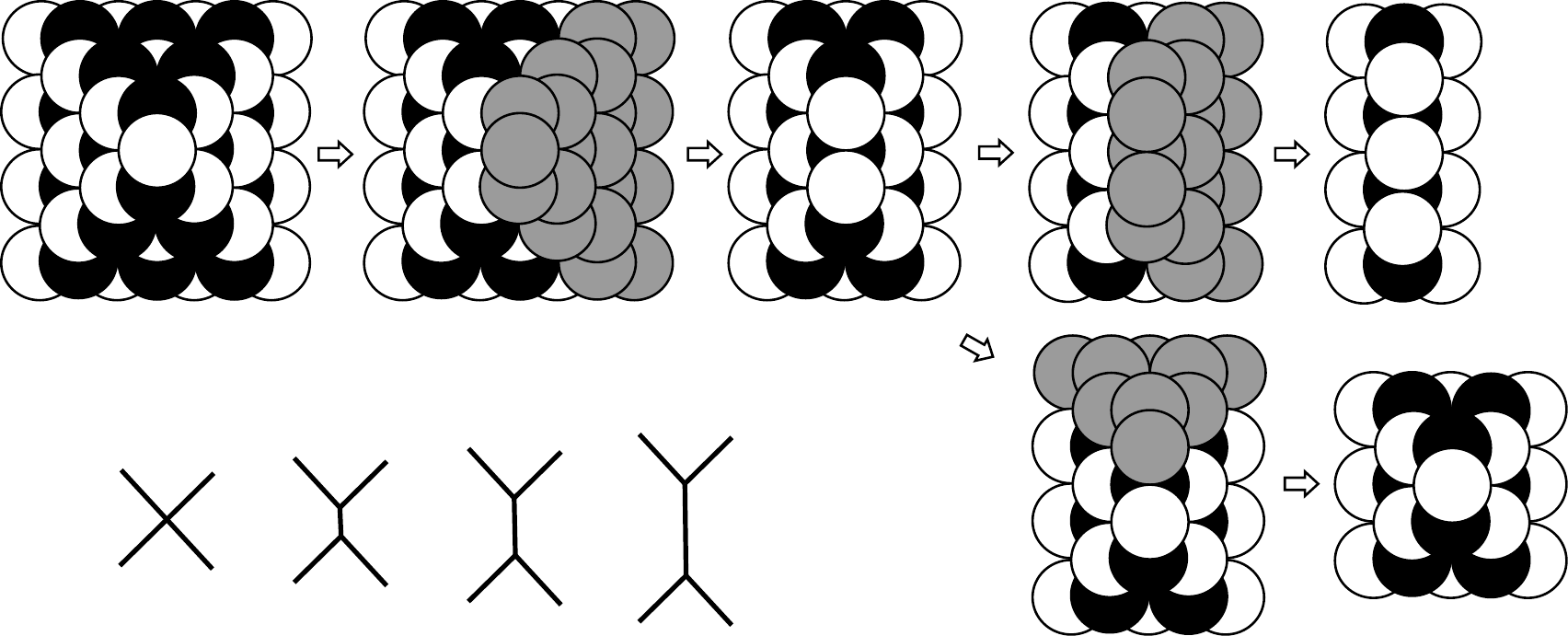}
    \caption{Peeling off the semi-infinite faces of the conifold crystal. This changes the length of the top row.}\label{figconifoldpeel}
\end{figure}

Let us take a closer look at the bicoloured crystals for $\mathbb{C}\times\mathbb{C}^2/\mathbb{Z}_2$ and the conifold in the chambers $C_N$. As shown in Figure \ref{figconifoldpeel}, we can ``peel'' one semi-infinite face (in grey) off the crystal for the conifold. This then leads to the crystal for chamber $C_1$. Keep peeling the semi-infinite face on the same side, and we can reach the crystal for any $C_N$.
In the web diagram which corresponds to the ridges of the crystal, peeling the semi-infinite face is actually changing the length of the internal line, that is, varying the K\"ahler moduli. This indicates that removing a factor of $(1-x^kQ^{-1})^k$ in the partition function corresponds to peeling a semi-infinite face off the crystal. If we peel another semi-infinite face as in the second row in Figure \ref{figconifoldpeel}, we can see that this is going the opposite direction in the moduli space, and we get back to the NCDT chamber $C_0$ from $C_1$.

\begin{figure}[h]
    \centering
    \includegraphics[width=15cm]{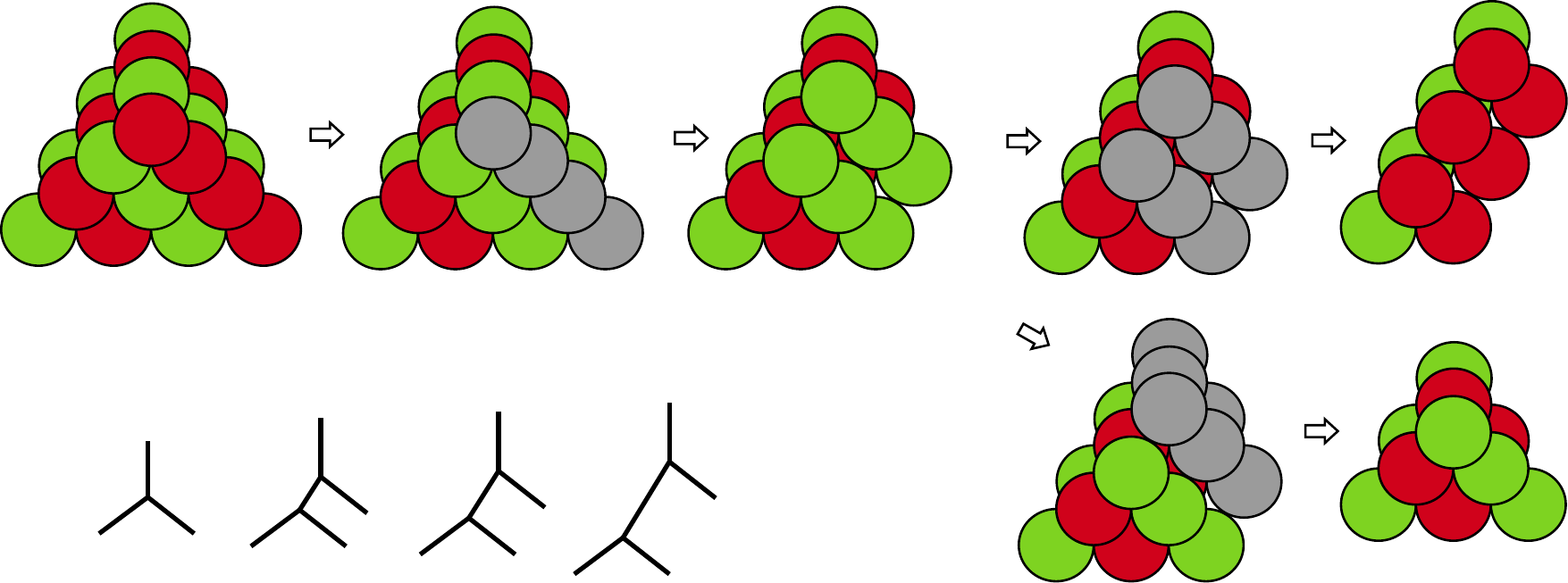}
    \caption{Peeling off the semi-infinite faces of the $\mathbb{C}\times\mathbb{C}^2/\mathbb{Z}_2$ crystal.}\label{figCC2Z2peel}
\end{figure}

Now we propose a similar construction for $\mathbb{C}\times\mathbb{C}^2/\mathbb{Z}_2$. In Figure \ref{figCC2Z2peel}, if we peel one semi-infinite ridge (in grey) off the crystal, we would reach the chamber $C_1$. Then keep peeling the semi-infinite face on the same side, and we can reach the crystal for any $C_N$.
In the web diagram, this is again changing the length of the internal line, that is, varying the K\"ahler moduli. This corresponds to removing a factor of $(1-x^kQ^{-1})^{-k}$ in the partition function. Similar to the conifold, the crystal partition function in this case should be given by
\begin{equation}
    Z_\text{crystal}\left(q_{0,N},q_{1,N}\right)=M(q_{0,N}q_{1,N})^2M\left(q_{0,N}^{N-1}q_{1,N}^N,q_{0,N}q_{1,N}\right)M_\wedge\left(q_{0,N}^{-(N-1)}q_{1,N}^{-N},q_{0,N}q_{1,N};N\right),
\end{equation}
with $q_{0,N}=q_0^Nq_1^{N-1}$ and $q_{1,N}=q_0^{-N+1}q_1^{-N+2}$.

\begin{figure}[h]
    \centering
    \includegraphics[width=15cm]{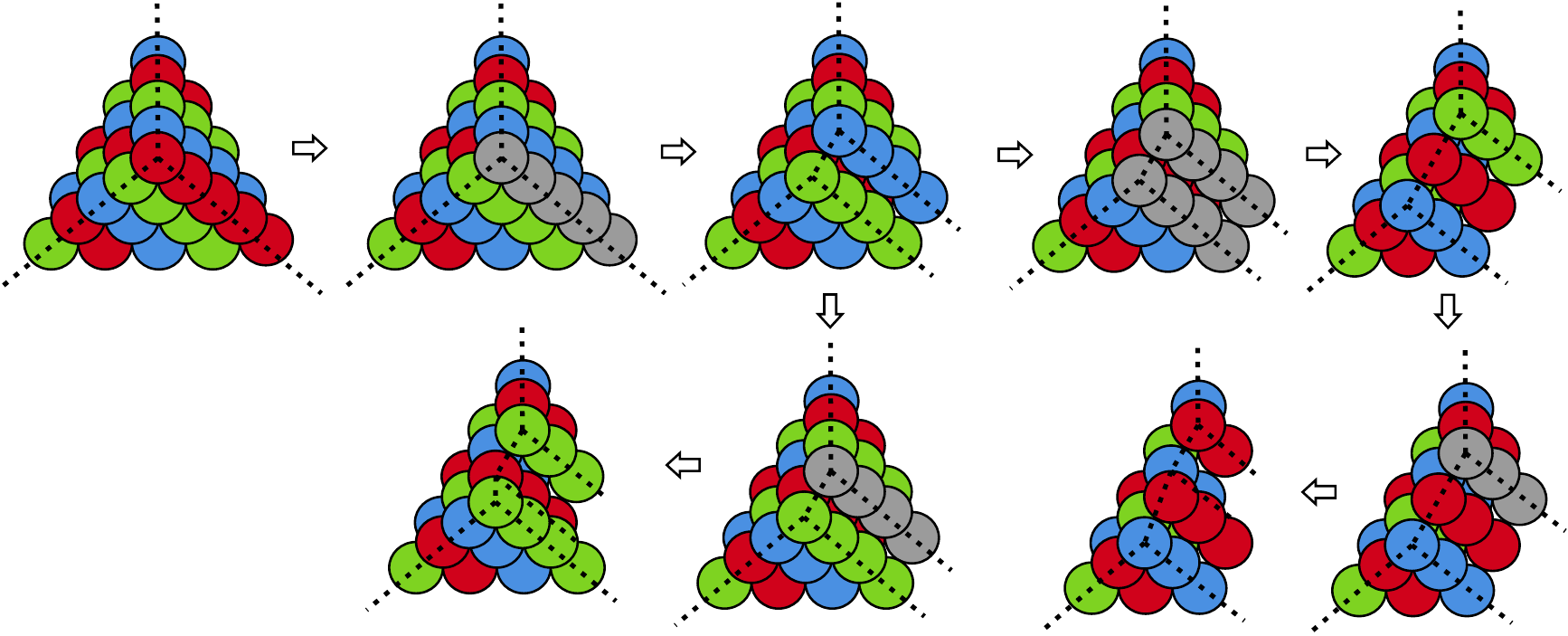}
    \caption{Peeling off the semi-infinite faces of the $\mathbb{C}\times\mathbb{C}^2/\mathbb{Z}_3$ crystal. The dashed brane webs is composed of the ridges.}\label{figCC2Z3peel}
\end{figure}

This peeling process can then be generalized to any toric CY. Every time we cross a wall, a semi-infinite face (with a ridge being a degenerate face) is peeled off the crystal. This corresponds to losing/obtaining a factor of $(1-x^k\prod Q_i^{-1})^{\pm k}$, where the sign in the power is determined by the curve ($\mathcal{O}(-2,0)$ or $\mathcal{O}(-1,-1)$) for the internal line in the web, or equivalently, the signs in $\sigma$. As an example, we illustrate several different ways of peeling for $\mathbb{C}\times\mathbb{C}^2/\mathbb{Z}_3$ in Figure \ref{figCC2Z3peel}.

In general, the initial atoms are at the intersections of (at least) two semi-infinite ridges. Moreover, these initial atoms do not have to lie at the same ``height'' in the crystal.

\subsection{Refined Partition Functions}\label{refined}
For any chamber $\mathtt{C}$, the refined BPS index/(protected) spin character is $\Omega(n_0,\bm{n_2};\mathtt{y};\mathtt{C})=\Tr_{\mathcal{H}_{n_0,\bm{n_2}}(\mathtt{C})}(-\mathtt{y})^{J_3}$, where $\mathcal{H}$ is the (reduced) Hilbert space of BPS states and $\mathtt{y}$ tracks the spin information $J_3$. In the limit $\mathtt{y}\rightarrow1$, one recovers the unrefined index. In the following, it would be more convenient to take $t_1=q\mathtt{y}$ and $t_2=q/\mathtt{y}$.

It is fairly straightforward to refine the partition functions discussed above:
\begin{equation}
    Z_\text{BPS}=M_R(t_1,t_2)^{m+n}\prod_{0<r\leq s<m+n}\widetilde{M}_R\left(\prod_{i=r}^sQ_i;t_1,t_2\right)^{(-1)^{\frac{\sigma_r-\sigma_{s+1}}{2}}}
\end{equation}
for the generalized conifold $xy=z^mw^n$ \cite{Iqbal:2007ii,Taki:2007dh} and
\begin{equation}
    Z_\text{BPS}=M_R(t_1,t_2)^4\prod_{I\in\mathcal{P}\{1,2,3\}}\widetilde{M}_R\left(\prod_{i\in I}Q_i;t_1,t_2\right)^{(-1)^{|I|}}
\end{equation}
for $\mathbb{C}^3/(\mathbb{Z}_2\times\mathbb{Z}_2)$ where $\mathcal{P}\{1,2,3\}$ is the power set of $\{1,2,3\}$. Here,
\begin{equation}
    \begin{split}
        &M_R(p;t_1,t_2)=\prod_{k,l=0}^\infty\frac{1}{1-pt_1^{k+1}t_2^l}=\text{PE}\left[\frac{pt_1}{(1-t_1)(1-t_2)}\right],\\ &M_R(t_1,t_2)=M_R(1;t_1,t_2),\quad \widetilde{M}_R(p;t_1,t_2)=M_R(p;t_1,t_2)M_R(p^{-1};t_1,t_2)
    \end{split}
\end{equation}
are the refined (generalized) MacMahon functions. In terms of PE, we have $Z_\text{BPS}=\text{PE}[g]$, where
\begin{equation}
    g=\frac{t_1\left(m+n+\sum\limits_{0<r\leq s<m+n}(-1)^{\frac{\sigma_r-\sigma_{s+1}}{2}}\left(\prod\limits_{i=r}^sQ_i+\prod\limits_{i=r}^sQ_i^{-1}\right)\right)}{(1-t_1)(1-t_2)}
\end{equation}
for the generalized conifold and
\begin{equation}
    g=\frac{t_1\left(4+\sum\limits_{I\in\mathcal{P}\{1,2,3\}}(-1)^{|I|}\left(\prod\limits_{i\in I}Q_i+\prod\limits_{i\in I}Q_i^{-1}\right)\right)}{(1-t_1)(1-t_2)}
\end{equation}
for $\mathbb{C}^3/(\mathbb{Z}_2\times\mathbb{Z}_2)$.

Recall that in the unrefined case, one would obtain/lose a factor of $\left(1-x^N\mathtt{Q}^{\pm1}\right)^{\pm N}$ when crossing a wall of marginal stability, where $\mathtt{Q}=\left(\prod\limits_{i\in J}Q_i\right)$ for a set $J$ of indices whose combination would appear in $Z_\text{NCDT}$. Likewise, in the refinement, one would obtain/lose a factor of $\left(\prod\limits_{k+l+1=N}(1-t_1^{k+1}t_2^{l}\mathtt{Q}^{\pm1})\right)^{\pm1}$ every time we cross a wall\footnote{It is conjectured that there does not exist walls invisible to unrefined indices such that only refined indices would jump \cite{Dimofte:2010wxa}.}.

We can also directly compare the refined partition functions with the previous discussions on Kac polynomials and Poincar\'e polynomials. Indeed, the refined $\mathbb{C}^3$ partition function is $M_R(t_1,t_2)$, which is exactly \eqref{C3Kac} under the change of variables $x=t_1$ and $t^{-1}=t_2$. One may also check that the results for all the affine quiver cases (in \S\ref{CC2Zn} and \S\ref{nontoric}) still hold. The partition function is
\begin{equation}
    \chi_r=\text{PE}\left[\frac{t_1}{(1-t_1)(1-t_2)}\left(r+\sum_{\bm{\alpha}\in\Psi}\bm{Q}^{\bm{\alpha}}\right)\right].
\end{equation}
This is precisely a factor of $P_Q^0(t,\bm{q})\widetilde{P}_Q^0(t,\bm{q})$ under $x=\bm{q}^{\bm{\delta}}=t_1$, $t^{-1}=t_2$ and $q_i=Q_i$.

\section{A Comment on D4-D2-D0 Bound States}\label{D4D2D0}
\begin{figure}[h]
    \centering
    \includegraphics{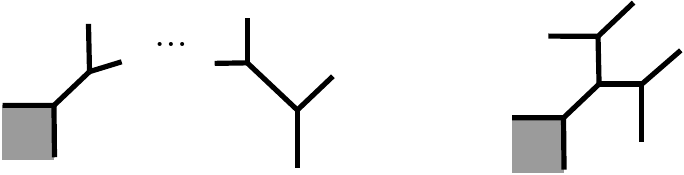}
    \caption{The non-compact toric divisor where a single D4 is supported for generalized conifolds and $\mathbb{C}^3/(\mathbb{Z}_2\times\mathbb{Z}_2)$.}\label{figdivisor}
\end{figure}

Based on \cite{Nishinaka:2010qk,Nishinaka:2010fh,Nishinaka:2011sv,Nishinaka:2011is,Nishinaka:2013mba}, it would be straightforward to write the generating functions for D4-D2-D0 brane bound states similar to the above discussions. Mathematically, they are related to curve counting on surfaces in the CY 3-fold \cite{Cirafici:2012qc,Gholampour:2013ifa}. Given a toric CY 3-fold, we consider a single non-compact D4-brane wrapping the shaded toric divisor as shown in Figure \ref{figdivisor}, with BPS D2-D0 branes bound to it\footnote{Here, we will only focus on this specific non-compact divisor that supports the single D4-brane. It would be nice to generalize this to other 4-cycles and investigate how these partition functions are related to each other.}.

As argued in \cite{Nishinaka:2013mba}, the D4-D2-D0 bound states can be enumerated by 2-dimensional crystals as opposed to the 3d crystals for D6-D2-D0 bound states. As a result, they should be counted via 2d Young tableaux instead of 3d plane partitions. Therefore, it is natural to conjecture that the (generalized) MacMahon functions should be replaced by the inverse (generalized) Euler functions $\phi(x,q)^{-1}$ counting integer partitions, where the (generalized) Euler functions are
\begin{equation}
    \phi(p,q):=\prod_{k=1}^\infty(1-pq^k),\quad\phi(q):=M(1,q),\quad\widetilde{\phi}(p,q):=\phi(p,q)\phi(p^{-1},q).
\end{equation}
The partition function is then
\begin{equation}
    Z_\text{BPS}=\phi(x)^{-(m+n)}\prod_{0<r\leq s<m+n}\widetilde{\phi}\left(\prod_{i=r}^sQ_i,x\right)^{(-1)^{\frac{\sigma_r-\sigma_{s+1}}{2}+1}}
\end{equation}
for generalized conifolds and
\begin{equation}
    Z_\text{BPS}=\phi(x)^{-1}\prod_{I\in\mathcal{P}\{1,2,3\}}\widetilde{\phi}\left(\prod_{i\in I}Q_i,x\right)^{(-1)^{|I|+1}}
\end{equation}
for $\mathbb{C}^3/(\mathbb{Z}_2\times\mathbb{Z}_2)$. One may check that these expressions agree with the results of $\mathbb{C}^3$ orbifolds, conifold and SPP studied in \cite{Nishinaka:2013mba} up to wall crossings. In terms of PE, we have $Z_\text{BPS}=\text{PE}[g]$, where
\begin{equation}
    g=\frac{x\left(m+n+\sum\limits_{0<r\leq s<m+n}(-1)^{\frac{\sigma_r-\sigma_{s+1}}{2}}\left(\prod\limits_{i=r}^sQ_i+\prod\limits_{i=r}^sQ_i^{-1}\right)\right)}{1-x}
\end{equation}
for the generalized conifold and
\begin{equation}
    g=\frac{x\left(4+\sum\limits_{I\in\mathcal{P}\{1,2,3\}}(-1)^{|I|}\left(\prod\limits_{i\in I}Q_i+\prod\limits_{i\in I}Q_i^{-1}\right)\right)}{1-x}
\end{equation}
for $\mathbb{C}^3/(\mathbb{Z}_2\times\mathbb{Z}_2)$. Likewise, following \cite{Nishinaka:2010qk}, every time we cross a wall of marginal stability, we obtain/lose a factor of $\left(1-x^k\mathtt{Q}^{\pm1}\right)^{\pm1}$. Here, $\mathtt{Q}=\left(\prod\limits_{i\in J}Q_i\right)$ where $J$ refers to the set of indices for any possible combination of $Q_i$'s that would appear in the NCDT chamber.

\section{Outlook}\label{outlook}
In this paper, we discussed crystal and BPS partition functions for toric CY 3-folds without compact 4-cycles as well as some non-toric examples. There are still undoubtedly rich explorations for future research. Here, we shall list a very small subset of them.

One longstanding problem is to consider the CYs with compact 4-cycles. The D4-branes wrapping those compact divisors would then also become dynamical BPS particles. There have been some calculations counting D6-D4-D2-D0 bound states for some cases such as in \cite{Cirafici:2010bd,Mozgovoy:2020has,Descombes:2021snc}. It would be intriguing to study the methods in these papers and apply them to any general cases.

The refined partition functions are closely related to the motivic DT invariants and various quantum algebras \cite{Joyce:2008pc,Kontsevich:2008fj}. For instance, as proven in \cite{Davison:2016bjk}, the associated grading of COHA with respect to the perverse filtration is the symmetric algebra of (the product of) the BPS algebra (and an extra piece). In particular, the refined BPS invariants would now appear inside PE to give the generating function for $\text{gr}^{\cdot}\text{COHA}$. It would also be interesting to find a more systematic relation between quiver Yangians and other relevant algebras. For example, it was conjectured in \cite{Davison:2013nza} that the positive part of the Maulik-Okounkov Yangian \cite{maulik2019quantum} of a quiver is isomorphic to the COHA associated to its tripled quiver. Moreover, recent progress on related quantum algebras and crystals has been made including shifted Yangians, toroidal and (hyper)elliptic BPS algebras \cite{Rapcak:2020ueh,Galakhov:2021xum,Galakhov:2021vbo,Noshita:2021dgj}. In particular, it would be interesting to compare the crystals for different chambers in this paper with those studied in \cite{Galakhov:2021xum}. In another direction, we may also compare the crystal descriptions here with the discussions in \cite{Aganagic:2010qr}.

It is natural to investigate the BPS/CFT correspondence, which is also known as the AGT or 2d-4d correspondence \cite{Alday:2009aq,Nekrasov:2015wsu,Feigin:2018bkf}. One may consider the M-theory lift of the type IIA picture with an extra $S^1$. Then we have M5-branes wrapping the cylinder $\mathcal{M}_2=\mathbb{R}\times S^1$ and some 4-dimensional variety $\mathcal{M}_4$. The compactification on $\mathcal{M}_2$ would give rise to a 4d supersymmetric gauge theory on $\mathcal{M}_4$ while compactifying on $\mathcal{M}_4$ yields some chiral algebra or 2d CFT on $\mathcal{M}_2$. The vertex operator algebra (VOA) is expected to arise from the COHA acting on the equivariant cohomology of the moduli space of instantons. For the simplest $\mathbb{C}^3$ example, one may consider three stacks of M5-branes on the three $\mathbb{C}^2$ divisors with a B-field. This would lead to a family of VOAs known as the $Y$-algebras parametrized by three parameters $L,M,N$ \cite{Gaiotto:2017euk}. Such algebras can be viewed as the truncations of $\mathcal{W}_{1+\infty}$ algebra, so it would be curious to see its connection to the truncations of quiver Yangians in \cite{Li:2020rij}. The generating function for the $Y$-algebras would now enumerate 3d partitions restricted between the usual octant for plane partitions and the octant with origin at $(L,M,N)$. It would be helpful to find such partition functions. For instance, the one for $Y_{0,0,N}$ in terms of PE is PE$\left[\frac{1+t^{N+1}}{(1-t)^2}\right]$. We might also compare the gluing process discussed here with the one in \cite{Prochazka:2017qum}. More generally, given a web diagram with $n$ faces for any toric CY, one can consider the VOAs labelled by $n$ parameters. They should have intimate relations with COHAs, brane tilings and quivers.

\section*{Acknowledgments}
We are grateful to Boris Pioline, Miroslav Rap\v{c}\'{a}k and Masahito Yamazaki for enlightening discussions. In particular, we are indebted to Ben Davison for invaluable comments. JB is supported by a CSC scholarship. YHH would like to thank STFC for grant ST/J00037X/1. The research of AZ has been supported by
the French “Investissements d’Avenir” program, project ISITE-BFC (No. ANR-15-IDEX-0003), and EIPHI Graduate School (No. ANR-17-EURE-0002).

\appendix
\section{The Crystal-to-BPS Map}\label{crystal2bps}
Roughly speaking, the partition functions for crystal and BPS states are related by a change of variables $q=\pm\prod\limits_{j=0}^{n-1}q_j$, $Q_i=\pm Q_i$ for $n=|\mathcal{Q}_0|$ and $i=1,\dots,n-1$. To determine these signs, we first introduce the (Euler-)Ringel form\footnote{Recall that $X_{ab}$ denotes an arrow from node $a$ to node $b$. Moreover, the node $a_0$ corresponding to the initial atom always uses the variable $q_0$.}
\begin{equation}
    \langle\bm{d}_1,\bm{d}_2\rangle=\sum_{a\in\mathcal{Q}_0}d_{1,a}d_{2,a}-\sum_{X_{ab}\in\mathcal{Q}_1}d_{1,a}d_{2,b}
\end{equation}
for the dimension vectors $\bm{d}_{1,2}$. Then the sign of the term $\bm{q}^{\bm{m}}=\prod\limits_{i=0}^{n-1}q_i^{m_i}$ is given by $(-1)^{m_0+\langle\bm{m},\bm{m}\rangle}$ \cite{mozgovoy2010noncommutative}.

In general, we need to check the signs term by term. However, for toric diagrams without internal points, we simply have
\begin{equation}
    (-1)^{m_0+\langle\bm{m},\bm{m}\rangle}=(-1)^{\sum\limits_{a\in S}m_a},
\end{equation}
where $S=\{a_0\}\sqcup\{a|\nexists~X_{aa}\in\mathcal{Q}_1\}$. Notice the disjoint union sign here. This means the initial node is counted twice, one from $\{a_0\}$ and one from $\{a|\nexists~X_{aa}\in\mathcal{Q}_1\}$ (if it does not have a loop). Therefore, the signs of variables can be determined as follows:
\begin{equation}\label{crystal2BPS}
    q_0=\begin{cases}
    p_0,&\nexists~X_{aa}\in\mathcal{Q}_1\\
    -p_0,&\exists~X_{aa}\in\mathcal{Q}_1
    \end{cases};\qquad
    q_{i\neq0}=\begin{cases}
    p_i,&\exists~X_{aa}\in\mathcal{Q}_1\\
    -p_i,&\nexists~X_{aa}\in\mathcal{Q}_1
    \end{cases}.
\end{equation}
We shall call this the crystal-to-BPS map. Then we can obtain $Z_\text{BPS}(q,\bm{Q})$ via $q=\prod\limits_{j=0}^{n-1}p_j$, $Q_i=p_j$. For generalized conifolds, the crystal-to-BPS map can equivalently be determined by the triangulations of the toric diagrams as discussed in \S\ref{genC}. For convenience,  especially when writing the sign-changed expressions, we have simply denoted the crystal-to-BPS map as $q_j\rightarrow\pm q_j$ in the main context, with the understanding of the signs according to \eqref{crystal2BPS}.

\section{Asymptotic Behaviour}\label{asymptotic}
Given an analytic function $h(t)=\sum\limits_{k=0}^\infty h_kt^k$,
and the generating function 
\begin{equation}
f(t)=\text{PE}[h(t)]
= \prod_{k=1}^\infty\frac{1}{(1-t^k)^{h_k}}=
\sum_{n=0}^\infty f_n t^n, 
\end{equation}
we can obtain the asymptotic behaviour of $f_n$ following \cite{meinardus1953asymptotische,haselgrove1954asymptotic,richmond1994some,Feng:2007ur}. We have the Dirichlet series $D(s):=\sum\limits_{k=1}^\infty\frac{h_k}{k^s}$ with only one simple pole at $s=\alpha>0$ and residue $A$. Then for large $n$,
\begin{equation}
    f_n\sim C_1n^{C_2}\exp\left(n^{\frac{\alpha}{\alpha+1}}\left(1+\frac{1}{\alpha}\right)(A\Gamma(\alpha+1)\zeta(\alpha+1))^{\frac{1}{\alpha+1}}\right),
\end{equation}
where
\begin{equation}
    C_1=\frac{\text{e}^{D'(0)}}{\sqrt{2\pi(\alpha+1)}}(A\Gamma(\alpha+1)\zeta(\alpha+1))^{\frac{1-2D(0)}{2(\alpha+1)}},\quad C_2=\frac{D(0)-1-\frac{\alpha}{2}}{\alpha+1}.
\end{equation}

As an example, let us consider the crystal model for the conifold in \S\ref{conifold} but without colouring. Therefore, we have a univariate function
\begin{equation}
    f(t)=\text{PE}\left[\frac{t(1+t+3t^2+4t^3+3t^4+t^5+t^6)}{(1-t^4)^2}\right].
\end{equation}
Hence,
\begin{equation}
    h_k=\begin{cases}
    \frac{k}{2}&,\quad k\equiv2~(\text{mod }4)\\
    k&,\quad\text{otherwise}
    \end{cases}.
\end{equation}
This yields the Dirichlet series
\begin{equation}
    D(s)=\frac{2-2^s+4^s}{4^s}\zeta(s-1),
\end{equation}
which has a pole at $s=\alpha=2$ with residue $A=7/8$. As a result, the asymptotic behaviour is
\begin{equation}
    f_n\sim\frac{(7\zeta(3))^{\frac{2}{9}}}{\sqrt{3\pi}}2^{-\frac{25}{36}}n^{-\frac{13}{18}}\exp\left(\frac{2}{3}(7\zeta(3))^{\frac{1}{3}}\left(\frac{n}{2}\right)^{\frac{2}{3}}+2\zeta'(-1)\right).
\end{equation}

\linespread{0.8}\selectfont
\addcontentsline{toc}{section}{References}
\bibliographystyle{utphys}
\bibliography{ref}

\providecommand{\href}[2]{#2}\begingroup\raggedright\begin{thebibliography}{100}

\bibitem{Bogomolny:1975de}
E.~B. Bogomolny, ``{Stability of Classical Solutions},'' {\em Sov. J. Nucl.
  Phys.} {\bfseries 24} (1976) 449.

\bibitem{Prasad:1975kr}
M.~K. Prasad and C.~M. Sommerfield, ``{An Exact Classical Solution for the 't
  Hooft Monopole and the Julia-Zee Dyon},''
  \href{http://dx.doi.org/10.1103/PhysRevLett.35.760}{{\em Phys. Rev. Lett.}
  {\bfseries 35} (1975) 760--762}.

\bibitem{Okounkov:2003sp}
A.~Okounkov, N.~Reshetikhin, and C.~Vafa, ``{Quantum Calabi-Yau and classical
  crystals},'' \href{http://dx.doi.org/10.1007/0-8176-4467-9_16}{{\em Prog.
  Math.} {\bfseries 244} (2006) 597},
  \href{http://arxiv.org/abs/hep-th/0309208}{{\ttfamily arXiv:hep-th/0309208}}.

\bibitem{Iqbal:2003ds}
A.~Iqbal, N.~Nekrasov, A.~Okounkov, and C.~Vafa, ``{Quantum foam and
  topological strings},''
  \href{http://dx.doi.org/10.1088/1126-6708/2008/04/011}{{\em JHEP} {\bfseries
  04} (2008) 011}, \href{http://arxiv.org/abs/hep-th/0312022}{{\ttfamily
  arXiv:hep-th/0312022}}.

\bibitem{Ooguri:2009ijd}
H.~Ooguri and M.~Yamazaki, ``{Crystal Melting and Toric Calabi-Yau
  Manifolds},'' \href{http://dx.doi.org/10.1007/s00220-009-0836-y}{{\em Commun.
  Math. Phys.} {\bfseries 292} (2009) 179--199},
  \href{http://arxiv.org/abs/0811.2801}{{\ttfamily arXiv:0811.2801 [hep-th]}}.

\bibitem{Yamazaki:2010fz}
M.~Yamazaki, ``{Crystal Melting and Wall Crossing Phenomena},''
  \href{http://dx.doi.org/10.1142/S0217751X11051482}{{\em Int. J. Mod. Phys. A}
  {\bfseries 26} (2011) 1097--1228},
  \href{http://arxiv.org/abs/1002.1709}{{\ttfamily arXiv:1002.1709 [hep-th]}}.

\bibitem{Dimofte:2010wxa}
T.~D. Dimofte, \href{http://dx.doi.org/10.7907/Q6WF-D678}{{\em {Refined BPS
  Invariants, Chern-Simons Theory, and the Quantum Dilogarithm}}}.
\newblock PhD thesis, Caltech, 2010.

\bibitem{Nakajima:1994nid}
H.~Nakajima, ``{Instantons on ALE spaces, quiver varieties, and Kac-Moody
  algebras},'' \href{http://dx.doi.org/10.1215/S0012-7094-94-07613-8}{{\em Duke
  Math. J.} {\bfseries 76} no.~2, (1994) 365--416}.

\bibitem{Douglas:1996sw}
M.~R. Douglas and G.~W. Moore, ``{D-branes, quivers, and ALE instantons},''
  \href{http://arxiv.org/abs/hep-th/9603167}{{\ttfamily arXiv:hep-th/9603167}}.

\bibitem{Kontsevich:2008fj}
M.~Kontsevich and Y.~Soibelman, ``{Stability structures, motivic
  Donaldson-Thomas invariants and cluster transformations},''
  \href{http://arxiv.org/abs/0811.2435}{{\ttfamily arXiv:0811.2435 [math.AG]}}.

\bibitem{kiem2012categorification}
Y.-H. Kiem and J.~Li, ``Categorification of donaldson-thomas invariants via
  perverse sheaves,'' \href{http://arxiv.org/abs/1212.6444}{{\ttfamily
  arXiv:1212.6444 [math.AG]}}.

\bibitem{Gaiotto:2015aoa}
D.~Gaiotto, G.~W. Moore, and E.~Witten, ``{Algebra of the Infrared: String
  Field Theoretic Structures in Massive ${\cal N}=(2,2)$ Field Theory In Two
  Dimensions},'' \href{http://arxiv.org/abs/1506.04087}{{\ttfamily
  arXiv:1506.04087 [hep-th]}}.

\bibitem{Gaiotto:2015zna}
D.~Gaiotto, G.~W. Moore, and E.~Witten, ``{An Introduction To The Web-Based
  Formalism},'' \href{http://arxiv.org/abs/1506.04086}{{\ttfamily
  arXiv:1506.04086 [hep-th]}}.

\bibitem{Hanany:2005ve}
A.~Hanany and K.~D. Kennaway, ``{Dimer models and toric diagrams},''
  \href{http://arxiv.org/abs/hep-th/0503149}{{\ttfamily arXiv:hep-th/0503149}}.

\bibitem{Kontsevich:2010px}
M.~Kontsevich and Y.~Soibelman, ``{Cohomological Hall algebra, exponential
  Hodge structures and motivic Donaldson-Thomas invariants},''
  \href{http://dx.doi.org/10.4310/CNTP.2011.v5.n2.a1}{{\em Commun. Num. Theor.
  Phys.} {\bfseries 5} (2011) 231--352},
  \href{http://arxiv.org/abs/1006.2706}{{\ttfamily arXiv:1006.2706 [math.AG]}}.

\bibitem{Li:2020rij}
W.~Li and M.~Yamazaki, ``{Quiver Yangian from Crystal Melting},''
  \href{http://dx.doi.org/10.1007/JHEP11(2020)035}{{\em JHEP} {\bfseries 11}
  (2020) 035}, \href{http://arxiv.org/abs/2003.08909}{{\ttfamily
  arXiv:2003.08909 [hep-th]}}.

\bibitem{Szendroi:2007nu}
B.~Szendroi, ``{Non-commutative Donaldson\textendash{}Thomas invariants and the
  conifold},'' \href{http://dx.doi.org/10.2140/gt.2008.12.1171}{{\em Geom.
  Topol.} {\bfseries 12} no.~2, (2008) 1171--1202},
  \href{http://arxiv.org/abs/0705.3419}{{\ttfamily arXiv:0705.3419 [math.AG]}}.

\bibitem{Young:2008hn}
B.~Young and J.~Bryan, ``{Generating functions for colored 3D Young diagrams
  and the Donaldson-Thomas invariants of orbifolds},''
  \href{http://dx.doi.org/10.1215/00127094-2010-009}{{\em Duke Math. J.}
  {\bfseries 152} (2010) 115--153},
  \href{http://arxiv.org/abs/0802.3948}{{\ttfamily arXiv:0802.3948 [math.CO]}}.

\bibitem{Cirafici:2010bd}
M.~Cirafici, A.~Sinkovics, and R.~J. Szabo, ``{Instantons, Quivers and
  Noncommutative Donaldson-Thomas Theory},''
  \href{http://dx.doi.org/10.1016/j.nuclphysb.2011.08.002}{{\em Nucl. Phys. B}
  {\bfseries 853} (2011) 508--605},
  \href{http://arxiv.org/abs/1012.2725}{{\ttfamily arXiv:1012.2725 [hep-th]}}.

\bibitem{Cirafici:2012qc}
M.~Cirafici and R.~J. Szabo, ``{Curve counting, instantons and McKay
  correspondences},''
  \href{http://dx.doi.org/10.1016/j.geomphys.2013.03.020}{{\em J. Geom. Phys.}
  {\bfseries 72} (2013) 54--109},
  \href{http://arxiv.org/abs/1209.1486}{{\ttfamily arXiv:1209.1486 [hep-th]}}.

\bibitem{Aganagic:2003db}
M.~Aganagic, A.~Klemm, M.~Marino, and C.~Vafa, ``{The Topological vertex},''
  \href{http://dx.doi.org/10.1007/s00220-004-1162-z}{{\em Commun. Math. Phys.}
  {\bfseries 254} (2005) 425--478},
  \href{http://arxiv.org/abs/hep-th/0305132}{{\ttfamily arXiv:hep-th/0305132}}.

\bibitem{Iqbal:2004ne}
A.~Iqbal and A.-K. Kashani-Poor, ``{The Vertex on a strip},''
  \href{http://dx.doi.org/10.4310/ATMP.2006.v10.n3.a2}{{\em Adv. Theor. Math.
  Phys.} {\bfseries 10} no.~3, (2006) 317--343},
  \href{http://arxiv.org/abs/hep-th/0410174}{{\ttfamily arXiv:hep-th/0410174}}.

\bibitem{Mozgovoy:2020has}
S.~Mozgovoy and B.~Pioline, ``{Attractor invariants, brane tilings and
  crystals},'' \href{http://arxiv.org/abs/2012.14358}{{\ttfamily
  arXiv:2012.14358 [hep-th]}}.

\bibitem{mozgovoy2021donaldson}
S.~Mozgovoy and M.~Reineke, ``Donaldson-thomas invariants for 3-calabi-yau
  varieties of dihedral quotient type,''
  \href{http://arxiv.org/abs/2104.13251}{{\ttfamily arXiv:2104.13251
  [math.AG]}}.

\bibitem{macmahon2001combinatory}
P.~A. MacMahon, {\em Combinatory Analysis, Volumes I and II}, vol.~137.
\newblock American Mathematical Soc., 2001.

\bibitem{Benvenuti:2006qr}
S.~Benvenuti, B.~Feng, A.~Hanany, and Y.-H. He, ``{Counting BPS Operators in
  Gauge Theories: Quivers, Syzygies and Plethystics},''
  \href{http://dx.doi.org/10.1088/1126-6708/2007/11/050}{{\em JHEP} {\bfseries
  11} (2007) 050}, \href{http://arxiv.org/abs/hep-th/0608050}{{\ttfamily
  arXiv:hep-th/0608050}}.

\bibitem{Feng:2007ur}
B.~Feng, A.~Hanany, and Y.-H. He, ``{Counting gauge invariants: The Plethystic
  program},'' \href{http://dx.doi.org/10.1088/1126-6708/2007/03/090}{{\em JHEP}
  {\bfseries 03} (2007) 090},
  \href{http://arxiv.org/abs/hep-th/0701063}{{\ttfamily arXiv:hep-th/0701063}}.

\bibitem{fulton2013representation}
W.~Fulton and J.~Harris, {\em Representation theory: a first course}, vol.~129.
\newblock Springer Science \& Business Media, 2013.

\bibitem{florentino2021plethystic}
C.~A. Florentino, ``Plethystic exponential calculus and characteristic
  polynomials of permutations,''
  \href{http://arxiv.org/abs/2105.13049}{{\ttfamily arXiv:2105.13049
  [math.CO]}}.

\bibitem{kac1980infinite}
V.~G. Kac, ``Infinite root systems, representations of graphs and invariant
  theory,'' {\em Inventiones mathematicae} {\bfseries 56} no.~1, (1980) 57--92.

\bibitem{Franco:2005rj}
S.~Franco, A.~Hanany, K.~D. Kennaway, D.~Vegh, and B.~Wecht, ``{Brane dimers
  and quiver gauge theories},''
  \href{http://dx.doi.org/10.1088/1126-6708/2006/01/096}{{\em JHEP} {\bfseries
  01} (2006) 096}, \href{http://arxiv.org/abs/hep-th/0504110}{{\ttfamily
  arXiv:hep-th/0504110}}.

\bibitem{Franco:2005sm}
S.~Franco, A.~Hanany, D.~Martelli, J.~Sparks, D.~Vegh, and B.~Wecht, ``{Gauge
  theories from toric geometry and brane tilings},''
  \href{http://dx.doi.org/10.1088/1126-6708/2006/01/128}{{\em JHEP} {\bfseries
  01} (2006) 128}, \href{http://arxiv.org/abs/hep-th/0505211}{{\ttfamily
  arXiv:hep-th/0505211}}.

\bibitem{Feng:2005gw}
B.~Feng, Y.-H. He, K.~D. Kennaway, and C.~Vafa, ``{Dimer models from mirror
  symmetry and quivering amoebae},''
  \href{http://dx.doi.org/10.4310/ATMP.2008.v12.n3.a2}{{\em Adv. Theor. Math.
  Phys.} {\bfseries 12} no.~3, (2008) 489--545},
  \href{http://arxiv.org/abs/hep-th/0511287}{{\ttfamily arXiv:hep-th/0511287}}.

\bibitem{Yamazaki:2008bt}
M.~Yamazaki, ``{Brane Tilings and Their Applications},''
  \href{http://dx.doi.org/10.1002/prop.200810536}{{\em Fortsch. Phys.}
  {\bfseries 56} (2008) 555--686},
  \href{http://arxiv.org/abs/0803.4474}{{\ttfamily arXiv:0803.4474 [hep-th]}}.

\bibitem{mozgovoy2010noncommutative}
S.~Mozgovoy and M.~Reineke, ``On the noncommutative donaldson--thomas
  invariants arising from brane tilings,'' {\em Advances in mathematics}
  {\bfseries 223} no.~5, (2010) 1521--1544,
  \href{http://arxiv.org/abs/0809.0117}{{\ttfamily arXiv:0809.0117 [math.AG]}}.

\bibitem{Ooguri:2009ri}
H.~Ooguri and M.~Yamazaki, ``{Emergent Calabi-Yau Geometry},''
  \href{http://dx.doi.org/10.1103/PhysRevLett.102.161601}{{\em Phys. Rev.
  Lett.} {\bfseries 102} (2009) 161601},
  \href{http://arxiv.org/abs/0902.3996}{{\ttfamily arXiv:0902.3996 [hep-th]}}.

\bibitem{Harvey:1996gc}
J.~A. Harvey and G.~W. Moore, ``{On the algebras of BPS states},''
  \href{http://dx.doi.org/10.1007/s002200050461}{{\em Commun. Math. Phys.}
  {\bfseries 197} (1998) 489--519},
  \href{http://arxiv.org/abs/hep-th/9609017}{{\ttfamily arXiv:hep-th/9609017}}.

\bibitem{schiffmann2013cherednik}
O.~Schiffmann and E.~Vasserot, ``Cherednik algebras, w-algebras and the
  equivariant cohomology of the moduli space of instantons on a 2,'' {\em
  Publications math{\'e}matiques de l'IH{\'E}S} {\bfseries 118} no.~1, (2013)
  213--342, \href{http://arxiv.org/abs/1202.2756}{{\ttfamily arXiv:1202.2756
  [math.QA]}}.

\bibitem{maulik2019quantum}
D.~Maulik and A.~Okounkov, ``Quantum groups and quantum cohomology,''
  \href{http://arxiv.org/abs/1211.1287}{{\ttfamily arXiv:1211.1287 [math.AG]}}.

\bibitem{Tsymbaliuk2017affine}
A.~Tsymbaliuk, ``The affine yangian of $\mathfrak{gl}_1$ revisited,''
  \href{http://dx.doi.org/10.1016/j.aim.2016.08.041}{{\em Advances in
  Mathematics} {\bfseries 304} (Jan, 2017) 583–645},
  \href{http://arxiv.org/abs/1404.5240}{{\ttfamily arXiv:1404.5240 [math.RT]}}.

\bibitem{Prochazka:2015deb}
T.~Proch\'azka, ``{$ \mathcal{W} $ -symmetry, topological vertex and affine
  Yangian},'' \href{http://dx.doi.org/10.1007/JHEP10(2016)077}{{\em JHEP}
  {\bfseries 10} (2016) 077}, \href{http://arxiv.org/abs/1512.07178}{{\ttfamily
  arXiv:1512.07178 [hep-th]}}.

\bibitem{Rapcak:2021hdh}
M.~Rapcak, ``{Branes, Quivers and BPS Algebras},''
  \href{http://arxiv.org/abs/2112.13878}{{\ttfamily arXiv:2112.13878
  [hep-th]}}.

\bibitem{Galakhov:2020vyb}
D.~Galakhov and M.~Yamazaki, ``{Quiver Yangian and Supersymmetric Quantum
  Mechanics},'' \href{http://arxiv.org/abs/2008.07006}{{\ttfamily
  arXiv:2008.07006 [hep-th]}}.

\bibitem{Rapcak:2018nsl}
M.~Rapcak, Y.~Soibelman, Y.~Yang, and G.~Zhao, ``{Cohomological Hall algebras,
  vertex algebras and instantons},''
  \href{http://dx.doi.org/10.1007/s00220-019-03575-5}{{\em Commun. Math. Phys.}
  {\bfseries 376} no.~3, (2019) 1803--1873},
  \href{http://arxiv.org/abs/1810.10402}{{\ttfamily arXiv:1810.10402
  [math.QA]}}.

\bibitem{bozec2017number}
T.~Bozec, O.~Schiffmann, and E.~Vasserot, ``On the number of points of
  nilpotent quiver varieties over finite fields,''
  \href{http://arxiv.org/abs/1701.01797}{{\ttfamily arXiv:1701.01797
  [math.RT]}}.

\bibitem{schiffmann2018kac}
O.~G. Schiffmann, ``Kac polynomials and lie algebras associated to quivers and
  curves,'' in {\em Proceedings of the International Congress of
  Mathematicians: Rio de Janeiro 2018}, pp.~1393--1424, World Scientific.
\newblock 2018.
\newblock \href{http://arxiv.org/abs/1802.09760}{{\ttfamily arXiv:1802.09760
  [math.RT]}}.

\bibitem{schiffmann2017cohomological}
O.~Schiffmann and E.~Vasserot, ``On cohomological hall algebras of quivers:
  Yangians,'' \href{http://arxiv.org/abs/1705.07491}{{\ttfamily
  arXiv:1705.07491 [math.RT]}}.

\bibitem{borel1960homology}
A.~Borel and J.~C. Moore, ``Homology theory for locally compact spaces.'' {\em
  Michigan Mathematical Journal} {\bfseries 7} no.~2, (1960) 137--159.

\bibitem{davison2016integrality}
B.~Davison, ``The integrality conjecture and the cohomology of preprojective
  stacks,'' \href{http://arxiv.org/abs/1602.02110}{{\ttfamily arXiv:1602.02110
  [math.AG]}}.

\bibitem{Behrend:2009dc}
K.~Behrend, J.~Bryan, and B.~Szendroi, ``{Motivic degree zero Donaldson-Thomas
  invariants},'' \href{http://arxiv.org/abs/0909.5088}{{\ttfamily
  arXiv:0909.5088 [math.AG]}}.

\bibitem{Davison:2013nza}
B.~Davison, ``{The critical CoHA of a quiver with potential},''
  \href{http://dx.doi.org/10.1093/qmath/haw053}{{\em Quart. J. Math. Oxford
  Ser.} {\bfseries 68} no.~2, (2017) 635--703},
  \href{http://arxiv.org/abs/1311.7172}{{\ttfamily arXiv:1311.7172 [math.AG]}}.

\bibitem{stanley1997enumerative}
R.~Stanley and S.~Fomin, {\em Enumerative Combinatorics: Volume 2}.
\newblock Cambridge Studies in Advanced Mathematics. Cambridge University
  Press, 1997.

\bibitem{wright1931asymptotic}
E.~Wright, ``Asymptotic partition formulaei. plane partitions,'' {\em The
  Quarterly Journal of Mathematics} no.~1, (1931) 177--189.

\bibitem{arnol1975critical}
V.~I. Arnol'd, ``Critical points of smooth functions and their normal forms,''
  {\em Russian Mathematical Surveys} {\bfseries 30} no.~5, (1975) 1.

\bibitem{bridgeland2001mckay}
T.~Bridgeland, A.~King, and M.~Reid, ``The mckay correspondence as an
  equivalence of derived categories,'' {\em Journal of the American
  Mathematical Society} {\bfseries 14} no.~3, (2001) 535--554,
  \href{http://arxiv.org/abs/math/9908027}{{\ttfamily arXiv:math/9908027}}.

\bibitem{kobayashi2013note}
M.~Kobayashi, M.~Mase, and K.~Ueda, ``A note on exceptional unimodal
  singularities and k3 surfaces,'' {\em International Mathematics Research
  Notices} {\bfseries 2013} no.~7, (2013) 1665--1690,
  \href{http://arxiv.org/abs/1107.2169}{{\ttfamily arXiv:1107.2169 [math.AG]}}.

\bibitem{He:2010mh}
Y.-H. He, ``{On Fields over Fields},''
  \href{http://arxiv.org/abs/1003.2986}{{\ttfamily arXiv:1003.2986 [hep-th]}}.

\bibitem{schiffmann2020cohomological}
O.~Schiffmann and E.~Vasserot, ``On cohomological hall algebras of quivers:
  generators,'' {\em Journal f{\"u}r die reine und angewandte Mathematik
  (Crelles Journal)} {\bfseries 2020} no.~760, (2020) 59--132,
  \href{http://arxiv.org/abs/1705.07488}{{\ttfamily arXiv:1705.07488
  [math.RT]}}.

\bibitem{young2009computing}
B.~Young, ``Computing a pyramid partition generating function with dimer
  shuffling,'' {\em Journal of Combinatorial Theory, Series A} {\bfseries 116}
  no.~2, (2009) 334--350, \href{http://arxiv.org/abs/0709.3079}{{\ttfamily
  arXiv:0709.3079 [math.CO]}}.

\bibitem{Davison:2018zyc}
B.~Davison, J.~Ongaro, and B.~Szendroi, ``{Enumerating coloured partitions in 2
  and 3 dimensions},'' \href{http://dx.doi.org/10.1017/S0305004119000252}{{\em
  Math. Proc. Cambridge Phil. Soc.} {\bfseries 169} no.~3, (2020) 479--505},
  \href{http://arxiv.org/abs/1811.12857}{{\ttfamily arXiv:1811.12857
  [math.AG]}}.

\bibitem{Gaberdiel:2017hcn}
M.~R. Gaberdiel, W.~Li, C.~Peng, and H.~Zhang, ``{The supersymmetric affine
  Yangian},'' \href{http://dx.doi.org/10.1007/JHEP05(2018)200}{{\em JHEP}
  {\bfseries 05} (2018) 200}, \href{http://arxiv.org/abs/1711.07449}{{\ttfamily
  arXiv:1711.07449 [hep-th]}}.

\bibitem{Gaberdiel:2018nbs}
M.~R. Gaberdiel, W.~Li, and C.~Peng, ``{Twin-plane-partitions and
  $\mathcal{N}=2$ affine Yangian},''
  \href{http://dx.doi.org/10.1007/JHEP11(2018)192}{{\em JHEP} {\bfseries 11}
  (2018) 192}, \href{http://arxiv.org/abs/1807.11304}{{\ttfamily
  arXiv:1807.11304 [hep-th]}}.

\bibitem{Li:2019nna}
W.~Li and P.~Longhi, ``{Gluing two affine Yangians of $\mathfrak{gl}_1$},''
  \href{http://dx.doi.org/10.1007/JHEP10(2019)131}{{\em JHEP} {\bfseries 10}
  (2019) 131}, \href{http://arxiv.org/abs/1905.03076}{{\ttfamily
  arXiv:1905.03076 [hep-th]}}.

\bibitem{Gaberdiel:2015wpo}
M.~R. Gaberdiel and R.~Gopakumar, ``{String Theory as a Higher Spin Theory},''
  \href{http://dx.doi.org/10.1007/JHEP09(2016)085}{{\em JHEP} {\bfseries 09}
  (2016) 085}, \href{http://arxiv.org/abs/1512.07237}{{\ttfamily
  arXiv:1512.07237 [hep-th]}}.

\bibitem{Katz:1996ht}
S.~H. Katz, D.~R. Morrison, and M.~R. Plesser, ``{Enhanced gauge symmetry in
  type II string theory},''
  \href{http://dx.doi.org/10.1016/0550-3213(96)00331-8}{{\em Nucl. Phys. B}
  {\bfseries 477} (1996) 105--140},
  \href{http://arxiv.org/abs/hep-th/9601108}{{\ttfamily arXiv:hep-th/9601108}}.

\bibitem{Aganagic:2010qr}
M.~Aganagic and K.~Schaeffer, ``{Wall Crossing, Quivers and Crystals},''
  \href{http://dx.doi.org/10.1007/JHEP10(2012)153}{{\em JHEP} {\bfseries 10}
  (2012) 153}, \href{http://arxiv.org/abs/1006.2113}{{\ttfamily arXiv:1006.2113
  [hep-th]}}.

\bibitem{nagao2008derived}
K.~Nagao, ``Derived categories of small toric calabi-yau 3-folds and counting
  invariants,'' \href{http://arxiv.org/abs/0809.2994}{{\ttfamily
  arXiv:0809.2994 [math.AG]}}.

\bibitem{Nagao:2009rq}
K.~Nagao and M.~Yamazaki, ``{The Non-commutative Topological Vertex and Wall
  Crossing Phenomena},''
  \href{http://dx.doi.org/10.4310/ATMP.2010.v14.n4.a3}{{\em Adv. Theor. Math.
  Phys.} {\bfseries 14} no.~4, (2010) 1147--1181},
  \href{http://arxiv.org/abs/0910.5479}{{\ttfamily arXiv:0910.5479 [hep-th]}}.

\bibitem{Benini:2009gi}
F.~Benini, S.~Benvenuti, and Y.~Tachikawa, ``{Webs of five-branes and N=2
  superconformal field theories},''
  \href{http://dx.doi.org/10.1088/1126-6708/2009/09/052}{{\em JHEP} {\bfseries
  09} (2009) 052}, \href{http://arxiv.org/abs/0906.0359}{{\ttfamily
  arXiv:0906.0359 [hep-th]}}.

\bibitem{Gaiotto:2009we}
D.~Gaiotto, ``{N=2 dualities},''
  \href{http://dx.doi.org/10.1007/JHEP08(2012)034}{{\em JHEP} {\bfseries 08}
  (2012) 034}, \href{http://arxiv.org/abs/0904.2715}{{\ttfamily arXiv:0904.2715
  [hep-th]}}.

\bibitem{Acharya:2021jsp}
B.~Acharya, N.~Lambert, M.~Najjar, E.~E. Svanes, and J.~Tian, ``{Gauging
  Discrete Symmetries of $T_N$-theories in Five Dimensions},''
  \href{http://arxiv.org/abs/2110.14441}{{\ttfamily arXiv:2110.14441
  [hep-th]}}.

\bibitem{Lawrence:1998ja}
A.~E. Lawrence, N.~Nekrasov, and C.~Vafa, ``{On conformal field theories in
  four-dimensions},''
  \href{http://dx.doi.org/10.1016/S0550-3213(98)00495-7}{{\em Nucl. Phys. B}
  {\bfseries 533} (1998) 199--209},
  \href{http://arxiv.org/abs/hep-th/9803015}{{\ttfamily arXiv:hep-th/9803015}}.

\bibitem{kac1990infinite}
V.~G. Kac, {\em Infinite-dimensional Lie algebras}.
\newblock Cambridge university press, 1990.

\bibitem{gholampour2009counting}
A.~Gholampour and Y.~Jiang, ``{Counting invariants for the ADE McKay
  quivers},'' \href{http://arxiv.org/abs/0910.5551}{{\ttfamily arXiv:0910.5551
  [math.AG]}}.

\bibitem{Mozgovoy:2011ps}
S.~Mozgovoy, ``{Motivic Donaldson-Thomas invariants and McKay
  correspondence},'' \href{http://arxiv.org/abs/1107.6044}{{\ttfamily
  arXiv:1107.6044 [math.AG]}}.

\bibitem{Aganagic:2009kf}
M.~Aganagic, H.~Ooguri, C.~Vafa, and M.~Yamazaki, ``{Wall Crossing and
  M-theory},'' {\em Publ. Res. Inst. Math. Sci. Kyoto} {\bfseries 47} (2011)
  569, \href{http://arxiv.org/abs/0908.1194}{{\ttfamily arXiv:0908.1194
  [hep-th]}}.

\bibitem{Chuang:2009crq}
W.-y. Chuang and D.~L. Jafferis, ``{Wall Crossing of BPS States on the Conifold
  from Seiberg Duality and Pyramid Partitions},''
  \href{http://dx.doi.org/10.1007/s00220-009-0832-2}{{\em Commun. Math. Phys.}
  {\bfseries 292} (2009) 285--301},
  \href{http://arxiv.org/abs/0810.5072}{{\ttfamily arXiv:0810.5072 [hep-th]}}.

\bibitem{Kenyon:2003uj}
R.~Kenyon, A.~Okounkov, and S.~Sheffield, ``{Dimers and amoebae},''
  \href{http://arxiv.org/abs/math-ph/0311005}{{\ttfamily
  arXiv:math-ph/0311005}}.

\bibitem{Iqbal:2007ii}
A.~Iqbal, C.~Kozcaz, and C.~Vafa, ``{The Refined topological vertex},''
  \href{http://dx.doi.org/10.1088/1126-6708/2009/10/069}{{\em JHEP} {\bfseries
  10} (2009) 069}, \href{http://arxiv.org/abs/hep-th/0701156}{{\ttfamily
  arXiv:hep-th/0701156}}.

\bibitem{Taki:2007dh}
M.~Taki, ``{Refined Topological Vertex and Instanton Counting},''
  \href{http://dx.doi.org/10.1088/1126-6708/2008/03/048}{{\em JHEP} {\bfseries
  03} (2008) 048}, \href{http://arxiv.org/abs/0710.1776}{{\ttfamily
  arXiv:0710.1776 [hep-th]}}.

\bibitem{Nishinaka:2010qk}
T.~Nishinaka and S.~Yamaguchi, ``{Wall-crossing of D4-D2-D0 and flop of the
  conifold},'' \href{http://dx.doi.org/10.1007/JHEP09(2010)026}{{\em JHEP}
  {\bfseries 09} (2010) 026}, \href{http://arxiv.org/abs/1007.2731}{{\ttfamily
  arXiv:1007.2731 [hep-th]}}.

\bibitem{Nishinaka:2010fh}
T.~Nishinaka, ``{Multiple D4-D2-D0 on the Conifold and Wall-crossing with the
  Flop},'' \href{http://dx.doi.org/10.1007/JHEP06(2011)065}{{\em JHEP}
  {\bfseries 06} (2011) 065}, \href{http://arxiv.org/abs/1010.6002}{{\ttfamily
  arXiv:1010.6002 [hep-th]}}.

\bibitem{Nishinaka:2011sv}
T.~Nishinaka and S.~Yamaguchi, ``{Statistical model and BPS D4-D2-D0
  counting},'' \href{http://dx.doi.org/10.1007/JHEP05(2011)072}{{\em JHEP}
  {\bfseries 05} (2011) 072}, \href{http://arxiv.org/abs/1102.2992}{{\ttfamily
  arXiv:1102.2992 [hep-th]}}.

\bibitem{Nishinaka:2011is}
T.~Nishinaka and Y.~Yoshida, ``{A Note on statistical model for BPS D4-D2-D0
  states},'' \href{http://dx.doi.org/10.1016/j.physletb.2012.03.071}{{\em Phys.
  Lett. B} {\bfseries 711} (2012) 132--138},
  \href{http://arxiv.org/abs/1108.4326}{{\ttfamily arXiv:1108.4326 [hep-th]}}.

\bibitem{Nishinaka:2013mba}
T.~Nishinaka, S.~Yamaguchi, and Y.~Yoshida, ``{Two-dimensional crystal melting
  and D4-D2-D0 on toric Calabi-Yau singularities},''
  \href{http://dx.doi.org/10.1007/JHEP05(2014)139}{{\em JHEP} {\bfseries 05}
  (2014) 139}, \href{http://arxiv.org/abs/1304.6724}{{\ttfamily arXiv:1304.6724
  [hep-th]}}.

\bibitem{Gholampour:2013ifa}
A.~Gholampour, A.~Sheshmani, and R.~Thomas, ``{Counting curves on surfaces in
  Calabi-Yau 3-folds},''
  \href{http://dx.doi.org/10.1007/s00208-014-1035-5}{{\em Math. Ann.}
  {\bfseries 360} (2014) 67--78},
  \href{http://arxiv.org/abs/1309.0051}{{\ttfamily arXiv:1309.0051 [math.AG]}}.

\bibitem{Descombes:2021snc}
P.~Descombes, ``{Cohomological DT invariants from localization},''
  \href{http://arxiv.org/abs/2106.02518}{{\ttfamily arXiv:2106.02518
  [math.AG]}}.

\bibitem{Joyce:2008pc}
D.~Joyce and Y.~Song, ``{A Theory of generalized Donaldson-Thomas
  invariants},'' \href{http://arxiv.org/abs/0810.5645}{{\ttfamily
  arXiv:0810.5645 [math.AG]}}.

\bibitem{Davison:2016bjk}
B.~Davison and S.~Meinhardt, ``{Cohomological Donaldson-Thomas theory of a
  quiver with potential and quantum enveloping algebras},''
  \href{http://arxiv.org/abs/1601.02479}{{\ttfamily arXiv:1601.02479
  [math.RT]}}.

\bibitem{Rapcak:2020ueh}
M.~Rapcak, Y.~Soibelman, Y.~Yang, and G.~Zhao, ``{Cohomological Hall algebras
  and perverse coherent sheaves on toric Calabi-Yau 3-folds},''
  \href{http://arxiv.org/abs/2007.13365}{{\ttfamily arXiv:2007.13365
  [math.QA]}}.

\bibitem{Galakhov:2021xum}
D.~Galakhov, W.~Li, and M.~Yamazaki, ``{Shifted quiver Yangians and
  representations from BPS crystals},''
  \href{http://dx.doi.org/10.1007/JHEP08(2021)146}{{\em JHEP} {\bfseries 08}
  (2021) 146}, \href{http://arxiv.org/abs/2106.01230}{{\ttfamily
  arXiv:2106.01230 [hep-th]}}.

\bibitem{Galakhov:2021vbo}
D.~Galakhov, W.~Li, and M.~Yamazaki, ``{Toroidal and Elliptic Quiver BPS
  Algebras and Beyond},'' \href{http://arxiv.org/abs/2108.10286}{{\ttfamily
  arXiv:2108.10286 [hep-th]}}.

\bibitem{Noshita:2021dgj}
G.~Noshita and A.~Watanabe, ``{Shifted Quiver Quantum Toroidal Algebra and
  Subcrystal Representations},''
  \href{http://arxiv.org/abs/2109.02045}{{\ttfamily arXiv:2109.02045
  [hep-th]}}.

\bibitem{Alday:2009aq}
L.~F. Alday, D.~Gaiotto, and Y.~Tachikawa, ``{Liouville Correlation Functions
  from Four-dimensional Gauge Theories},''
  \href{http://dx.doi.org/10.1007/s11005-010-0369-5}{{\em Lett. Math. Phys.}
  {\bfseries 91} (2010) 167--197},
  \href{http://arxiv.org/abs/0906.3219}{{\ttfamily arXiv:0906.3219 [hep-th]}}.

\bibitem{Nekrasov:2015wsu}
N.~Nekrasov, ``{BPS/CFT correspondence: non-perturbative Dyson-Schwinger
  equations and qq-characters},''
  \href{http://dx.doi.org/10.1007/JHEP03(2016)181}{{\em JHEP} {\bfseries 03}
  (2016) 181}, \href{http://arxiv.org/abs/1512.05388}{{\ttfamily
  arXiv:1512.05388 [hep-th]}}.

\bibitem{Feigin:2018bkf}
B.~Feigin and S.~Gukov, ``{VOA[$M_4$]},''
  \href{http://dx.doi.org/10.1063/1.5100059}{{\em J. Math. Phys.} {\bfseries
  61} no.~1, (2020) 012302}, \href{http://arxiv.org/abs/1806.02470}{{\ttfamily
  arXiv:1806.02470 [hep-th]}}.

\bibitem{Gaiotto:2017euk}
D.~Gaiotto and M.~Rap\v{c}\'ak, ``{Vertex Algebras at the Corner},''
  \href{http://dx.doi.org/10.1007/JHEP01(2019)160}{{\em JHEP} {\bfseries 01}
  (2019) 160}, \href{http://arxiv.org/abs/1703.00982}{{\ttfamily
  arXiv:1703.00982 [hep-th]}}.

\bibitem{Prochazka:2017qum}
T.~Proch\'azka and M.~Rap\v{c}\'ak, ``{Webs of W-algebras},''
  \href{http://dx.doi.org/10.1007/JHEP11(2018)109}{{\em JHEP} {\bfseries 11}
  (2018) 109}, \href{http://arxiv.org/abs/1711.06888}{{\ttfamily
  arXiv:1711.06888 [hep-th]}}.

\bibitem{meinardus1953asymptotische}
G.~Meinardus, ``Asymptotische aussagen {\"u}ber partitionen,'' {\em
  Mathematische Zeitschrift} {\bfseries 59} no.~1, (1953) 388--398.

\bibitem{haselgrove1954asymptotic}
C.~Haselgrove and H.~Temperley, ``Asymptotic formulae in the theory of
  partitions,'' in {\em Mathematical Proceedings of the Cambridge Philosophical
  Society}, vol.~50, pp.~225--241, Cambridge University Press.
\newblock 1954.

\bibitem{richmond1994some}
L.~B. Richmond, ``Some general problems on the number of parts in partitions,''
  {\em Acta Arithmetica} {\bfseries 66} no.~4, (1994) 297--313.

\end{thebibliography}\endgroup

\end{document}